\documentclass[iop,]{emulateapj}
\usepackage{lscape}

\usepackage{graphicx}

\newcommand{\flux}{erg~cm$^{-2}$s$^{-1}$}
\newcommand{\lx}{erg~s$^{-1}$}
\newcommand{\nh}{cm$^{-2}$}

\slugcomment{Published in The Astrophysical Journal, Volume 716, Issue 2, pp. 1217-1240 (2010).}

\shorttitle{SMC Deep Fields}
\shortauthors{Laycock et al.}

\begin{document}


\title{Exploring the Small Magellanic Cloud to the Faintest X-ray Fluxes: \\
Source Catalog, Timing and Spectral Analysis}


\author{Silas Laycock \altaffilmark{1}, Andreas Zezas \altaffilmark{2,3,4}, Jaesub Hong\altaffilmark{2}, Jeremy J. Drake\altaffilmark{2}, Valsamo Antoniou\altaffilmark{2,3}}
\altaffiltext{1}{Gemini Observatory, 670 North A'ohoku Place, Hilo, HI, 96720, USA}
\altaffiltext{2}{Harvard-Smithsonian Center for Astrophysics, 60 Garden St, Cambridge, MA, 02138, USA}
\altaffiltext{3}{Physics Department, University of Crete, 71003, Heraklion, Greece} 
\altaffiltext{4}{IESL, Foundation for Research and Technology, 71110, Heraklion, Greece}



\begin{abstract} 
We present the results of a pair of 100 ksec {\it Chandra} observations in the Small Magellanic Cloud to survey high mass X-ray binaries (HMXBs), stars and LMXBs/CVs down to $L_x$ = 4.3$\times$ 10$^{32}$ \lx. The two SMC Deep Fields are located in the most active star forming region of the bar, with Deep Field-1 positioned at the most pulsar-rich location identified from previous surveys. Two new pulsars were discovered in outburst: CXOU J004929.7-731058 (P=892s),  CXOU J005252.2-721715  (P=326s), and 3 new HMXB candidates were identified. Of 15 Be-pulsars now known in the field, 13 were detected, with pulsations seen in 9 of them. Ephemerides demonstrate that 6 of the 10 pulsars known to exhibit regular outbursts were seen outside their periastron phase, and quiescent X-ray emission at $L_X=$10$^{33-34}$\lx is shown to be common. Comparison with ROSAT {\it ASCA} and {\it XMM-Newton} catalogs resulted in positive identification of several previously ambiguous sources. Bright optical counterparts exist for 40 of the X-ray sources, of which 33 are consistent with early-type stars ($M_{V}<$-2, $B-V<$0.2), and are the subject of a companion paper. The results point to an underlying HMXB population-density up to double that of active systems. The full catalog of 394 point-sources is presented along with detailed analyses of timing and spectral properties. 
\end {abstract}


\section{Introduction}

The Small Magellanic Cloud (SMC) has proven to be an incomparable laboratory for astrophysics. The SMC is one of the dwarf irregular satellites of the Milk Way, and unlike most other members of that group (excepting the LMC) is experiencing an era of intense star  formation. The star formation rate (SFR) of the SMC is in the range 0.05-0.4$M_{\odot}$~yr$^{-1}$ \citep{shtykovskiy2005}, where the upper bound is about 150 times greater than for the Milky Way.  Situated in a part of the sky unobstructed by the galactic plane, the SMC affords a ringside seat to the unfolding drama. The combination of low extinction ($N_{H}\sim$10$^{21} cm^{-2}$) and a small depth/size in relation to its distance from earth ($D_{SMC}$=60 kpc, \citealt{hilditch2005}) effectively puts the entire population at a common distance. A substantial fraction of the SMC can be observed simultaneously thanks to its compact size, which facilitates population studies.  The recent star-formation episode in the SMC has led to a large population of massive stars, and high mass X-ray binaries (HMXB) which are the relics of the short-lived upper end of the initial mass function. The young population, and the majority of the known X-ray binaries, are concentrated in the SMC's bar.

HMXB pulsars are rotating neutron stars in binary systems with Be-type (spectral type 09-B2, luminosity class V-III) or super-giant stellar companions. Most HMXBs are of the Be type which account for 70\% in the Milky way and $\geq98$\% in the SMC \citep{coe2005}. The Be-star equatorial disk provides a reservoir of matter that can be accreted onto the neutron star during periastron passage (most known systems have large orbital eccentricity) or during large-scale disk ejection episodes. This scenario leads to strings of X-ray outbursts with typical luminosities $L_{X}\sim$10$^{36-37}$\lx, spaced at the orbital period, plus infrequent giant outbursts of greater duration and luminosity (see \citealt{negueruela1998} for a review).

Monitoring surveys of the SMC with e.g.  {\it RXTE} (\citealt{laycock2005} --L05, \citealt{galache2008} --G08) see X-ray pulsars in outburst at $>$10$^{36}$\lx and have counted $\sim$50 to date. The above scenario suggests most X-ray binary pulsars spend the majority of the time in quiescence. In order to enumerate the overall population, it is necessary to observe and count X-ray pulsars in their low luminosity state as well. The {\it ROSAT} and {\it ASCA} missions detected many faint X-ray point sources (e.g. \citealt{hs2000}), but the typical positional uncertainties frequently made positive identification difficult. Recent studies using {\it XMM-Newton} (\citealt{haberl2008}, \citealt{hp2004}) and {\it Chandra} (\citealt{antoniou2009}, \citealt{edge2004}) have now pinpointed the optical counterparts of many of these HMXBs.  The literature contains measurements of many previously unidentified X-ray sources (e.g from {\it ROSAT} and ASCA) that can now be localized, to provide a valuable historical record of variability which is a hallmark of X-ray binaries, even in the absence of pulsations. For this study we targeted the most densely populated regions of the SMC Bar with sufficient sensitivity (10$^{33}$\lx) to detect quiescent HMXB systems, with good positional accuracy. 
 
X-ray binaries with large circular orbits, and those containing fast-rotating neutron stars, do not fit the classic pattern of periodic activity, due to the generally low luminosity of the former and the infrequent occurrence of outbursts in the latter.  In systems with low eccentricity, no periastron-passage events can occur and the Be-star disk remains tidally truncated well within the orbital radius of the neutron-star \citep{okazaki2001}. It is widely recognized that accretion onto the magnetic neutron-star faces a centrifugal barrier that scales rapidly with spin frequency (see e.g. \citealt{frank}). Fast rotating pulsars remain in the so-called propellor regime \citep{sunyaev} until the mass transfer rate from their companion exceeds the barrier or threshold value. For spin periods below $\sim$10s, the resulting accretion-powered luminosity is $>$10$^{38}$\lx and such outbursts occur only rarely.

The primary aim of our deep (100 ksec) {Chandra} survey is to provide a complete census of X-ray binaries (XRBs), with greatly reduced luminosity bias compared to the existing sample within a limited survey region. Stellar population synthesis models will greatly benefit from hard numbers upon which to base the relative probabilities of various evolutionary paths. For example in modelling the supernovae that occur in formation of XRBs, one needs to reproduce the actual distribution of orbital periods and eccentricities. If current catalogs under represent systems with large orbital separations and low eccentricity the input to these models will be flawed.

A second motivation of the Deep Fields project is to  investigate the apparently universal correlation between star formation rate (SFR) and the integrated luminosity of HMXB found by \cite{grimm2003}. The absolute number normalization of this relationship remains unknown for two reasons. Sufficiently faint fluxes have not been probed to discover the underlying numbers of HMXBs, and the transient nature of most HMXBs makes it difficult to determine the total population. \cite{shtykovskiy2005} found that although the number of HMXB in the SMC fits the \cite{grimm2003} relation based on optical estimates of SFR, if $H\alpha$, FIR and UV indicators are used instead, HMXBs are overabundant by up to a factor of 10 compared to the Milky Way. 

We present the complete X-ray source catalog as an electronic table accompanying this paper. We describe below in detail the properties of the brightest sources, and an analysis of the timing and spectral properties of the sample, focussing on X-ray pulsars and sources coincident with known objects. An optical survey is being conducted in parallel (\citealt{antoniou2010}, in preparation) to enable more robust and complete classifications of X-ray binaries, cataclysmic variables (CVs), stellar coronae and background active galactic nuclei (AGN).  The further  goals of this project are as follows, and are being pursued in a series of forthcoming papers:

(1) Classify all the sources by finding and studying their optical counterparts. Discriminating the different source classes is a requisite for aims 3-5 below. Our optical survey is being carried out in parallel with the X-ray survey using the Hubble Space Telescope and Magellan 6.5m telescope. With high resolution imaging/photometry to R,I$<$26 we will be able to independently classify the optical counterparts to most of the sample.  The optical counterpart identifications will enable items 2-4 below. 

(2) Construct the luminosity distribution down to faint X-ray fluxes, this will test whether multiple point source populations may be present, vs a monolithic HMXB ensemble.

(3) Search for coronal X-ray emission from the most active stars in the SMC.

(4) Constrain the incidence of LMXBs and CVs in the SMC. CVs should be present throughout the SMC which is dominated by an old population of red giants. Outbursts (Novae and dwarf novae) are extremely infrequent and fainter in X-rays than HMXB. Thus they will be harder to find.  No LMXBs or CVs are known in the SMC, and extrapolation from the total mass in stars predicts only a handful of LMXBs, thus an
observational constraint (or positive discoveries of LMXB)  will provide important input to population synthesis models.

\section{Observations and Data Reduction}
The two SMC fields (Table~\ref{tab:obs}) are located in the most active star forming region of the Bar. Each field was observed for a total of 100 ksec with {\it Chandra}'s Advanced CCD Imaging Spectrometer (ACIS), as detailed in Table~\ref{tab:obs}. For mission operational reasons the observations could not be scheduled as contiguous blocks, a circumstance anticipated in advance. DF1 was observed in two observations of $\sim$ 50 ksec each, over the course of two days. DF2 was split into 3 observations spread over three days. The spacecraft aim-point and roll-angle were identical for the observations making up each field, so as to simplify the data reduction. The ACIS-I array was selected for its large (16$\times$16') contiguous square field imaged by 4 front-illuminated CCDs (I0-I3). Two of the "S" array CCDs (S5, S6) were also active. Timed exposure, faint-mode was used with the default "most-efficient" frame time of 3.2sec, and the unfiltered energy range (0.1-15 keV).

Deep Field-1 (DF1) is  positioned at the most pulsar-rich location identified from previous surveys. The DF1 coordinates also coincide with the primary position used by the {\it RXTE} monitoring program (L05), for weekly pulsar observations of 5-10 ksec duration for the past 10 years, DF2 was also frequently observed by {\it RXTE}.  Figure~\ref{fig:surveyfields} shows all {\it Chandra} observations in the SMC Bar, which was targeted by an earlier shallow survey (\citealt{taylor2004}, \citealt{antoniou2009}).  DF1 and DF2 were previously observed as Field-5  and Field-4 of the shallow survey. The earlier observations were 10 ksec ACIS-I observations which discovered the pulsars SXP503 and SXP138 \citep{edge2004}, and a bright supernova remnant \citep{taylor2004} coincident with the {\it ROSAT} source PSPC 461= HRI 38 \citep{sasaki2000}. 

The data were stacked and reduced using the processing pipeline developed at CfA for the {\it Chandra} Multi-wavelength Project (ChaMP, \citealt{kim2004} and ChaMPlane, \citealt{hong2005}). The pipeline performs all standard {\it Chandra} calibrations and corrections including GTI filtering, exposure map correction, cosmic-ray afterglow removal, aspect solution, charge-transfer inefficiency, effective area, and time-dependent energy response (due to contaminant buildup on the ACIS optical blocking filter).  Source detection was performed with the {\it wavdetect} algorithm, on images extracted in three energy bands (Bx=0.3-8 keV, Sx=0.3-2.5 keV, Hx=2.5-8 keV) at spatial scales of 1, 2, 4, 8, 16 pixels. 
X-ray source properties were measured using the aperture photometry procedure described by  \cite{hong2005}, which accounts for neighboring sources with overlapping source or background regions. The full position-dependent instrument sensitivity is accounted for by application of the exposure maps and response matrices (RMFs and ARFs) generated by the ChaMP pipeline.  Positional error circles are assigned to each source based on net counts and distance from the aim-point, according to the standard model of \cite{hong2005} which is based on detailed simulations of the full {\it Chandra} optical path and ACIS characteristics. Source detection and photometry were performed both on the individual exposures in Table~\ref {tab:obs} and the complete stacked data-set for each field.  The resulting object lists were then combined to yield a catalog of unique sources, with the best measurement of each source used to define its coordinates and error circle. In nearly all cases the stacked Bx image provided the best measurement with the exception of variable sources which may be detected at better S/N in a particular observation, and particularly hard or soft sources which may achieve the best S/N in the Sx or Hx images. 

We present the full list of detected sources on the ACIS-I array in Table~\ref{tab:catalog} of the online edition of the Journal. The print edition contains a subset of sources ranked by signal-to-noise ratio (S/N), cut to exclude sources with  S/N$<$10. The catalog lists co-ordinates, error-radii (95\% statistical error), X-ray fluxes in 3 conventional energy bands ($Bc$ 0.5-8.0 keV,  $Sc$ 0.5-2.0 keV, $Hc$ 2.0-8.0 keV), and energy Quantiles.  The energy bands are chosen to match the convention used for ROSAT, and quantiles were generated according to the definitions of \cite{hong2004}.  Background subtraction is applied to each source, to compute net-counts, by subtracting the count rate in an annular sky region surrounding the source extraction region. The aperture photometry routine uses extraction regions tied to the (spatially varying) 90\% encircled energy radius, which results in slightly different count rates than are reported by wavedetect.    

The flux values in Table~\ref{tab:catalog} are generated using a varying spectral model, whose parameters are selected according to the quantile value for each source. This approach is introduced by \cite{hong2009} and frees the flux estimation from systematic errors inherent in assuming a single spectral model for all sources. The electronic version of Table~\ref{tab:catalog} also contains flux values derived for fixed power-law ($\Gamma$=1.4) and thermal plasma ($kT$=1 keV) spectral models. The quantile based fluxes are derived from one of 3 power law models that are defined by the natural groupings of sources in the quantile diagram (C.F. Section~\ref{sect:quantiles}). The group to which each source is assigned for flux calculation is given in Table~\ref{tab:catalog}. For field DF1 the models are: group 1: $N_H$=2.8, $\Gamma$=2.35, group 2: $N_H$=2.5, $\Gamma$=1.42, group 3: $N_H$=20.0, $\Gamma$=1.52  (absorption column density in units of 10$^{21}$\nh). For field DF2 the models are similar, group 1: NH=0.6, $\Gamma$=2.94, group 2: $N_H$=4.9, $\Gamma$=1.38, group 3: $N_H$=27.0, $\Gamma$=1.91. In both fields the algorithm identifies quantitatively similar groupings, corresponding to a very soft spectrum with typical (or sub) SMC absorption (group 1), a hard XRB-like spectrum with SMC absorption (group 2), and a heavily absorbed AGN-like spectrum (group 3). The 2 fixed models are provided as they will be more accurate if the nature of an individual source is known by other means.

In Figures~\ref{fig:imageDF1} \& ~\ref{fig:imageDF2}, we present adaptively smoothed images of the stacked data for the two fields. After creating co-added, filtered level-2 event files the broad-band (0.3-8 keV) data were adaptively smoothed with the CIAO {\it csmooth} tool with the significance scale set to 3-5$\sigma$ (resulting in minimum and maximum scales for the smoothing kernel of 1-31 pixels).  On these images we overlay the detected HMXB candidates as well as previously known HMXB.

\section{Timing Analysis}
\label{sect:timing}
Light-curves were extracted in units of net count-rate (i.e background corrected counts per second) using good events from source and background regions of every sufficiently bright object detected.  The lightcurves were extracted following the usual CIAO analysis thread with the tool {\it dmextract} at the intrinsic ACIS frame-time resolution of 3.2s.
Power density spectra (PDS) were computed using the Lomb-Scargle (LS) method \citep{scargle1982} and all periods above 90\% significance were flagged for further investigation. We analyzed both the complete light-curves as well as the 2-3 individual segments per field listed in Table~\ref{tab:obs}. The number of counts needed to unambiguously detect a pulse period depends on the ratio of period to observation duration, sampling interval,  pulse fraction and pulse profile. Taking an optimistic cut at
 $>$100 net counts in 100 ksec, provides a lower threshold of approximately $10^{34}$ \lx for detecting periodicities in SMC pulsars. The threshold for detecting a point-source without regard to timing properties is about ten times fainter. For comparison the pulsation detection threshold (90\%) for the {\it RXTE} SMC pulsar monitoring program is $\gtrsim$10$^{36}$ \lx, and {\it only} pulsed sources can be detected.

The periodogram analysis was conducted as a blind-search (period search range 6-10,000s) on every source with more than 50 net counts per lightcurve segment, or 100 net counts for the stacked data. DF1 and DF2 contained 19 and 14 sources respectively with $>$100 net counts. Spurious periods are expected near the {\em Chandra} dither periods (1000s, 707s), their harmonics (principally 500s, 353s) and the ACIS sampling period (3.2s). Timing artifacts are only expected for sources that fall close to a CCD edge or node boundary. In this situation the dither offsets can periodically move all or part of the source region on and off the detector, resulting in a modulation of the count-rate. Examples of dither artifact can be seen in the PDS of several sources and are identified in the captions and/or accompanying text.  Significance levels for the periodograms were computed for all lightcurves  according to the prescription of \citealt{press1992}. A Monte Carlo approach was used to verify the significance levels in several cases noted in the discussion below. At each iteration the flux values were randomly redistributed among the (unmodified) time-bins and the periodogram computed. After 1000 iterations we constructed the cumulative distribution of the maximum spectral power, in order to to obtain the 95$^{th}$ percentile. This method accounts for the true distribution of flux values and the sampling pattern (e.g. gaps in the lightcurve).  

In order to validate the results of the blind period search by the
LS algorithm, we have also employed three other complimentary
techniques: epoch folding \citep{leahy1983}, $Z^2$ or Rayleigh
statistic \citep{buccheri1983}, and multi-Harmonic Periodogram (MHP,
\citealt{SZ}).  For epoch folding, we accumulate both
source and background photons in 10 phase bins at each trial period,
and calculate the $\chi^2$ statistic of the background subtracted
folded-light curve with respect to the assumed constant rate.  In our implementation of the $Z^2$
statistic, we calculate $n=1$  harmonics ($Z_1^2$) using all the photons
in the 95\% PSF of each source.  While the epoch folding method is
less dependent on the shape of modulation profile, the $Z_1^2$ harmonic approach
is relatively more sensitive to sinusodial modulation. The latter takes
advantage of the full timing resolution without resorting binning but
does not allow background subtraction, which is not a major concern
since the background is relatively small and presumably non-periodic.

For each source with a positive period-detection, the data were folded at the most probable period and grouped into 10 phase bins. Our folding code automatically establishes the epoch of peak flux by folding the data using the first data-point as a trial epoch. The data are then folded a second time using the derived epoch, resulting in profiles with the peak flux ($p_{max}$) at phase $\phi\sim0$. This procedure greatly reduces systematic error caused by arbitrary choice of the number and placement of bins. Pulsed flux ($f_{p}$) and pulse-fraction ($PF$) were computed from these pulse-profiles according to the prescription of \cite{bildsten1997}. Our implementation is shown in equations~1-3, where $\bar{f}$ is the mean flux, and the phase interval $\phi$=0-1 is typically divided into $n$=10 equal width bins. 

\begin{center}
\begin{eqnarray}
f_{P} = \int^{1}_{0} [f(\phi)-f_{min}]d\phi \\
          \simeq \frac{\sum_{i=1}^{n} (f_{i}-f_{min})}{n} \\
          PF = \frac{f_{p}}{\bar{f}}
\end{eqnarray}
\end{center}

We note that $f_{p}$ and hence $PF$ are sensitive to errors on the minimum flux ($f_{min}$), although this is always better constrained than $p_{max}$. The phase-folded data-points were also processed with the {\it loess} algorithm, a non parametric smoother \citep{cleveland1981} to obtain an un-binned visualization of the pulse profile. A best-fitting sine-wave is also overplotted.

Several of the {\it Chandra} sources were firmly identified as known HMXB pulsars by their coordinates. In these cases the pulse period is well known and it is not appropriate to use a blind-search approach to period detection. Instead a narrower period range centered on the known pulse-period must be examined, following the approach used in the {\it RXTE} pulsar monitoring pipeline (L05) we chose 5\% either side of the literature period. If a periodogram peak is detected at a period consistent with the historical record for that particular pulsar it constitutes a valid measurement of the pulsar. For this purpose we examined the decade-long {\it RXTE} pulse period history of \cite{galache2008}. In cases where no significant peak is found, the maximum periodogram value in the expected period-range may be interpreted as an upper limit on the pulsed flux.

At our detection limit of 4$\times$10$^{32}$ \lx (5 net counts at the ACIS-I aimpoint in 100 ksec, assuming $D_{SMC}$=60 kpc, $\Gamma$=1, $N_H$=5$\times$10$^{21}$\nh) we expect to detect the high luminosity end of the SMC populations of flaring stellar coronae and cataclysmic variables. Cataclysmic variables (accreting WDs) and Low-Mass X-ray binaries (LMXBs: accreting NS and BH binaries) exhibit aperiodic variability (flickering) over a wide range of timescales from milliseconds up. Such variability produces a characteristic power-law distribution (red noise) in the PDS which can be spotted by visual inspection.  No such features were detected in the bright sources selected for timing analysis. To search for other variability, lightcurves binned by 16 times the ACIS frame time were visually inspected for all sources with at least 50 counts in the stacked dataset. This search  turned up a foreground flare-star (CXOU J005428.9-723107), in DF1 which is the subject of separate analysis by  \cite{laycock2009}. None of the other 95 sources so examined showed any outstanding feature in their lightcurve.

\section{Spectral Analysis}
\label{sect:spectra}
X-ray spectral analysis requires a reasonable number of photons, and was performed for the 14 sources with more than 250 net counts in the stacked dataset. Spectra were extracted from the {\it ChaMP} pipeline source and background event files using CIAO 3.4, and grouped to give 10 source counts per spectral bin. Spectral fits were performed with XSPEC, the results of which are given in Table~\ref{tab:spectralfits}. Eight of these sources are pulsars and are well described by the power-law $\Gamma\sim1$ model typical of X-ray pulsars over this energy range. Photon indices are in the range 0.7--1.3  and derived neutral hydrogen column densities lie in the range  0.6 -- 10 $\times$10$^{21}$ \nh.  Two probable HMXBs (CXO J5057.1-731008 = RX J0050.9-7310, and CXO J005215.4-731915) also exhibit $\Gamma\sim$1 X-ray spectra; neither has been definitively identified with pulsations. Three of the remaining spectra are softer, showing $\Gamma\sim$1.8, suggestive of Active Galactic Nuclei (AGN). This picture is muddied by the detection of candidate periodicities in the X-ray lightcurves of 2 of them (Table~\ref{tab:pulsars}), although these could be spurious and due to aperiodic noise. Finally the softest spectrum belongs to a foreground flare-star. Pile-up (multiple photons contributing to a single ACIS charge-island during the 3.2s readout period) can be safely ignored for all our sources, the brightest of which has a count rate of 0.10 counts~s$^{-1}$, which would result in only a few percent pileup.

The majority of the sources in our catalog are too faint for spectral fitting, and we therefore turn to quantile analysis \citep{hong2004}. Quantile analysis is a method to optimally extract spectral information from low-count X-ray sources, without imposing arbitrary energy-band choices. The advantage over X-ray color indices is that all sources can be represented in the same parameter space, with a somewhat uniform error distribution. Energy band choices made in generating hardness ratios and color-indices inevitably lead to some sources having a gross imbalance in the number of counts in each band. Following the methods of \cite{hong2005}, background corrected quantile values ($QDx$ and $QDy$) were computed from each source's event list photon energy distribution, and spectral model grids were generated using the full {\it Chandra} ACIS-I response function (ARF and RMF). The quantile values, along with over-plotted model grids for absorbed power-law and thermal bremsstrahlung are plotted in Figure~\ref{fig:quantiles1}  for a series of cuts in signal-to-noise-ratio (S/N$>3$, $>$5, $>$10). These figures may be used in interpreting the quantile values presented for the deep-fields source catalog (Table~\ref{tab:catalog}).

X-ray luminosity is computed throughout the paper using $D_{SMC}$=60 kpc \citep{hilditch2005}.

\section{High Mass X-ray Binaries and Candidates}
\label{hmxb}
In order to identify High Mass X-ray Binaries (HMXBs), make associations with previously unclassified  sources and assess evidence of historical variability for new candidate systems,  we compared our {\it Chandra} deep-fields source list against a number of published X-ray source catalogs and works on SMC HMXBs.  The results of this analysis are presented below in the form of a source by source (in RA order) discussion of all HMXBs and candidates that were observed in our {\it Chandra} program. An additional group of sources with bright HMXB-like optical counterparts are summarized in section~\ref{sect:optical}, these objects lack conclusive timing and spectral information and are not associated with literature X-ray sources.

Prior to our Chandra observations there were 12 known X-ray pulsars covered by these observations: seven in Field-1 and five in Field-2. We saw strong pulsations in five of the known pulsars (SXP756, SXP326, SXP172, SXP138, SXP59) and weak pulsations in two others (SXP8.88, SXP15.3).  Three known pulsars (SXP9.13, SXP7.78, SXP 46.6) were detected as point-sources but without pulsations being discernible, and two known pulsars were not detected: SXP82.4 and SXP4.78.

The position plotted for the un-detected pulsar SXP4.78 in Figure~\ref{fig:imageDF1} is that of the proposed optical counterpart, the emission line star [MA93]537. The RXTE position for SXP4.78 \citep{laycock2003} is subject to a $\sim$1' uncertainty which lies wholly within (but close to the edge of) DF1. We can therefore say that no 4.78s pulsar was detected in the error box of SXP4.78. Subject to the optical counterpart being correct, the {\it Chandra} non-detection places an upper limit of  $L_{X}<$1.1$\times$10$^{33}$ \lx (equivalent to 10 net-counts at the position of [MA93]537, scaled by the normalized exposure map value 0.83). 
 
Most of the known Be/X-ray binaries in the SMC are seen regularly by the {\it RXTE} monitoring program (L05, G08) and as a result have known X-ray ephemerides. We calculated the orbital phase for each detected pulsar in order to search for quiescent emission in systems far from periastron; this phase information is given in Table~\ref{tab:phases}. As a {\it caveat} we remind the reader that X-ray ephemerides are based on the spacing of X-ray outbursts, and thus may not exactly reflect the orbital parameters;  for example there might be shifts between outburst peak and periastron. The {\it RXTE} light-curves themselves provide additional context because they record the general state of activity (and pulse period) in each pulsar around the times of the DF1 and DF2 observations.    

We refer to the  SMC X-ray Pulsar (SXP) compilation for {\it RXTE} monitoring results \citep{galache2008}, the {\it ROSAT} source catalogs of \cite{haberl2000} and \cite{sasaki2000} (for the PSPC and HRI instruments respectively), and the catalog of \cite{yokogawa2003} (Y03) for  {\it ASCA} results. Published  observations with {\it XMM-Newton} of HMXBs in the SMC Bar were also consulted, principally \cite{haberl2008} which became available while the present work was in progress. 

Possible {\it ROSAT} counterparts were identified by combining the {\it Chandra} and {\it ROSAT} error radii in quadrature. The Y03 {\it ASCA} catalog does not provide positional error estimates, so we relied instead on their tabulated cross identifications with HRI sources. The complete Y03 catalog contains 38 ASCA sources with HRI counterparts. Examination of the distribution of the separations (column "Sep2" in Y03) for these objects, shows that 95\% lie within Sep2$ <=$ 52", which we adopted as our search radius.

11 {\it Chandra} sources match 7 PSPC sources (2 of the PSPC sources have 3 {\it Chandra} matches each). 3 {\it Chandra} sources have (unique) HRI counterparts. 44 {\it Chandra} sources match 13 {\it ASCA} sources (multiple matches due to the relatively large positional uncertainties). In many cases the multiple matches can be resolved by the detection of pulsations or by YO3's identifications with {\it ROSAT} sources. The 10 confirmed HMXBs are separated into Part 1 of the table, the remainder (Part 2) are a combination of chance positional matches, and partially resolved supernova remnants.

We searched for optical counterparts consistent with HMXBs using the emission-line star catalog of \cite{MA93} (hereafter MA93), and the Magellanic Clouds Photometric Catalog (hereafter MCPS) of \cite{mcps}. In the case of the literature X-ray sources, most of these counterparts have been previously reported, unless otherwise stated in the discussion.

The results of our literature search appear in Table~\ref{tab:pulsars} for the SXP pulsars, and Tables~\ref{tab:HRI}~\ref{tab:PSPC},~\ref{tab:ASCA}, \&~\ref{tab:XMM}  for HRI, PSPC, {\it ASCA}  and {\it XMM-Newton} catalogs respectively.  Optical emission-line counterparts are given in Table~\ref{tab:MA93}, and MCPS stars in Table~\ref{tab:MCPS}. For completeness we list all matches, resulting in some objects appearing more than once in the tables. All such duplications are disambiguated in the source-by-source discussion that follows. 

In summary we detected 13 of the 15 pulsars now known in the fields, including 2 new pulsars, and 5 additional HMXB candidates based on timing, spectra and association with existing X-ray and optical catalogs. Using the MA93 and MCPS catalogs we identified a total of 40 {\it Chandra} sources with bright (V$<$17) optical counterparts, of which $\sim$half are confirmed HMXBs and all but 6 of the remainder are considered candidates (see Section~\ref{sect:optical}).

\subsection{CXOU J004913.5-731138  -- SXP9.13}
We detected both of the two {\it ROSAT} sources (see Table~\ref{tab:PSPC}) that have been proposed by \cite{filipovic2000} as candidates of the {\it ASCA} pulsar AX J0049-732 = SXP9.13. The source preferred by \cite{filipovic2000} (\#427, RX J0049.5-7310 = CXOU J004929.7-731058 ) as being the {\it ASCA} pulsar based on a positionally coincident emission-line star (MA93 No. 300), is found by us to be a P=892 s pulsar (See Figure~\ref{fig:imagesxp892} and Table~\ref{tab:pulsars}).  Both X-ray  sources were detected by \cite{hp2004} in {\it XMM-Newton} observations from October 2000, at which time pulsations were not detected in either source. The discovery of SXP892 appears to partially resolve the ambiguous identity of AX J0049-732 as there is now only one candidate (\#430 RX J0049.2-7311 = CXOU J004913.5-731138). The possibility remains that the ASCA pulsar is not this object and the case will be fully resolved only if the {\it Chandra} source shows 9.13s pulsations.
CXOU J004913.5-731138 was relatively bright (1178 net ACIS counts) however pulsations were not detected, despite the time resolution of {\it Chandra} ACIS (3.2 sec) being sufficient to critically sample the 9.13s pulse period. The periodogram (Figure~\ref{fig:sxp9.13}) evaluated over a period range covering P=9.13s$\pm$5\% contains many noise peaks of similar height, and the pulse fraction must therefore be very small at this time.  The orbital phase (G08 ephemeris, Table~\ref{tab:phases}) was 0.82, outside the range of the normal outbursts seen in SXP9.13, and therefore presumably far from periastron, which can account for the lack of pulsations. We measured a luminosity of L$_{x}=8.5\times10^{34}$ \lx ($B_C$ band).  This source was again detected by {\it XMM-Newton} on 2007 March 12 by \cite{haberl2008}, the {\it XMM-Newton} and {\it Chandra} coordinates are separated by 1.29" within an error budget of 2.01" (quadrature sum of 95\% statistical and systematic errors for both telescopes). 

\subsection{CXOU J004824.0-731918} 
This source lies 5.7" away from {\it ROSAT} source RXJ0048.4-7319 (PSPC \#454, see Table~\ref{tab:PSPC}).The {\it ROSAT} object was associated with the supernova remnant  SNR 0046-73.5 by \cite{haberl2000}, and is indicated  on the {\it Chandra} DF2 image (Figure~\ref{fig:imageDF2}). The object is at outer edge of the {\it Chandra} ACIS-I array at an off-axis distance of $\sim$8 arcmin. Consequently the object could be either a bright region of the SNR or point-source, the lack of a bright-star counterpart rules out an HMXB and favors the former explanation.
 
\subsection{CXOU J004929.7-731058 -- SXP892} 
This source is detected in DF2 with 1434 net counts. This is a  new pulsar with an 892s period. The pulsation is detected at very high significance as shown in Figure~\ref{fig:sxp892}. The derived luminosity is $L_x$ = 9.3$\times$10$^{34}$ \lx, consistent with a faint type-I (normal) outburst from a Be X-ray binary pulsar. The ACIS spectrum from 0.35--8 keV is well fit by a power-law with photon index $\Gamma$=1  and very low extinction (Table~\ref{tab:spectralfits}).  
DF2 was observed by {\it Chandra} in 2002 for 10 ksec, when SXP892 was not detected, confirming its transient nature. Had it been active at the same level it would have registered $\sim$143 counts. The fact that SXP892 has never been detected above the pulsed flux limit of {\it RXTE}  ($\sim$10$^{36}$\lx), despite being frequently in the field of view indicates that its outbursts are always faint. The position of SXP892 coincides with the {\it ROSAT} \& {\it XMM-Newton} source RX J0049.5-7310 \citep{hp2004}, as shown in Table~\ref{tab:PSPC} and Figure~\ref{fig:imagesxp892}. The separation between the {\it Chandra} and HRI coordinates is 2.82" with a total error budget of 5.52". An emission-line star (\#300 in MA93) was associated with RX J0049.5-7310 by \cite{filipovic2000}. The star is also coincident with our {\it Chandra} position (Table~\ref{tab:MA93}) and is in the MCPS catalog with V=16.15 mag, B-V=0.20 mag (see Table~\ref{tab:MCPS}; Antoniou et al., 2010, in prep). The discovery of an outburst and 892s pulse period confirms the object as a Be-HMXB pulsar.

\subsection{CXOU J004942.0-732314 -- SXP756}
We detected SXP756 with 4116 counts in DF2.  Figure~\ref{fig:sxp756} shows the power spectrum and folded pulse-profile, which exhibits one prominent narrow peak per cycle defined as $\phi$=0, plus a much smaller (and possibly broader) peak at $\phi$=0.5. The pulse period was 746s which is at the shorter end of the period range (746s-767s) seen by G08 with RXTE.  The X-ray spectrum is well described by an absorbed power-law ($\Gamma$=1.0, $n_H$=5$\times$10$^{21}$~cm$^{-2}$) with luminosity$L_X$=7.7$\times$10$^{35}$\lx. According to the G08 ephemeris ({Table~\ref{tab:phases}) the orbital phase of the system was 0.78 in the quiescent portion of the long-term {\it RXTE} lightcurve. The pulsed flux from the {\it RXTE} data folded at the orbital period of 389.9d shows that in addition to regular outbursts, activity at $L_{X}\sim$10$^{36}$\lx has occurred sporadically throughout most of the cycle. This persistent pulsed emission is confirmed by the {\it Chandra} results, which show a low pulsed fraction ($\sim$25\%).

\subsection{CXOU J005044.6-731605 -- SXP323}
We detected a bright source with pulse period 316s, located at the coordinates of SXP323 which is likely the same pulsar, although the period differs significantly from the discovery value of 323s \citep{yokogawa2003}. The {\it RXTE} monitoring program detected the system in outburst on MJD 52960 with pulse period 317s (100\% period-detection significance, uncertainty$<$0.1s), following an erratic multi-year trend of decreasing period (spin-up). \cite{galache2008} attributed the MJD 52960 outburst to a Type II (super-outburst) event as it did not follow the ephemeris derived from previous and subsequent outbursts. At the time of the DF2 observation (MJD 54061) the orbital phase of SXP323 was 0.8 (G08 ephemeris, Table~\ref{tab:phases}) and the pulsar was far from its presumed periastron. During the $\sim$3 yrs between the super-outburst and our {\it Chandra} observation, the pulsar has not slowed, but has rather continued the gradual spin-up trend. \cite{haberl2008} demonstrate that {\it ASCA} and {\it XMM} data confirm long-term spin-up of SXP323 and point out that {\it RXTE}'s turbulent period history could be contaminated by frequencies arising from nearby SXP326 (Section~\ref{sect:SXP326}).

\subsection{CXOU J005057.1-731008 -- RX J0050.9-7310}
A {\it ROSAT} source consistent with CXOU J005057.1-731008 appears in the HRI (\#36) and PSPC (\#421) catalogs at very small positional offsets. Both source coordinates agree at the  $\sim$2" level, within a total error budget of $\sim$4". There is also an {\it ASCA} counterpart (Table~\ref{tab:ASCA}, \#35 in \citealt{yokogawa2003}), identified with the same {\it ROSAT} source (AX J0050.8-7310 = RX J0050.9-7310), an association backed up by XMM results (\citealt{hp2004} \&~\citealt{shtykovskiy2005}. 

 \cite{hs2000} proposed this source to be a Be-XRB, based on its association with an emission-line star (\#414 in MA93), which is confirmed by our {\it Chandra} error circle (Table~\ref{tab:MA93}). The star also appears in MCPS with V=14.35 mag, B-V=0.08 mag.  The {\it Chandra} ACIS spectrum in the DF2 data-set is well fit (reduced $\chi ^2$=0.86) by an absorbed power law with photon index $\Gamma$=1.07$\pm$0.1 and absorption column density $n_H$=5.20$\pm$1.48$\times$10$^{21}$\nh which is typical for HMXB pulsars in the SMC.  The source shows strong variability between the ROSAT, ASCA, and the 2006 {\it Chandra} observations.  Together these lines of evidence strongly support the Be-XRB classification, although pulsations have not been detected.

\subsection{CXOU J005151.9-731033 --  SXP172}
SXP172 was detected with pulse period 171.8s (Figure~\ref{fig:sxp172}) in DF2 (MJD 54061) with 2-10 keV luminosity $L_{x}$= 1.53$\times$10$^{35}$\lx. The pulsar has historically shown a prolonged episode (MJD 51500-52500) of steady spin-up 
(see G08, Figure 30). The period detected here is in line with the end of that active episode, indicating the pulsar has not continued to spin up during the $\sim$1000 days since its activity dropped below the detection threshold of RXTE, nor has it spun down appreciably. The orbital phase (Table~\ref{tab:phases}) for SXP172 on MJD 54061 was 0.32 (L05 ephemeris), placing it reasonably far from peak-phase. This pulsar has historically been seen in outburst at times uncorrelated with the X-ray and optical period. 

\subsection{CXOU J005153.2-723148 -- SXP8.88}
The {\it Chandra} source has a {\it ROSAT} counterpart  (PSPC \#265) in Table~\ref{tab:PSPC} (RXJ0051.8-7231) and is a known Be X-ray pulsar.   
The CXOU coordinates are consistent with the optical counterpart (\#506 in MA93), which also appears in the MCPS (V=14.38 mag, B-V=0.41 mag). SXP8.88 was detected in our data with 183.8 net ACIS-I counts, exhibiting a hard spectrum and $L_{X}$=(1.46$\pm$0.13)$\times$10$^{34}$\lx. The median photon energy ($E_{50}$=2.32 keV) and quantiles  ($QDx$=-0.45, $QDy$=1.06) imply a power-law index $\Gamma\sim$1. The pulse period history of SXP8.88 is known to high accuracy from {\it RXTE} studies (G08) Accordingly we searched the period range 8.5s-9.5s, finding the highest peak at P=8.89909 seconds,  within the 0.889s-0.915s variation seen by RXTE.  The periodogram and folded light-curve appear in Figure~\ref{fig:sxp8.88}. The orbital phase was 0.82 according to the G08 ephemeris (Table~\ref{tab:phases}) placing this detection outside of the normal range of outbursts, which accounts for the faint flux.

\subsection{CXOU J005205.6-722604 -- SXP7.78 -- SMC X-3}
The source was detected in DF1 at coordinates consistent with SMC X-3. With 72 net counts the point-source detection is highly significant and yields accurate position and quantile information. Neither the Lomb-Scargle nor the 3 alternative periodograms (Z-statistic, epoch-folding, MHP) show any evidence for 7.78s pulsations. Instead they show a peak at 448s which however has a significance below 95\% based on Mont-carlo simulations or the \cite{press1992} criteria.  Figure~\ref{fig:smcx3} shows the periodogram for the expected period range (a 5\% range either side of 7.78s), with the narrow limits of the range seen by G08 with {\it RXTE} indicated. 

The association of SXP7.78 with SMC X-3 was made by \cite{edge2004} from a shallow (10 ksec) {\it Chandra} observation of the Bar on July 20, 2002 that detected pulsations with the same period as the {\it RXTE} pulsar SXP7.78. The more accurate {\it Chandra} position confirmed the association of this source with the emission-line star MA93 \#53 (\citealt{edge2004}, which lies at the same coordinates as the SMC X-3 optical counterpart proposed by \cite{crampton1978}. Our 2006 detection is consistent with the same counterpart which is also an MCPS star (Table~\ref{tab:MCPS}).

As shown in Table~\ref{tab:HRI}, the {\it ROSAT} HRI catalog \citep{sasaki2000} contains an entry (\#43) for SMC X-3 that matches a different {\it Chandra} Deep Fields source (CXOU J005209.2-722553). According to \cite{edge2004} SMC X-3 lies at RA=00:52:05.7, Decl=-72:26:04 (error radius=0.62") and is thus consistent only with CXOU J005205.6-722604  and not CXOU J005209.2-722553. The {\it ROSAT} HRI source \#43 was probably SMC X-3 despite the positional offset, owing to its strong variability and the relatively large positional uncertainty (15.9"). 
 
The orbital phase (G08 ephemeris, Table~\ref{tab:phases}) was 0.62 which is within the normal range of outbursts for SXP7.78

\subsection{CXOU J005209.2-722553}
This source appears in Table~\ref{tab:HRI} as a spurious positional coincidence to a {\it ROSAT} HRI detection of SMC X-3. CXOU J005209.2-722553 is an unrelated source lying several arcseconds away from the established position of SMC X-3, based on the identification of its optical counterpart \citep{edge2004}.
CXOU J005209.2-722553 was detected in the DF1 data set with 14.4 ACIS-I net-counts, corresponding to a flux $F_{Bc}$=(2.4$\pm$0.9)$\times$10$^{-15}$\flux. The median photon energy $E50$=1.56 keV and quantiles ($QDX$= -0.71, $QDY$=1.1) suggest a spectrum consistent with power-law photon index $\Gamma$=1.5-2 keV and moderate absorption $N_{H}\sim$5$\times$10$^{21}$~cm$^-2$  (Figure~\ref{fig:quantiles1}). There is no optical counterpart \citep{antoniou2010} in MA93 or MCPS, which together with the X-ray properties suggests a background AGN.

\subsection{CXOU J005214.0-731918 -- SXP15.3}
DF2 registered a clear detection of a point-source at the coordinates of the SXP15.3 optical counterpart identified by \cite{edge2005}. We detect 147 net ACIS-I counts, corresponding to a luminosity $L_{x}=6.77\times10^{33}$\lx. There is no X-ray ephemeris for SXP15.3, but \cite{edge2005} produced an ephemeris (P=75d) from optical photometry of the counterpart. Accordingly the orbital phase was 0.13 on MJD 54061.   

CXOU J005214.0-731918 matches the {\it ROSAT} HRI source \#44 \citep{sasaki2000}  and {\it ASCA} source \#43 \cite{yokogawa2003}, both of which are identified as the HMXB pulsar RXJ0052.1-7319 = SXP15.3. Table~\ref{tab:HRI} shows this object was also associated with PSPC source \#453 by \citealt{sasaki2000} based on the overlapping error radii of the two {\it ROSAT} instruments. The PSPC source does not appear in Table~\ref{tab:PSPC} as the {\it Chandra} coordinates lie beyond the combined PSPC and {\it Chandra} 95\% positional uncertainty.

Having confirmed CXOU J005214.0-731918 as SXP15.3 on positional grounds, we identified weak pulsations at P=15.239s. The range of periods seen by G08 with {\it RXTE} is indicated in Figure~\ref{fig:sxp15.3} by 2 vertical lines. The minimum pulse-period recorded by {\it RXTE} was 15.23s, on MJD$\sim$53950,  approximately 110 days before the {\it Chandra} observation. Prior to this {\it RXTE} has shown a steady 3-year long spin-up, with the longest period (P=15.27s) recorded on MJD$\sim$52900. The peak detected in our {\it Chandra} lightcurve at P=15.239s is thus in close agreement with the most contemporaneous {\it RXTE} detection. Alternatively the {\it Chandra} result can be regarded as an upper limit on the pulsed flux.

Another separate source, CXOU J005215.4-731915 was detected just 7.5 arcseconds away from SXP15.3,  as can be seen in Figure~\ref{fig:imageDF2}.  \cite{haberl2008} have reported an {\it XMM-Newton} detection of SXP15.3 which actually better matches the nearby source. The situation is clarified by 
Figure~\ref{fig:imagesxp15}.

Our {\it Chandra} coordinates for this object place it 7.5 arcseconds away from an {\it XMM-Newton} source identified with SXP15.3 by \cite{haberl2008}. However as seen in Figure~\ref{fig:imagesxp15} the error-circle of the {\it XMM-Newton} source is inconsistent with CXOU J005214.0-731918. This in combination with the fact the latter shows tentative periodicity at P$\sim$15.24s and is associated with both an emission-line counterpart (\#552 in MA93) and the {\it ROSAT} HRI source \#44, strongly suggests that this is SXP15.3. Meanwhile, the nearby source CXOU J005215.4-731915 is positionally coincident with the {\it XMM-Newton} source, strongly suggesting that these represent an unrelated object.

\subsection{CXOU J005215.4-731915}
\label{sect:J005215.4}
The superb angular resolution and PSF of {\it Chandra} enabled a clean discovery of this source, which is a very close neighbor of SXP15.3. The two objects are separated by only 7.5 arcseconds. At the time of our {\it Chandra} observation, CXOU J005215.4-731915 was brighter than SXP15.3. This indicates that J005215.4 is strongly variable, and that it is active much less frequently than SXP15.3.  The {\it Chandra} coordinates are a positional match to an {\it XMM-Newton} source (Table~\ref{tab:XMM}) that was reported by \cite{haberl2008} who in the absence of pulsations, associated it with SXP15.3.  As Figure~\ref{fig:imagesxp15} shows, the {\it XMM-Newton} detection is more consistent with an independent discovery  of CXOU J005215.4-731915. There is also a PSPC source (\#453) that lies almost equidistant from both objects.

The source was detected with 286 net ACIS counts (c.f. 147 for SXP15.3) and its spectrum (Table~\ref{tab:spectralfits}) well fit by a power-law ($\Gamma$=0.93$\pm$0.20) with absorption column  ($n_H$=0.65$\pm$1.5 ~cm$^{-2}$) consistent with SMC pulsars.  The X-ray flux of 1.13$\times$10$^{-13}$\flux corresponds to $L_{X}$=4.86$\times$10$^{34}$\lx in the SMC. 

There is a bright optical counterpart in the MCPS catalog (V=15.90 mag, B-V=-0.14 mag) with magnitude and color consistent with an early B spectral type. The counterpart, spectrum and long-term variability all point to a HMXB. Further observations are required to either detect X-ray pulsations, or otherwise constrain its nature.

\subsection{CXOU J005252.2-721715 -- SXP326} 
\label{sect:SXP326}
This is a new pulsar discovered in DF1. The X-ray pulsar was in a bright outburst state with 9292 net ACIS-I counts, and consequently the 326s pulsation is detected at very high significance as shown in Fig.~\ref{fig:sxp326}. The derived luminosity is $L_x$ = 5.5$\times$10$^{35}$\lx, consistent with a moderate type-I (normal) outburst from a Be X-ray binary pulsar. The ACIS spectrum in the 0.3--8 keV band is well fit by a power-law with photon index $\Gamma$=1, with very low extinction (Table~\ref{tab:spectralfits}).  
DF1 is monitored regularly by RXTE, which has never detected SXP326, therefore either the pulsar is rarely in outburst,  or its normal outbursts are always below the {\it RXTE} pulsed flux detection limit of $\sim$10$^{36}$ \lx.
The source was also detected by  {\it XMM-Newton} and proposed as a Be-HMXB pulsar with period 325s  by \cite{haberl2008} (XMMU J005252.1-721715). The {\it XMM-Newton} and {\it Chandra} coordinates are separated by 0.66", within an error budget of 1.93" (quadrature sum of 95\% statistical and systematic errors for both telescopes).

SXP326 is associated with an early-type star in the MCPS catalog with photometric parameters consistent with those of Be-stars.

\subsection{CXOU J005323.8-722715 -- SXP138}
Based on its coordinates and the detection of 138s pulsations, this source is confirmed as SXP138. In addition to the previously known pulse-period of 138s we detected a number of other periods (Fig.~\ref{fig:sxp138}). 
We note that the 138s period contained the most power as shown in Fig~\ref{fig:sxp138}. Analysis of the combined light-curve of both observations revealed highly significant periods at 500.00, 333.33, 161.29, 138.88.  The exact factors (500, 333.33) of the {\it Chandra} Y-axis dither period (1000s) suggests at least some of these periods are due to the source being moved back and forth across a chip edge or node-boundary. Indeed CXOU J005323.8-722715 lies within 10'' of the CCD's edge. It is not easy to see why the remaining periods (972,161,138) should follow a separate geometric relation, unless one of them is a pulse period and the others are harmonics. Careful analysis of the two individual 50 ksec  sub-observations and the full lightcurve shows that although both segments alone show a 138s period modulated at $\sim$972s, the phase of the long periodicity is different in the two obsids and in consequence is totally suppressed in the full lightcurve. We conclude the additional periods are an interaction between the pulse period and Chandra's dither. 
At the time of the DF1 observation (MJD 53851) SXP138 was at orbital phase 0.005 (G08 ephemeris, Table~\ref{tab:phases}), which is the point of minimum flux, and assumed to be far from periastron. The luminosity of the pulsar at this presumable low-state is 2.25$\times$10$^{35}$\lx, with a pulse fraction PF=0.46.

\subsection{CXOU J005331.7-722240}
This source is sufficiently bright (386 net ACIS counts) for spectral fitting and timing analysis.  We found a (LS) periodogram peak at P=131.11s, that meets our initial criterion of the 90\% significance level from the formula of \cite{press1992} (see Figure~\ref{fig:p131}).  More accurate significance levels were then computed via Monte Carlo simulation (as described is section~\ref{sect:timing}) which accounts for the statistical and sampling properties of the light curve. The simulations yielded a significance  of $>$95\% for the 131.1s period. There are no other periodicities seen in the light curve making it unlikely that aliasing or beating of a slower modulation is responsible for the peak.  The 131.11s period was confirmed by the multi-harmonic periodogram, although the MHP highest peak was at P=262.234 which is exactly twice the LS period (MHP often finds the sub-harmonics \citealt{SZ}). 

The X-ray spectrum is well fit by a power-law (Table~\ref{tab:spectralfits}), with index $\Gamma$=1.84$\pm$0.22, absorbed by a column density $N_{H}\sim$2$\times$10$^{21}$ \nh. The X-ray flux (2-10 keV) was $f_{x}$=6.3$\times$10$^{-14}$\flux, which if located in the SMC corresponds to $L_{X}\sim$2.7$\times$10$^{34}$\lx.  There is no prior detection of an X-ray source at these coordinates in the literature, although if active at this level it could have been detected by {\it XMM-Newton} or {\it Chandra} (in the Shallow survey), indicating variability. 

We do not find any counterparts to this source in the MCPS catalog down to $V<$20 mag, similarly we do not find any counterparts in the MA93 catalog, which appears to rule out a HMXB.  As the periodicity does not reach 99\% significance, it could be a false period, driven by actual but aperiodic variability in the light-curve. The powerlaw index is suggestive of an AGN.

\subsection{CXOU J005352.5-722639}
\label{sect:J005352}
This new HMXB candidate appears in Table~\ref{tab:PSPC} as one of 3 positional counterparts for PSPC source \#242 (RXJ0053.9-7227). The separation is 21.52" with an error budget of 25.71" (the CXOU positional error is  0.51" as the source fell close to the {\it Chandra} aimpoint). The {\it ROSAT} source has been identified as the 46.6 s {\it RXTE} pulsar XTEJ0053-724 by CML98, which has since been pin-pointed by \cite{mcgowan2008} to a location that is consistent with CXOU J005355.3-722645 (see Section~\ref{sect:sxp46}  for SXP46.6).  Thus the association of PSPC \#242 with CXOU J005352.5-722639 is spurious, however there is more to the story.

There is a probable optical counterpart to CXOU J005352.5-722639, an emission-line star (\#717 in MA93) that is consistent with our {\it Chandra} error-circle. The separation between the MA93 and CXOU coordinates is 0.84", which is less than the quadrature sum of the 95\% positional error radius (0.513") and the {\it Chandra} aspect uncertainty (a total error budget of 0.91"). 

\cite{buckley2001} reported finding two Be stars in the ROSAT PSPC error circle of the source PSPC \#42 = SXP46.6.  These were referred to as "Eastern Component of Star A", and "Star B". The correct counterpart to SXP46.6, "Star B" was identified by \cite{mcgowan2008}, based on its optical variability period matching the RXTE ephemeris of the pulsar. 

We have discovered that CXOU J005352.5-722639 and [MA93]717 are both coincident with "Eastern Component of Star A", thus the deprecated counterpart turns out to be a separate HMXB candidate in its own right. There is no MCPS match to this object, but \cite{buckley2001} show it to be a B1e-B2e star with V=14.2mag and a variable infrared excess.

For this faint Chandra source (9.5 ACIS-I net-counts) we measured an X-ray flux $F_{Bc}$=(1.2$\pm$0.6)$\times$10$^{-15}$\flux, corresponding to $L_X$=5.2$\times$10$^{32}$\lx. The quantile values ($QDx=-0.2$, $QDy=0.9$, see Figure~\ref{fig:quantiles1} for model grid) and median photon energy ($E50$=3.2 keV) suggest a hard source, with little ($<$10$^{21}$ but poorly constrained) absorption. 

In summary, CXOU J005352.5-722639 is a new HMXB candidate based on its hard X-ray spectrum and Be star counterpart. It has not been detected in outburst and no pulsations have been seen.

\subsection{CXOU J005354.8-722722}
This source (with 17 ACIS net-counts) appears in Table~\ref{tab:PSPC} as one of 3 positional counterparts for PSPC source \#242 (RXJ0053.9-7227). The separation is 22.78" with an error budget of 25.70".
We measured an X-ray flux $F_{Bc}$=(2.3$\pm$0.7)$\times$10$^{-15}$\flux, corresponding to $L_X$=1.0$\times$10$^{33}$\lx. The quantile values ($QDx=-0.34$, $QDy=1.58$, see Figure~\ref{fig:quantiles1} for model grid) and median photon energy ($E50$=2.70 keV) suggest a hard source with high absorption column density ($N_{H}>$10$^{22} ~cm ^{-2}$). A likely interpretation is a background AGN.

\subsection{CXOU J005355.3-722645 -- SXP46.6}
\label{sect:sxp46}
\cite{mcgowan2008} have identified the frequently active 46.6s pulsar XTE J0053-724 \citep{marshall97} with one of 2 optical counterparts ("Star B") proposed by \cite{buckley2001}, using archival {\it Chandra} data (including the DF1 dataset). \cite{mcgowan2008} discovered that the optical counterpart of the {\it Chandra} source exhibits a photometric period of P=136d which matches the X-ray outburst based ephemeris derived by the RXTE monitoring project for SXP46.6 (L05, G08), although {\it Chandra} has not seen the 46.6s pulsations.

We detect this source with 63 net ACIS-I counts and $F_{Bc}$=(5.4$\pm$0.8)$\times$10$^{-15}$\flux   corresponding to $L_X$=2.3$\times$10$^{33}$\lx. The median photon energy ($E50$=1.61keV) and quantiles ($QDx$=-0.69, $QDy$=1.30) imply a medium-hard spectrum, consistent with power-law index $\Gamma =$2-2.5, and absorption column $n_H\sim$5$\times$10$^{21} ~cm ^{-2}$.  The power-law index is rather soft for an HMXB pulsar, however the identification is secure owing to the detection of a 136 day orbital modulation in the pulsar and the optical counterpart.
According to the well defined {\it RXTE} ephemeris of G08 the orbital phase of SXP46.6 was 0.336 at the time of the DF1 exposure. As has been noted by \cite{mcgowan2008} this would be consistent with low luminosity as the pulsar was far from periaston. 

Table~\ref{tab:PSPC} shows that the {\it ROSAT} catalog entry for SXP46.6 (PSPC \#242, RXJ0053.9-7227; \citealt{haberl2000}) has a large positional uncertainty, and consequently is also coincident with two other DF1 sources (CXOU J005354.8-722722 and CXOU J005352.5-722639) which are discussed separately.

\subsection{J005403.9-722633 -- XMMU J005403.8-722632 -- SXP342}
This source is coincident with a 342s Be-HMXB pulsar recently proposed by \cite{haberl2008} (XMMU J005403.8-722632). The separation between the CXOU and {\it XMM-Newton} coordinates is just 0.27", within a combined error budget of 1.93" (the quadrature sum of the 95\% statistical and systematic errors for both telescopes).  We detected SXP342 with only 36 net ACIS-I counts, which is insufficient for timing analysis. The 0.3-8 keV flux $F_{Bc}$=8.8$\times$10$^{-15}$\flux corresponds to a luminosity of $L_{X}$=3.8$\times$10$^{33}$\lx. The median photon energy ($E50$=3.4 keV) and quantiles ($QDx$=-0.17, $QDy$=0.89) imply a very hard spectrum, consistent with power-law index $\Gamma <$0.5, and low absorption column $N_{H}\leq$10$^{21} ~cm^{-2}$. Similarly \cite{haberl2008} reported the {\it XMM-Newton} spectrum during outburst was the hardest of any Be-XRB in the SMC. They also note that SXP342 cannot be unambiguously identified with the pulsar SXP348 tracked by the {\it RXTE} monitoring project (G08), which may be an amalgam of SXP342 and SAX J0103.2-7209 (P=345s), as both lie within the same {\it RXTE} field of view. 

\subsection{CXOU J005428.8-722810, CXOU J005433.0-722806, CXOU J005436.5-722816  -- RXJ0054.5-7228}
Three DF1 sources (Table~\ref{tab:PSPC}) fall in the positional error circle of {\it ROSAT} PSPC source \#248 (RXJ0054.5-7228).
\cite{hs2000} point out that six emission-line objects from the catalog of MA93 lie in the error circle. We found that none of the 3 {\it Chandra} sources has an MA93 counterpart, using a search radius equal to the {\it Chandra} 95\% error circle in quadrature with 0.75" aspect uncertainty and 5" tolerance on the MA93 coordinates. Thus no positive identification can be made with RXJ0054.5-7228.

\subsection{CXOU J005437.1-722637} 
This source was sufficiently bright (433 net ACIS counts) for spectral fitting and timing analysis.  We found a Lomb-Scargle periodogram peak at P=10.426s, that meets our initial criterion of $>$90\% significance (see Figure~\ref{fig:p10.4}).  More accurate significance levels were then computed via a Monte Carlo simulation (as described above) which accounts for the statistical and sampling properties of the light-curve. The simulations yielded a significance $>$95\% for the 10.426s period, which is not an integer multiple of the ACIS readout (3.2s). There are no other periodicities seen in the light-curve making it unlikely that aliasing or beating of a slower modulation is responsible for the peak.  Alternative methods including the Multi-Harmonic Periodogram \citep{SZ}, Epoch Folding, and Z-Statistic did not find this peak, making the detection questionable.

The X-ray spectrum is well fit by a moderate power-law, with index $\Gamma$=1.81$\pm$0.21, absorbed by a column density $N_{H}\sim$10$^{21}$\nh. The X-ray flux (0.3-8 keV) was $F_{Bc}$=(4.6$\pm$0.2)$\times$10$^{-14}$ \flux, which if located in the SMC corresponds to $L_{X}\sim$2$\times$10$^{34}$\lx.  There is no prior detection of an X-ray source at these coordinates in the literature, although if active at this level it could have been detected by ROSAT, {\it XMM-Newton} or Chandra, from which we infer it is variable. 

We do not find any stars in MCPS  down to V$\sim$20 mag, or emission-line objects in the MA93 catalog, within the {\it Chandra} error circle. This in combination with the marginal period detection and relatively soft X-ray spectrum (for an X-ray pulsar) suggests that the source is not an HMXB, and is likely an AGN.

\subsection{CXOU J005446.3-722523 --- Candidate HMXB P=4693s}
\label{sect:P4693}
This source lies in DF1, and was detected with 140 net counts. The median energy ($E50$=2.24 keV) indicates that the source is hard, and X-ray quantiles ($QDx$=-0.47, $QDy$=1.13) show the spectrum to be consistent with an absorbed power-law (photon index $\Gamma\sim2$, $N_H\sim$10$^{22}$\nh).  The flux $F_{Bc}$=(2.1$\pm$0.2)$\times$10$^{-14}$\flux results in $L_X$=9$\times$10$^{32}$\flux if located in the SMC.

Timing analysis (Figure~\ref{fig:sxp4693}) revealed a very long periodic modulation at P=4693s, which exceeded the 99\% significance level derived by Monte Carlo and the formula of \cite{press1992} for the Lomb-Scargle periodogram. Alternative methods (the Z-Statistic, epoch folding and multi-harmonic periodogram) confirmed the 4693s period. The periodogram and folded light-curve are shown in Figure~\ref{fig:sxp4693}. 

The source has a bright MCPS optical counterpart as shown in Table~\ref{tab:MCPS} which lies well within the {\it Chandra} 95\% error radius (0.94", the quadrature sum of 0.56" and the aspect uncertainty 0.75"), at a separation of 0.55". The counterpart has a magnitude V=15.36mag  and color indices B-V=0.14mag, U-B=-1.11mag, V-I=0.109mag. At the distance of the SMC (60 kpc) the absolute magnitude is $M_V$=-3.53, which together with the U-B color implies a spectral type of B0V-B1V. The emission line star [MA93]798 (Table~\ref{tab:MA93}) is also associated with the source. With this counterpart, the object is a likely Be-HMXB, which would make it the longest-period HMXB pulsar to be discovered. The implications of such a long period pulsar  are discussed in Section~\ref{sect:longperiod}.

\subsection{CXOU J005456.1-722648 -- SXP59}
 The source was detected in the {\it Chandra} observation of Field 1 at a luminosity of 1.83$\times$10$^{34}$ \lx with pulse period 58.8 s. 
 According to the G08 ephemeris, SXP59 was at orbital phase 0.65 during the DF1 observation on MJD 53851, which is within the range for normal outbursts seen in SXP59. We measured a pulse period of 58.85815s  which is longer than the last reported measurement by G08 on MJD 54000, but in line with the long term average value. SXP59 tends to spin-up when in outburst, and spin-down in quiescence.




\section{Deep Fields X-ray Sources with Bright Stellar Counterparts }
\label{sect:optical}
We searched for optical counterparts consistent with HMXBs using the emission-line star catalog of  MA93, and the Magellanic Clouds Photometric Catalog (MCPS).  The MA93 objective-prism survey catalog contains the counterparts to almost all known Be-HMXB in the SMC. The MCPS is complete for V$<$20, corresponding to $M_V$=1 at $D_{SMC}$=60 kpc, which encompasses the full brightness range of HMXB. According to \cite{mcbride2008} all HMXB counterparts in the SMC have spectral type earlier than B3, thus we restricted our MCPS catalog search to V$<$17, corresponding to $M_{V}<$-2. The MCPS counterpart search used the CXOU 95\% statistical error circle in quadrature with the aspect uncertainty (0.75" for 95\%). The MA93 counterpart search used the CXOU 95\% statistical error circle in quadrature with the aspect uncertainty (0.75" for 95\%) and a 2.5" tolerance on the MA93 coordinates. The MA93 tolerance was adopted by doubling the largest offset required for an HMXB with an established counterpart ($\sim$1.2" for SXP59). The fact emission line stars are rare and the MA93 plate material dates from the 1970's, justifies the larger search region.

There are 18 {\it Chandra} sources with MA93 counterparts listed in Table~\ref{tab:MA93}, of which 9 are confirmed HMXB pulsars, another 2 are {\it ROSAT} sources.  A further 7 objects are new discoveries, and are denoted by ``Be?" in Table~\ref{tab:MA93}  as their X-ray emission and emission-lines indicate they are probable Be-stars.  

There are 40 {\it Chandra} DF sources with MCPS counterparts brighter than $V<17$, all unique, listed in Table~\ref{tab:MCPS}. The MCPS coordinate accuracy is quoted as $rms\sim$0.3",  and we found that adding up to 1"  tolerance did not increase the number of matches beyond those contained in the CXOU 95\% error circles.  21 of our MCPS counterparts are either confirmed HMXBs and/or are associated with an MA93 star. The confirmed HMXBs and MA93 stars are all blue, having $B-V<$0.2 (excepting SXP8.88 which has $B-V$=0.4). There are a further 12 Chandra+MCPS counterparts (``C' in Table~\ref{tab:MCPS}) that lie in the same range of $B-V<0.2$, for which the reddest has $B-V$=0.12.  The remaining 7 objects are significantly redder, having $B-V$=0.57-1.89, and are labelled ``R" in Table~\ref{tab:MCPS}. 

This bi-modality of the optical color distribution strongly suggests that 33 objects are HMXBs, and the 7 `red outliers' are likely foreground stars and/or random matches. Caution should be exercised in interpretation of the new {\it Chandra} counterpart identifications as there are other (albeit far rarer) species known to be present in the SMC that could also meet the matching criteria used. In particular, Wolf-Rayet stars, colliding wind binaries and supergiant HMXBs all manifest as luminous, blue hard X-ray emitters. Accordingly a full analysis of the optical counterparts is presented in a companion paper \citep{antoniou2010}.

\section{Discussion}

\subsection{Pulsars in the Deep Fields: Number, Orbital Phase and Pulsed Flux}
There were 12 X-ray pulsars known in the survey region prior to these observations, 7 in DF1 and 5 in DF2. The new pulsars SXP326, SXP892 and SXP342 (a faint point-source in our data, identified as a pulsar by \cite{haberl2008}) have increased the sample of confirmed pulsars in the region to a total of 15. Most interesting is the fact that we were able to detect all the known pulsars in DF2 and most (5/7) in DF1. 

Our observed pulse profiles show a range of forms. The typical profile exhibits significant flux above the un-pulsed component for most of the pulsation cycle. A qualitative distinction exists between profiles that are broad and asymmetric (e.g. SXP326, Figure~\ref{fig:sxp326}), and those that are strongly peaked which tend to be narrower and more symmetric (e.g. SXP172, Figure~\ref{fig:sxp172}).  These differences are likely due to differences in geometry of the emitting regions (polar caps) of the neutron star. The literature documents several efforts to model X-ray pulse profiles in terms of combinations of fan and pencil beams (See for example \citealt{parmar1989}). Broad and cuspy profiles are attributed to fan beam geometry, where the X-ray beam has the form of a (somewhat irregular) hollow cone,  due to obscuration of the central part of the polar cap by the incoming accretion stream, or reflection/re-processing effects. Narrow pulses more closely resemble radio pulsars and are accordingly interpreted in the "light-house" model, as collimated beams (pencil-beams) emanating directly from the polar caps. 

One of the  primary objectives of this project was to find out if HMXB pulsars could be detected between outbursts, and to determine their quiescent luminosity.  Using the latest X-ray ephemerides produced by the {\it RXTE} monitoring program (G08) we calculated the (presumed orbital) phase of each pulsar at the time of the {\it Chandra} DF observations (Table~\ref{tab:phases}). The ephemerides were constructed from pulsed flux lighturves spanning about 10 years of weekly {\it RXTE} observations. As such they predict the most likely time of X-ray outburst, which is presumed to reflect orbital modulation of the mass-transfer rate from the mass-donor star to the neutron star (the actual orbital ephemeris is not known for any SMC pulsars except SMC X-1).   Of the 10 pulsars in our fields with X-ray ephemerides from RXTE, 6 were detected far from their normal range of outburst-phase and with luminosity well below outburst levels. Two pulsars (SXP59, SXP7.78) were detected during their normal orbital phase range for X-ray outbursts.

We found no correlation between orbital phase (Table~\ref{tab:phases}) and the pulsed flux or pulse fraction as reported in Table~\ref{tab:pfrac}.  There are 5 pulsars (SXP8.88, SXP59, SXP138, SXP323, SXP756) for which both pieces of information are available, their orbital periods range from 28.47 d (SXP8.88) to 389 d (SXP756). It is not surprising that such a small sample cannot reveal a link, as we expect the pulse-fraction to be a strong function of orbital phase only over intervals when the accretion rate is changing rapidly. These measurements are however valuable for establishing the baseline pulsed flux for the individual pulsars at known positions along their respective orbits. 

The X-ray luminosity upper limit for the two known pulsars that were not detected is 1.1$\times$10$^{33}$\lx for SXP4.78 and 1.3$\times$10$^{33}$\lx for SXP82  (10 counts at the aimpoint, divided by the normalized exposure map at the target location). One of these undetected pulsars (SXP82.4) has an X-ray ephemeris (G08) according to which its phase $\phi$=0.863  (Table~\ref{tab:phases}) indicates it was far from the usual time of outburst.  None of the known pulsars that were detected as point-sources without identifiable modulation at the expected pulse-periods lay within $\pm0.2$ of ephemeris phase $\phi$=0.5 which is defined as outburst peak by G08. This supports the accepted picture of outbursts occurring close to periastron in the bright, recurrent Be-HMXBs, it also demonstrates that appreciable ``quiescent" emission can be detected throughout the orbit.

\subsection{Pulsars with Very Long Periods}
\label{sect:longperiod}
Several long-period pulsars were detected in our 100ksec  {\it Chandra}  observations, including two new examples, SXP892, and CXOU J005446.3-722523. SXP892 is a confirmed X-ray pulsar with typical HMXB optical counterpart; while in Section~\ref{sect:P4693} we reported  a 4693s modulation in J005446.3-722523 which also has a bright, early-type optical companion. As the period is far longer than any known SMC pulsars we discuss the implications for the HMXB population and for accretion physics, of the existence of such a pulsar, and the growing number with P$>$500s. 

The longest confirmed pulse period in the SMC is  P=1323s for the Be-HMXB pulsar RX J0103.6-7201 \citep{hp2005}. A total of 9 SMC pulsars have been found with periods in excess of 500s: 1323s, 967s \citep{haberl2008}, SXP892, RXJ0049.7-7323: 756s, RXJ0105.9-7203: 726s \citep{egger2008}, XMMU J005517.9-723853: 701s \citep{haberl2004},   XMMU J005535.2-722906: 645s, CXOUJ005736.2-721934: 565s, CXOU J005455.6-724510: 504s {\citep{edge2004}. In the Milky Way two very long-period X-ray pulsars with periods $>$1000s are known, both of them Be-HMXBs: SAX J2239.3+6116 at P=1247s \citep{zand2001},  SAX J0146.9+6121 at P=1412s \citep{hellier1994}. The longest X-ray period of all belongs to 2S0114+650 at P=10,008s which has a B1 supergiant optical counterpart, although it is uncertain whether its X-ray period is the neutron-star spin period, or a periodic variation in the accretion rate, driven by tidal modulation of the stellar wind \citep{koenigsberger2006}. 

X-ray pulsars are the product of close binary systems containing  a pair of massive stars. In the widely accepted model (e.g. \citealt{bh1991}), following the supernova, the newly formed neutron star (NS) evolves through three distinct epochs (ejector, propellor, accretor) characterized by changes to its spin-rate, through interaction with the wind of the companion.  Initially the NS spin period ($<<$1s) lengthens rapidly with energy losses dominated by magnetic dipole (and higher order) radiation (ejector phase). In the presence of a mass-losing companion the spin period lengthens far beyond the ($<$8.5s) regime \citep{young1999} occupied by radio pulsars, as the NS enters the propellor phase. Matter intercepting the NS with insufficient angular momentum to be accreted is flung away as it enters the magnetosphere resulting in a braking torque on the NS. Once the NS has spun-down to a critical value, accretion begins and the NS emerges as an X-ray pulsar. \cite{corbet1984} showed that most X-ray pulsars occupy a spin-orbit equilibrium, which is different for wind-fed and disk-fed systems. Monitoring studies (e.g \citealt{bildsten1997}) reveal that HMXB pulsars alternate between spin-up and spin-down states around their long-term equilibrium value. 

The maximum spin-period depends on the NS magnetic field, mass transfer rate and duration of the propellor phase, and thus involves stellar evolution calculations. \cite{urpin1998} find that for any realistic HMXB model, there exists a maximum NS spin-period of $P_{max}$=500s that can be reached in the main sequence lifetime of the mass-donor. The existence of long-period X-ray pulsars challenges accretion theory to explain periods that greatly exceed this limit. 
  
A theoretical investigation into long-period X-ray pulsars by \cite{iksanov2006} postulated a modification to NS spin-evolution that divides the propellor phase into supersonic and subsonic propellor phases. They report that this refinement removes the $\sim$500s barrier in specific circumstances, and reconciles the standard theory of close-binary evolution with the undoubted existence of long period HMXB pulsars. \cite{iksanov2006} also gives two scenarios that lead to X-ray pulsars with periods $>$1000s.  Firstly if the NS accretes spherically the spin-orbit equilibrium period is an order of magnitude longer than for disk-fed accretion. Secondly if the mass-donor star has recently increased its mass-loss rate, a slow-rotating NS companion (that had spun-down to equilibrium spin in a more tenuous wind) will be re-activated as a long period X-ray pulsar. 

Pulsars with periods longer than 1000s are difficult to identify for two main reasons, they are likely to be faint due to low accretion rates, and their wide orbits mean that outbursts will be infrequent. In addition long ($>>$1000s) observations are needed to unambiguously discern such long periods even in bright sources. Although {\it RXTE} has found that SXP756 has regular outbursts presumably linked to its orbital period (L05), we do not yet know if the much fainter long-period pulsars found by {\it XMM-Newton} and {\it Chandra} are periodically active.  The pulsars SXP892 (Fig.~\ref{fig:sxp892}), SXP756 (Fig.~\ref{fig:sxp756}), and the putative P=4693 source (Fig~\ref{fig:sxp4693}), exhibit pulse profiles consistent with pencil-beam dominated emission geometry (e.g. \citealt{SWR}) expected at low accretion rates.  SXP756 does show some fine sub-structure in the form of a interpulse $\sim$1/10th the strength of the main peak.  In contrast several of the 100-300s pulsars have wide peaks with complex structures that can exceed the un-pulsed flux level for most of the pulse cycle. 

Whether CXOU J005446.3-722523 can be confirmed as a $\sim$4000s pulsar or not, it is clear that sensitive observations with {\it Chandra} and {\it XMM-Newton} are providing a new window onto the long-period tail of the spin distribution. Continued monitoring and optical counterpart studies will be important to reveal whether the spin-orbit equilibrium for trans-$P_{max}$ pulsars differs from  pulsars in the $<$500s range, testing the suggestion of \cite{iksanov2006} that different physics comes into play. A large number of hard point-sources in the SMC are still unidentified and could plausibly be X-ray pulsars. The difficulty in detecting long-period systems is sufficient to suspect that the period distibution of un-discovered pulsars could depart significantly from the known examples. 

\subsection{Population Overview from X-ray Quantiles}
\label{sect:quantiles}
According to our quantile diagrams (Figure~\ref{fig:quantiles1}), the majority of the confirmed pulsars, (with and without detected pulsation in the {\it Chandra} data) cluster around power-law $\Gamma$=1, and neutral hydrogen column density $N_{H}\leq$10$^{21}$ ~cm$^{-2}$.  Several unidentified sources lie in this region also,  and are therefore consistent with being additional X-ray binary pulsars. In particular there are two objects with S/N$>$10 (one in Field 1 and one in Field 2) whose quantiles are indistinguishable from specific confirmed pulsars within the errors, CXOU J005057.1-731008 (RX J0050.9-7310) and CXOU J005215.4-731915. The nature of these sources in the context of their multi-wavelength counterparts and spectral properties is discussed in \cite{antoniou2010}. The two stark exceptions to this rule are SXP46.6 ($\Gamma=$2-2.5) and SXP342 ($\Gamma<$0.5), the latter is also unusual in being a bright gamma-ray source detected by {\it Integral} \citep{mcbride2007}. 

The upper left region of the quantile diagram contains a group of soft sources that extend outside the power-law grid, and are consistent with thermal models appropriate for stellar coronae.  In the figures we show a model grid for thermal bremsstrahlung, and the sources occupy a region approximately bounded by plasma temperatures in the range $kT$ = 0.4 -- 2 keV and $N_{H} <$10$^{21}$ \nh. We note that grids can be constructed for more sophisticated stellar coronal models such as Raymond-Smith or MEKAL, but the additional parameters become degenerate in the quantile space, and for low count sources they provide no additional information.

A group of sources with very large extinction occupy the upper right corner of the power-law grid. These are likely obscured active galactic nuclei (AGN) with significant internal extinction. The quantile parameters for these AGN-candidates are consistent with $\Gamma \sim$1.7 power-law spectra, typical of the class.

After accounting for stars, obscured AGN, and confirmed pulsars, we are still left with the majority of our sources. The unknown sources are not randomly distributed in the quantile-plane but lie predominantly in the region bounded by $N_{H}\geq$10$^{21}$ ~cm$^{-2}$ and $\Gamma \simeq$1--2 suggesting they are colliding-wind binaries, shocked-wind early-type stars and accretion powered sources  in the SMC (X-ray binaries, CVs) and beyond (AGN). X-ray binary pulsars can exhibit 
relatively soft quantiles with $\Gamma \sim 2$, see for example Figure 3 in \cite{mcgowan2007}, so it is plausible for some of these objects to be pulsars observed in a soft state. Cataclysmic Variables (CVs) and Low-Mass X-ray binaries (LMXBs) are also expected to appear in this region of the quantile diagram. Additionally CVs and LMXBs are likely to be faint with 0.5-10 keV  luminosities in the range 10$^{31-35}$, which is consistent with many of these sources. 
However, the greater proportion of the objects in this group are likely to be background AGN, which are typically characterized by power-law spectra with $\Gamma \sim$1.7 and low intrinsic absorption. The expected absorption distribution for background AGN follows the line of sight column through the SMC, which is $N_H$=1--5$\times$10$^{21}$\nh.

In order to provide a preliminary overview of the content of  our {\it Chandra} SMC DF source catalog (Table~\ref{tab:catalog}), we computed an estimate of the proportions and absolute numbers of occupying the above-discussed regions of the quantile diagram (Figure~\ref{fig:quantiles1}). The principle is to count the number of high signal-to-noise ratio (S/N$>$5) sources populating regions defined for: (a) thermal sources with $kT<1$KeV (b) hard-power-law sources ($\Gamma<$1.2), and (c) soft-power-law sources ($\Gamma>$1.5). In this way we obtain the fraction of high S/N sources consistent with spectral models that roughly correspond to stars, HMXB (and colliding wind stars), and AGN. Absolute numbers for the fields are then computed by multiplying this fraction by the total size of the catalog down to the completeness limit of our dataset. Uncertainties are computed by taking  $\sqrt{N}$ of the number of 5$\sigma$ sources and propagating them to the full sample. We defined a flux limit corresponding 10 count source at the Chandra/ACIS-I aimpoint in our 100 ksec stacked dataset, to approximate the completeness limit of our sample. 
We do this calculation using a flux limited sample because our flux calculations include correction for Chandra's changes in sensitivity and spectral response with distance from the aim-point. If a count limit were used instead, it would not be uniform across the field. Each source has a 0.5-8 keV flux computed based on its quantile values and the exposure map; thus a 10 count limit at the aimpoint translates to a flux limit $F_{x} >  $10$^{-15}$\flux over the whole field. 

In DF1 we find 10.7\% thermal sources, 53.3\% soft-power-law sources (of which 9\% are obscured), and 35.7\% hard power-law sources. Multiplying these percentages by the number (131) of sources down to our completeness limit  yields 14$\pm$6 thermal, 71$\pm$17 soft-power-law, and 46$\pm$13 hard power-law sources. 

The same analysis of DF2 yields 10.5\% thermal sources, 50\% soft-power-law (of which 7.9\% are obscured), and 39.5\% hard-power-law sources.  Multiplying these percentages by the number (107) of sources down to our completeness limit  yields 11$\pm$6 thermal, 54$\pm$16 soft-power-law, and 42$\pm$14 hard-power-law.  

An underlying assumption in the extrapolating the ratios of high S/N sources to absolute numbers, is that the luminosity functions of all 3 populations are the same over that range. Although this fact this is unlikely to be true, we note that the fainter (S/N$>$3) sources (top panels in Figure~\ref{fig:quantiles1}) appear to follow the same groupings as the high S/N examples. Therefore we quote these quantile-based population numbers as a guide only. Determination of the nature of each individual source is being sought by a multi-wavelength survey, which will enable the actual luminosity functions to be determined for each species (HMXB, Stars, AGN etc), which is beyond the scope of this paper. 

For comparison with the quantile results, we estimated the expected number of background AGN in our survey based on the work of \cite{kim2007}. For the same flux limit of $F_{x} > $10$^{-15}$\flux, \cite{kim2007} Figure 4 shows of order 10$^3$ AGN per sq. degree. Thus for an ACIS-I field of view (16'$\times$16') one expects to see $\sim$71 AGN per field.  The purpose of this comparison is to test whether the quantile results are reasonable in the light of what is known about the background AGN population.
The quantile-derived AGN number is remarkably consistent with our prediction based on the AGN luminosity function of  \cite{kim2007} . We note that DF2 contains a large X-ray bright supernova remnant which likely explains the slightly (23\%) lower AGN numbers in this field. Systematic errors were estimated by varying the boundary of the region in the quantile plane used to classify the AGN. Moving the boundary between $\Gamma$=1.2 --1.5  changed AGN vs XRB fractions by an insignificant amount compared to the statistical error.  
We note that Cataclysmic variables and LMXB are expected to occupy the same quantile parameter-space as the AGN, due to their soft power-law and/or hot thermal spectra. Given that our AGN-count is consistent with predictions, there is not a great deal of room for additional source populations. Thus we can tentatively constrain the total number of LMXB and CVs with soft spectra to be within the error bars of the AGN estimate. 

Based on the above results, we find that the DF {\it Chandra} observations have detected a large population of point-sources in the SMC, significantly in excess of the number of background AGN.  

\section{Conclusions}

We have presented the first results from deep (100 ksec) {\it Chandra} observations of two fields in the SMC Bar. These observations reach a detection limit of $\sim$10$^{-15}$\flux (for 5 ACIS-I net-counts). At the distance of the SMC (60kpc) this corresponds to a  luminosity of $L_{X}$=4.3$\times$10$^{32}$\lx (for $\Gamma$=1, $N_H$=5$\times$10$^{21}$\nh). 

We detected 19 High Mass X-ray Binaries (HMXB) in the SMC {\it Chandra} Deep Fields, including 11 previously known pulsars, 2 new pulsars and 4 other HMXB candidates (this number includes only objects with X-ray variability, and/or association with ROSAT sources, and optical counterparts).  The previously ambiguous identification of two close pairs of HMXBs  (1. SXP892=RX J0049.5-7310 and  RX J0049.2-7311.  2.  RXJ0052.1-7319=SXP15.3 and CXOU J005215.4-731915)  have been resolved thanks to Chandra's spatial resolution. The 2 newly discovered pulsars were in outburst, and their periods were measurable to a high degree of accuracy. We have therefore assigned the names SXP892 \& SXP326 following the convention used by the SMC {\it RXTE} monitoring project (L05, G08).  Three other sources showed periodic modulation (significance $>$95\%), one of these (CXOU J005446.3-722523) has a period of $\sim$4600s, a bright early-type optical counterpart, and hard X-ray quantiles, strongly suggesting an HMXB.  Additional HMXB candidates were detected with sufficient S/N to place them unambiguously on the quantile diagram (Figure~\ref{fig:quantiles1}) in the region inhabited by pulsars, in tight agreement with the power-law spectral model ($\Gamma$=1, nH$\sim$10$^{21}$) characteristic of SMC pulsars. We are confident in the identification of the 3 HMXB candidates (CXOU J005057.1-731008 = RX J0050.9-7310, CXOU J005215.4-731915, CXOU J005352.5-722639) which were identified by the combined evidence of X-ray spectral properties, other X-ray catalogs and positional association with bright early-type stars and emission-line objects (MA93).  

There are in total 40 {\it Chandra} sources  (17 in DF1, 23 in DF2) with bright stellar counterparts. Currently 21 out of these 40 have been identified as confirmed or probable HMXB, due to pulsation, X-ray variability (the 19 HMXBs mentioned above), and/or optical emission lines (providing 2 additional candidates). The published photometric data show that a further 12 counterparts occupy the same range of optical magnitude and $B-V$ color as the confirmed HMXBs, and are consistent with early-type ($<$B3) stars.  These objects likely include HMXBs, shocked-wind early-type stars and colliding wind binaries, pending a comprehensive analysis to be published in a companion paper \citep{antoniou2010}.  A comparable study by \cite{Naze2004} using  {\it Chandra} and {\it XMM-Newton} observations of NGC 346, a young open star-cluster which lies on the outskirts of the Bar of the SMC's Bar, revealed 97 point-sources. Based on hardness ratios and bright-star counterparts the majority were  proposed to be HMXB, although none has yet been seen in a bright outburst state.

By detecting 8 of the 10 pulsars for which there are ephemerides, with 6 of them outside of their normal orbital phase range for X-ray outbursts, we have demonstrated that a significant fraction (at least 60\%) of these systems have appreciable accretion driven X-ray emission during quiescence. To these six confirmed "quiescent-activity" pulsars we can add the 3 candidates as quiescent or persistent low-luminosity HMXBs,  whose 2-10 keV luminosities are substantially lower than normal Be-pulsar outbursts. The faintest confirmed Be-pulsar was SXP342, at $L_X$=3.8$\times$10$^{33}$\lx, while the candidate Be-HMXB CXO J005352.5-722693 (Section~\ref{sect:J005352}) was detected at $L_X$=5.2$\times$10$^{33}$\lx. Future efforts should be directed at increasing the number of HMXBs with orbital ephemerides, and extending the sample to cover the persistent low-luminosity population. 

The new sample of pulsars is consistent with a population about twice the size of the known ``active" pulsar population regularly seen by {\it RXTE}. It is  possible that some of these pulsars are HMXBs in a perpetually low luminosity state, similar to the Galactic HMXB X Persei and its brethren \citep{delgado2001}. The X-ray luminosity of X Per varies, but is seldom above $L_x$=10$^{35}$ \lx.  The theoretical work of \cite{okazaki2001} informs us that Be/NS systems with low orbital eccentricity accrete at a very low, almost constant rate. The Be star's circumstellar disk in such systems is permanently tidally truncated (at the 3:1 resonance) well within the neutron star's orbital radius, except during large mass-ejections from the star (the assumed cause of type II "giant" outbursts). X-ray pulsars with very short (less than a few seconds) periods that remain in the propellor regime probably also contribute to the quiescent population.  

The complete source catalog (Table~\ref{tab:catalog}) is available on-line in electronic form, it contains positions, count-rates, fluxes and quantiles for 394 X-ray sources (211 in DF1 \& 183 in DF2). 

This is the first of a series of papers that address the very faint X-ray source population in the SMC, in particular the X-ray luminosity function of HMXBs, and a search for CVs, stars and LMXBs in this low metallicity dwarf galaxy.

\section{Acknowledgements}
We thank J. McDowell,  V. Kalogera, and M. Smith for their expert input which made the {\it Chandra} observations possible.
This work has been partly supported by NASA LTSA Grant NAG5-13056, and {\it Chandra} grant GO6-7087A. SL thanks Gemini Observatory for supporting this research. Gemini is operated by the Association of Universities 
for Research in Astronomy, Inc., on behalf of the international Gemini partnership of Argentina, 
 Australia, Brazil, Canada, Chile, the United Kingdom, and the United States of America.

\clearpage 

\begin{deluxetable}{lllllll}
\tabletypesize{\scriptsize} 
\tablecaption{Chandra Observations \label{tab:obs}} 
\tablewidth{0pt}
\tablehead{\colhead{Field} & \colhead{ObsID} &  \colhead{Aimpoint} & \colhead{Roll} & \colhead{Start} &  & \colhead{Exposure} \\
\colhead{}         &  \colhead{}           & \colhead{RA, Dec}   & \colhead {degrees}      & \colhead{UT}    &    \colhead{MJD}  & \colhead{$\times10^3$ sec} }
\startdata
DF1 & 7155 & 00:53:34.50 -72:26:43.20   & 161  & Apr 25 2006 05:32 & 53850.23  & 50 \\
                        & 7327 & 00:53:34.50, 72:26:43.20   & 161 & Apr 26 2006 15:14 & 53851.63   & 50 \\
DF2 & 8479 & 00:50:41.40 -73:16:10.30 & 317 & Nov 21 2006 12:21 &  54060.51   & 45 \\
                        & 7156 & 00:50:41.40 -73:16:10.30 & 317 & Nov 22 2006 19:06 &  54061.80  & 39 \\ 
                        & 8481 & 00:50:41.40  -73:16:10.30 & 317 & Nov 23 2006 16:06 &  54062.67  & 16 \\
\enddata
\end{deluxetable}

\clearpage
\begin{landscape}
\begin{deluxetable}{llll|lcl|lll|l}
\tabletypesize{\scriptsize} 
\tablewidth{0pt}
\tablecaption{Chandra SMC Deep Fields Source Catalog \label{tab:catalog}} 
\tablewidth{0pt}
\tablehead{   
\colhead{Name} & \colhead{RA} & \colhead{Dec} & \colhead{r95} & \colhead{$F_{Bc}$}       & \colhead{$F_{Sc}$}                & \colhead{$F_{Hc}$}                  &    \colhead{}         &    \colhead{Quantiles}  &  \colhead{}           & \colhead{S/N} \\ 
\colhead{CXOU} & \colhead{}      & \colhead{}        & \colhead{}       & \colhead{0.5-8 keV} & \colhead{0.5-2.0 keV}       & \colhead{2.0-8 keV}            & \colhead{E50}    & \colhead{QDx}               & \colhead{QDy}    & \colhead{}  \\
\colhead{}            &\colhead{deg}& \colhead{deg} & \colhead {arcsec}    & \colhead{}                   & \colhead{10$^{-14}$\flux} & \colhead{}                            & \colhead{keV}     &     \colhead{}                   &  \colhead{}           & \colhead{} }
\startdata
J005252.2-721715 & 13.217712 & -72.287619 & 0.48 & 141.40$\pm$1.52 & 23.92$\pm$0.39 & 134.60$\pm$1.92 & 2.37$\pm$0.03 & -0.43$\pm$0.0077 & 1.03$\pm$0.01 & 95.1 \\ 
J004942.0-732314 & 12.425030 & -73.387348 & 0.45 & 64.23$\pm$1.04 & 7.62$\pm$0.21 & 63.66$\pm$1.28 & 2.85$\pm$0.04 & -0.31$\pm$0.0093 & 1.06$\pm$0.02 & 63.5 \\ 
J005151.9-731033 & 12.966380 & -73.176053 & 0.44 & 65.83$\pm$1.21 & 11.03$\pm$0.29 & 50.67$\pm$1.32 & 2.01$\pm$0.04 & -0.55$\pm$0.014 & 0.93$\pm$0.02 & 57.1 \\ 
J005044.6-731605 & 12.686208 & -73.268118 & 0.29 & 32.68$\pm$0.70 & 4.17$\pm$0.15 & 31.07$\pm$0.85 & 2.70$\pm$0.07 & -0.34$\pm$0.017 & 1.02$\pm$0.02 & 48.0 \\ 
J004929.7-731058 & 12.374103 & -73.182874 & 0.48 & 22.43$\pm$0.63 & 2.38$\pm$0.12 & 23.50$\pm$0.80 & 2.98$\pm$0.07 & -0.27$\pm$0.016 & 1.09$\pm$0.02 & 37.3 \\ 
J005323.8-722715 & 13.349580 & -72.454326 & 0.30 & 30.00$\pm$0.88 & 4.45$\pm$0.22 & 31.22$\pm$1.18 & 2.71$\pm$0.08 & -0.34$\pm$0.02 & 0.98$\pm$0.03 & 35.1 \\ 
J004913.5-731138 & 12.306384 & -73.193905 & 0.53 & 20.97$\pm$0.65 & 2.23$\pm$0.13 & 21.96$\pm$0.82 & 2.91$\pm$0.08 & -0.29$\pm$0.019 & 1.13$\pm$0.03 & 33.7 \\ 
J005057.1-731008 & 12.737953 & -73.168905 & 0.46 & 11.00$\pm$0.44 & 1.41$\pm$0.09 & 10.42$\pm$0.54 & 2.60$\pm$0.10 & -0.37$\pm$0.026 & 1.02$\pm$0.04 & 26.0 \\ 
J004948.2-732211 & 12.451046 & -73.369782 & 0.54 & 7.23$\pm$0.36 & 1.32$\pm$0.09 & 5.05$\pm$0.39 & 1.86$\pm$0.07 & -0.6$\pm$0.023 & 1.13$\pm$0.05 & 20.7 \\ 
J005437.1-722637 & 13.654878 & -72.443838 & 0.41 & 4.21$\pm$0.21 & 1.64$\pm$0.10 & 2.93$\pm$0.27 & 1.63$\pm$0.04 & -0.68$\pm$0.018 & 1.25$\pm$0.10 & 20.7 \\ 
J005331.7-722240 & 13.382157 & -72.378046 & 0.37 & 5.01$\pm$0.27 & 1.20$\pm$0.08 & 3.27$\pm$0.29 & 1.73$\pm$0.05 & -0.64$\pm$0.018 & 1.13$\pm$0.08 & 19.6 \\
J005456.1-722648 & 13.734005 & -72.446703 & 0.50 & 5.59$\pm$0.30 & 1.06$\pm$0.08 & 4.82$\pm$0.37 & 2.03$\pm$0.17 & -0.54$\pm$0.052 & 0.94$\pm$0.04 & 19.5 \\ 
J005428.9-723107 & 13.620442 & -72.518645 & 0.53 & 2.69$\pm$0.17 & 1.36$\pm$0.10 & 0.79$\pm$0.16 & 1.25$\pm$0.04 & -0.85$\pm$0.022 & 1.42$\pm$0.07 & 16.6 \\ 
J005215.4-731915 & 13.064582 & -73.320960 & 0.61 & 4.86$\pm$0.30 & 0.78$\pm$0.07 & 3.88$\pm$0.34 & 2.04$\pm$0.15 & -0.53$\pm$0.048 & 0.95$\pm$0.07 & 16.3 \\ 
J005445.3-722358 & 13.688995 & -72.399571 & 0.55 & 2.97$\pm$0.23 & 0.70$\pm$0.07 & 1.96$\pm$0.26 & 1.68$\pm$0.08 & -0.66$\pm$0.031 & 1.00$\pm$0.08 & 13.7 \\ 
J005352.4-723159 & 13.468608 & -72.533165 & 0.53 & 2.46$\pm$0.21 & 0.40$\pm$0.05 & 2.41$\pm$0.27 & 2.41$\pm$0.17 & -0.42$\pm$0.048 & 1.13$\pm$0.05 & 13.0 \\ 
J004910.7-731717 & 12.294808 & -73.288102 & 0.62 & 3.60$\pm$0.30 & 0.07$\pm$0.03 & 3.92$\pm$0.34 & 3.82$\pm$0.21 & -0.075$\pm$0.048 & 1.73$\pm$0.07 & 13.0 \\ 
J005337.8-722409 & 13.407884 & -72.402501 & 0.35 & 3.53$\pm$0.30 & 0.84$\pm$0.09 & 2.30$\pm$0.33 & 1.75$\pm$0.07 & -0.64$\pm$0.026 & 1.14$\pm$0.10 & 12.6 \\ 
J005045.0-731539 & 12.687698 & -73.261022 & 0.32 & 2.20$\pm$0.19 & 0.46$\pm$0.05 & 1.27$\pm$0.19 & 1.64$\pm$0.07 & -0.68$\pm$0.026 & 1.25$\pm$0.10 & 12.3 \\ 
J005153.2-723148 & 12.972059 & -72.530146 & 0.95 & 3.02$\pm$0.27 & 0.51$\pm$0.07 & 2.86$\pm$0.35 & 2.32$\pm$0.20 & -0.45$\pm$0.057 & 1.06$\pm$0.07 & 11.9 \\ 
J005351.2-721818 & 13.463641 & -72.305261 & 0.84 & 2.75$\pm$0.24 & 0.73$\pm$0.08 & 1.51$\pm$0.26 & 1.58$\pm$0.06 & -0.7$\pm$0.025 & 1.19$\pm$0.13 & 11.6 \\ 
J005446.3-722523 & 13.693035 & -72.423111 & 0.56 & 2.14$\pm$0.20 & 0.38$\pm$0.05 & 1.93$\pm$0.25 & 2.24$\pm$0.22 & -0.47$\pm$0.062 & 1.28$\pm$0.09 & 11.6 \\ 
J005214.0-731918 & 13.058339 & -73.321724 & 0.71 & 2.51$\pm$0.23 & 0.33$\pm$0.05 & 2.34$\pm$0.28 & 2.40$\pm$0.21 & -0.42$\pm$0.057 & 1.07$\pm$0.09 & 11.5 \\ 
J005457.2-723021 & 13.738400 & -72.506080 & 0.76 & 2.18$\pm$0.21 & 0.53$\pm$0.06 & 1.40$\pm$0.23 & 1.67$\pm$0.13 & -0.67$\pm$0.049 & 1.18$\pm$0.13 & 11.4 \\ 
J005448.9-722544 & 13.704161 & -72.429125 & 0.58 & 1.90$\pm$0.18 & 0.53$\pm$0.06 & 0.90$\pm$0.18 & 1.54$\pm$0.08 & -0.72$\pm$0.031 & 1.15$\pm$0.15 & 11.3 \\ 
J005411.5-723513 & 13.548300 & -72.586961 & 1.00 & 1.75$\pm$0.16 & 0.66$\pm$0.08 & 1.28$\pm$0.22 & 1.59$\pm$0.08 & -0.7$\pm$0.033 & 1.30$\pm$0.13 & 11.2 \\ 
J004905.1-731411 & 12.271583 & -73.236528 & 0.74 & 1.86$\pm$0.23 & 0.34$\pm$0.06 & 1.31$\pm$0.22 & 1.94$\pm$0.16 & -0.57$\pm$0.053 & 1.03$\pm$0.15 & 11.1 \\ 
J005235.0-722516 & 13.146195 & -72.421379 & 0.47 & 1.76$\pm$0.18 & 0.38$\pm$0.05 & 1.31$\pm$0.21 & 1.82$\pm$0.16 & -0.61$\pm$0.054 & 1.08$\pm$0.10 & 10.7 \\ 
J005233.6-722437 & 13.140237 & -72.410445 & 0.50 & 1.88$\pm$0.20 & 0.29$\pm$0.05 & 1.92$\pm$0.26 & 2.39$\pm$0.19 & -0.43$\pm$0.052 & 1.27$\pm$0.13 & 10.6 \\ 
J005504.3-722230 & 13.768110 & -72.375004 & 0.89 & 1.94$\pm$0.20 & 0.45$\pm$0.06 & 1.32$\pm$0.23 & 1.75$\pm$0.10 & -0.63$\pm$0.035 & 1.15$\pm$0.11 & 10.5 \\ 
\enddata
\tablecomments{This table contains a subset of the catalog ranked by S/N and truncated at S/N$>$10. The complete catalog in electronic form is available in the online edition of the journal. Column definitions: {\em RA, Dec}: Coordinates incorporating the full {\it Chandra} aspect solution. {\em r95}: Positional uncertainty for 95\% confidence radius, {\em $F_{Bc}$, $F_{Sc}$, $F_{Hc}$}: Energy flux in conventional broad, soft and hard bands using source-appropriate spectral model based on quantiles. {\em Quantiles}: Median photon energy (E50) and X,Y values for source in quantile parameter space \citep{hong2004}. {\em S/N}: Signal/Noise ratio for broad-band detection (net source counts divided by uncertainty)}
\end{deluxetable}
\clearpage
\end{landscape}

\begin{deluxetable}{lllllllll}
\tabletypesize{\scriptsize} 
\tablecaption{Pulsars Detected in the {\it Chandra} Data \label{tab:pulsars}}
\tablewidth{0pt}
\tablehead{ \colhead{Name} & \colhead{Source (CXOU)} &\colhead{Period(s)}   & \colhead{Signif(\%)} & \colhead{Net Counts}  & \colhead{RA\degr} &   \colhead{Dec\degr}  &   \colhead{r95\%('')}  }
\startdata 
\cutinhead{New Pulsars in Outburst}
SXP326 & J005252.2-721715   & 326.79$\pm$0.16   & 100.0 & 9292 & 13.217712  & -72.287619 & 0.484   \\
SXP892 & J004929.7-731058   & 894.36$\pm$0.98    & 100.0 & 1433 & 12.374103 & -73.182874 & 0.485   \\
\cutinhead{Known Pulsars Detected With Pulsations}
SXP138  & J005323.8-722715 & 138.925$\pm$0.028       & - & 1232 & 13.349580   & -72.454325 & 0.296    \\
SXP59    & J005456.1-722648 & 58.8336$\pm$0.0051    & - & 398    & 13.734005   & -72.446703 & 0.504     \\
SXP756  & J004942.0-732314  & 746.24$\pm$0.68          & - & 4116  & 12.425030  & -73.387348 & 0.451        \\
SXP323  & J005044.6-731605 & 317.26$\pm$0.12            & - & 2311 & 12.686208  & -73.268118 & 0.289     \\
SXP172  & J005151.9-731033 & 171.851$\pm$0.036       & - & 3322 & 12.966380  & -73.176053 & 0.443        \\                         
SXP8.88  & J005153.3-723148 &  8.89909$\pm$0.00008 & - &  184  & 12.972413  & -72.530032 &  0.949     \\
SXP15.3  & J005214.0-731918 &  15.239 $\pm$0.01        &  - & 147   & 13.058339  &  -73.321724 &   0.711 \\
\cutinhead{Known Pulsars Detected Without Pulsations:}
SXP46.6          & J005355.3-722645                      & -     &  -        & 63     & 13.480710  & -72.4460110   &  0.366 \\
SXP7.78          &  J005205.6-722604                     &  -    & -         & 76     & 13.023633  & -72.4344790   &   0.769     \\
SXP9.13         &  J004913.5-731138                      & -     & -         &1178 & 12.306384 & -73.1939050    &    0.527   \\
SXP342          & J005403.9-722633                       &   -    & -        &   36   & 13.516329 & -72.4425030   &  0.428    \\
\cutinhead{Known Pulsars in the Field Not Detected:}
SXP4.78         &                               &            &            &                     &                         &                         &       \\
SXP82.4        &                               &             &            &                     &                          &                         &      \\
\cutinhead{Other Periodic Candidates}
		        & J005446.3-722523 & 4693$\pm$20                       & 99.0 & 140 & 13.693000 & -72.423111	 & 0.562    \\ 
                          & J005331.7-722240  & 131.114$\pm$0.02      & 91.5 & 386  & 13.382157 & -72.378046 & 0.373    \\
                          & J005437.1-722637 & 10.4258$\pm$0.01  & 90.4 & 433  & 13.654878 & -72.443838 & 0.415     \\
\enddata
\tablecomments{$r95$ is the radius of the 95\% statistical confidence region on the source position. It does not include the 0.75" systematic aspect uncertainty.\\
Significance is computed for a blind-search in period range 6.5-10,000s, it is not quoted for pulsars with known periods.} 
\end{deluxetable}

\begin{deluxetable}{lllllll}
\tabletypesize{\scriptsize} 
\tablecaption{Spectral Fits (Sources with $>$250 net counts) \label{tab:spectralfits}} 
\tablewidth{0pt}
\tablehead{\colhead{CXOU} & \colhead{$n_H$}         & \colhead{$\Gamma$} & \colhead{Spectrum} & \colhead{$\chi^{2}_{\nu}$}  & \colhead{$f_{X}$(2-10 keV)}  & \colhead{Remarks} \\
\colhead{}  &   \colhead{10$^{21}$~~cm$^{-2}$}    & \colhead{}                     & \colhead{bins}          & \colhead{}                               & \colhead{$\times$10$^{-13}$\flux} & \colhead{}}
\startdata
J004913.5-731138 & 10.13 $\pm$ 1.33 & 1.34 $\pm$ 0.12 & 67 & 1.27 & 4.635     & SXP9.13 \\
J004929.7-731058 & 7.69   $\pm$ 1.01 & 0.96 $\pm$ 0.10 & 81 & 0.96 & 6.958     & SXP892\\
J004942.0-732314 & 5.95   $\pm$ 0.49 & 1.04 $\pm$ 0.05 & 197 & 0.98 & 17.97  & SXP756\\
J004948.2-732211 & 4.95   $\pm$ 2.14 & 1.73 $\pm$ 0.24 & 27 & 1.46 & 1.111     & AGN (1)  \\
J005044.6-731605 & 3.49   $\pm$ 0.59 & 0.89 $\pm$ 0.07 & 123 & 0.98 & 10.33  & SXP323\\
J005057.1-731008 & 5.20   $\pm$ 1.48 & 1.07 $\pm$ 0.1 & 42 & 0.86 & 3.012      & RXJ0050.9\\
J005151.9-731033 & 1.58   $\pm$ 0.29 & 1.14 $\pm$ 0.05 & 159 & 1.02 & 10.24 & SXP172 \\
J005215.4-731915 & 0.65   $\pm$ 1.50 & 0.93 $\pm$ 0.20 & 20 & 1.21 & 1.133     & HMXB? (2)\\
J005252.2-721715 & 3.44   $\pm$ 0.23 & 1.14 $\pm$ 0.03 & 322 & 1.25 & 29.34 & SXP326 \\
J005323.8-722715 & 2.30   $\pm$  0.77 & 0.76 $\pm$ 0.09 & 71 & 0.77 & 5.241  & SXP138 \\
J005331.7-722240 & 2.39   $\pm$ 1.52 & 1.84 $\pm$ 0.22 & 23 & 1.34 & 0.629   & (3)\\
J005428.9-723106 & 1.65   $\pm$ 0.94 & 2.91 $\pm$ 0.30 & 17 & 0.24 & 0.149   & Flare Star (4) \\
J005437.1-722637 & 1.57   $\pm$ 1.25 & 1.81 $\pm$ 0.21 & 26 & 1.4 & 0.667     & (3)  \\
J005456.1-722648 & 1.69   $\pm$ 1.54 & 0.90 $\pm$ 0.21 & 24 & 1.39 & 1.369   & SXP59 \\
\enddata
\tablecomments{(1) AGN based on spectral fit, lack of counterpart or periodicity. (2) HMXB based on spectrum, Optical and XMM \citep{haberl2008} counterpart, 
see Section~\ref{sect:J005215.4}. (3) AGN-like spectrum, no counterpart, candidate period see Table~\ref{tab:pulsars}. (4) \cite{laycock2009} }
\end{deluxetable}


\begin{deluxetable}{llllll}
\tabletypesize{\scriptsize} 
\tablecaption{Orbital Phase of Known Pulsars during {\it Chandra} Observations \label{tab:phases}} 
\tablewidth{0pt}
\tablehead{\colhead{Name} & \colhead{Field} & \colhead{Mid-Time} & \colhead{Period} & \colhead{Epoch} & \colhead{Phase} \\
\colhead{}  & \colhead{} & \colhead{MJD} & \colhead{days} & \colhead{MJD} & ($\phi$) }
\startdata
SXP138$^{a}$  & 1     & 53851  &   103.6     &  52400.5  & 0.001 \\
SXP59$^{a}$   &   1   &  53851 &   122.10   &  52306.1  & 0.653\\
SXP756$^{a}$  &   2   &  54061 &   389.9     &  52196.1  &  0.783 \\
SXP323$^{a}$  &  2    &  54061 &   116.6     &  52336.9  & 0.786 \\
SXP172$^{a}$  &   2   &  54061  &   -              &   -                &     -      \\                      
SXP8.88$^{a}$ &   1   &  53851   &  28.47     & 52392.2   & 0.240\\
SXP46.6$^{b}$   &   1  &  53851  & 137.36   & 52293.9      & 0.336 \\
SXP15.3$^{a}$  &  2   &  54061  &        -        &      -             &    -        \\
SXP7.78$^{b}$  &   1  &  53851   & 44.92     & 52250.9    & 0.621 \\
SXP9.13$^{b}$  &    2  &  54061   &  77.2      &  52380.5    & 0.768 \\
SXP4.78$^{c}$   &   1  &  53851  &    -            &     -                 &     -     \\
SXP82.4$^{c}$  &  1   &  53851  &  362.3    &  52089.0      & 0.863 \\         
\enddata
\tablecomments{Phases computed according to the ephemerides of the RXTE monitoring project \citep{galache2008}. The phase convention places the X-ray outburst peak at $\phi$=0.5, which is presumed to reflect the time of periastron. $a$: Pulsations identified.  $b$: Point source without pulsations. $c$: Not detected.}
\end{deluxetable}

\begin{deluxetable}{lllllllll}
\tabletypesize{\scriptsize}
\tablecaption{ROSAT HRI Counterparts from Sasaki 2000 catalog \label{tab:HRI}}
\tablewidth{0pt}
\tablehead{\colhead{CXOU} & \colhead{SXP} & \colhead{$r_{CXO}$/"} & \colhead{PSPC} & \colhead{HRI} & \colhead{$r_{HRI}$/"} & \colhead{Sep/"}  & \colhead{$r_{Total}$/"}  & \colhead{Remarks} }
\startdata
J005057.1-731008 &                   &   0.46   &  421 &  36   &  3.3    &   2.21  & 3.33      &   HMXB, AXJ0050.8-7310  \\
J005209.2-722553 &                   &   1.59   &  -      &  43    &  15.9 &   7.53  & 15.98    &   HRI:SMC X-3 $\neq$ CXO\\
J005214.0-731918 &  SXP15.3 &   0.71   &  453 &  44   &  1.1   &  0.46    &  1.31     &   RXJ0052.1-7319\\
\enddata
\end{deluxetable}

\begin{deluxetable}{llllllll}
\tabletypesize{\scriptsize} 
\tablecaption{ROSAT PSPC Counterparts from Haberl 2000 catalog  \label{tab:PSPC}} 
\tablewidth{0pt}
\tablehead{\colhead{CXOU} & \colhead{SXP} & \colhead{$r_{CXO}$/"} & \colhead{PSPC} & \colhead{$r_{PSPC}$/"} & \colhead{Sep/"}  & \colhead{$r_{Total}$/"}   & \colhead{Remarks} }
\startdata
 J004824.0-731918  &                    &  4.14  &    454   &   10.8  &  5.70    &     11.57  &   RXJ0048.4-7319, SNR 0046-73.5   \\
J004913.5-731138  &                    &  0.53  &    430  &    6.1  &   4.73   &     6.12   &   RXJ0049.2-7311, SXP9.13?\\
J004929.7-731058  & SXP892    &  0.49  &    427  &   5.5   &  2.82    &     5.52   &   RXJ0049.5-7310 \\
 J005057.1-731008  &                   &  0.46  &    421  &   4.5   &  2.45    &     4.52   &   RXJ0050.9-7310, AXJ0050.8-7310\\
 J005153.2-723148  & SXP8.89  &  0.95 &    265  &   45.8  &  20.31   &     45.81  &   RXJ0051.8-7231 \\
 J005352.5-722639  &                    &  0.51 &    242   &  25.7   & 21.52    &    25.71   & Be, HMXB?   \\
J005354.8-722722  &                    &  0.47 &     242  &   25.7  &  22.78   &     25.70  &  AGN?  \\
 J005355.3-722645  &  SXP46.6 &  0.37 &    242   &  25.7   & 17.57    &    25.70   &  RXJ0053.9-7227, XTEJ0053-724  \\
 J005428.8-722810  &                   &  1.96 &    248   &  45.7   & 19.76    &    45.74   &  RXJ0054.5-7228\\
J005433.0-722806  &                    &  1.83 &    248   &  45.7   & 1.23     &    45.74   &  RXJ0054.5-7228\\
J005436.5-722816  &                   &  2.29 &    248   &  45.7   & 17.63    &    45.76   &  RXJ0054.5-7228 \\
\enddata
\tablecomments{See text for discussion of multiple associations}
\end{deluxetable}

\begin{deluxetable}{llllll}
\tabletypesize{\scriptsize} 
\tablecaption{ASCA Counterparts from Yokogawa 2003 catalog  \label{tab:ASCA}} 
\tablewidth{0pt}
\tablehead{\colhead{CXOU} & \colhead{ASCA} & \colhead{Sep/"} & \colhead{PSPC} & \colhead{HRI} & \colhead{Remarks} }
\startdata
\cutinhead{Part 1: Identified HMXBs}
J004913.5-731138        &   26  &  31.5  &      430 &   -  &    AX J0049-732, RX J0049.0-7314    [SXP9.13?] \\                                                             
J004929.7-731058        &   27  &  16.8  &      427 &   -  &    RX J0049.4-7310                  [SXP892]   \\ 
J004942.0-732314        &   30  &  34.5  &      468 &   -  &     AX J0049.5-7323,RX J0049.7-7323  [SXP756]   \\           
J005044.6-731605        &   32  &  26.3  &      444 &   34 &    AX J0051-733,RX J0050.8-7316     [SXP323]   \\                   
J005057.1-731008        &   35  &  15.7  &      421 &   36 &    RX J0050.9-7310                  [HMXB]     \\                   
J005151.9-731033        &   40  &  32.2  &      424 &   41 &    AX J0051.6-7311                  [SXP172]   \\                   
J005214.0-731918        &   43  &  37.2  &      453 &   44 &    RX J0052.1-7319                  [SXP15.3]  \\                                      
J005355.3-722645        &   47  &  3.7   &      242 &   51 &    1WGA J0053.8-7226                [SXP46]    \\                                    
J005403.9-722633        &   47  &  40.4  &      242 &   51 &    1WGA J0053.8-7226                [SXP342]   \\                                                        
J005456.1-722504        &   51  &  51.4  &      241 &   58 &    XTE J0055-724, 1SAX J0054.9-7226 [SXP59]    \\                                    
\cutinhead{Part 2: Positional matches without positive identification.}
J004845.4-731842        &   23  &  37.5 &       454 &    -  &   SNR 0046-735              \\                                                                   
J0049-0.0-731314        &   25  &  36.5 &       437 &    -  &   SNR 0047-735              \\                                                                          
J004927.1-731211        &   26  &  39.1 &       430 &    -  &   AX J0049-732              \\                                                
J004929.8-731150        &   26  &  50.3 &       430 &    -  &   AX J0049-732              \\                                              
J004934.7-731049        &   27  &  36.6 &       427 &    -  &   RX J0049.4-7310           \\
J004937.4-732228        &   30  &  25.6 &       468 &    -  &   AX J0049.5-7323           \\    
J004948.2-732211        &   30  &  37.1 &       468 &    -  &   AX J0049.5-7323           \\
J005037.2-731530        &   32  &  20.9 &       444 &   34  &   AX J0051-733              \\                 
J005044.8-731618        &   32  &  37.2 &       444 &   34  &   AX J0051-733              \\                       
J005045.0-731539        &   32  &  20.6 &       444 &   34  &   AX J0051-733              \\                       
J005045.0-731519        &   32  &  33.0 &       444 &   34  &   AX J0051-733               \\                      
J005047.6-731619        &   32  &  45.2 &       444 &   34  &   AX J0051-733               \\                      
J005047.8-731736        &   34  &  23.7 &           &   -   &                              \\                                                                                         
J005048.7-732115        &   36  &  45.3 &       461 &   38  &   SNR 0049-736               \\                                                                                 
J005049.6-730926        &   35  &  44.1 &       421 &   36  &   RX J0050.9-7310            \\                                     
J005049.7-731608        &   32  &  45.9 &       444 &   34  &   AX J0051-733               \\                      
J005052.4-732135        &   36  &  47.2 &       461 &   38  &   SNR 0049-736               \\                                                                                 
J005054.3-731741        &   34  &  30.2 &           &   -   &                              \\                                                                                         
J005054.5-731027        &   35  &  21.1 &       421 &   36  &   RX J0050.9-7310            \\            
J005055.0-731802        &   34  &  50.1 &           &   -   &                              \\                                                                                         
J005055.4-732036        &   36  &  21.6 &       461 &   38  &   SNR 0049-736               \\                                                                                 
J005059.0-732055        &   36  &   4.2 &       461 &   38  &   SNR 0049-736               \\                                                                                 
J005102.9-732126        &   36  &  37.7 &       461 &   38  &   SNR 0049-736               \\                                                                                 
J005103.4-732139        &   36  &  49.6 &       461 &   38  &   SNR 0049-736               \\                                                                                 
J005104.4-732108        &   36  &  30.5 &       461 &   38  &   SNR 0049-736               \\                                                                                 
J005145.1-731038        &   40  &   5.2 &       424 &   41  &   AX J0051.6-7311             \\                                            
J005215.4-731915        &   43  &  30.2 &       453 &   44  &   RX J0052.1-7319            \\                                            
J005218.9-731940        &   43  &  37.3 &       453 &   44  &   RX J0052.1-7319            \\                                             
J005225.9-731921        &   43  &  22.7 &       453 &   44  &   RX J0052.1-7319            \\                                             
J005352.5-722639        &   47  &  12.3 &       242 &   51  &   1WGA J0053.8-7226          \\                                                 
J005354.8-722722        &   47  &  40.5 &       242 &   51  &   1WGA J0053.8-7226          \\                                                
J005357.6-722605        &   47  &  38.4 &       242 &   51  &   1WGA J0053.8-7226          \\                                                 
J005358.5-722614        &   47  &  31.0 &       242 &   51  &   1WGA J0053.8-7226          \\                                                
J005406.0-722650        &   47  &  49.7 &       242 &   51  &   1WGA J0053.8-7226          \\  
\enddata
\tablecomments{{\it Chandra} and {\it ASCA} coordinates  were matched using a 52" search radius, determined from the 95$^{th}$ percentile of the distribution of separations between {\it ASCA} and {\it ROSAT} HRI coordinates in the complete ASCA catalog of the SMC \citep{yokogawa2003}. Many of the {\it ASCA} sources have multiple {\it Chandra} positional coincidences, those which are not definitely identified with specific HMXBs are listed in Part 2 of the table. The PSPC, HRI, and Remarks columns are drawn directly from the \cite{yokogawa2003} catalog, modified in the case of Part 1, with the SXP designation added if known. The PSPC and HRI source numbers derive from Haberl et al. (2000) and Sasaki et al (2000), and can be compared with Tables~\ref{tab:PSPC}, \&~\ref{tab:HRI}. }
\end{deluxetable}

\begin{deluxetable}{llllll}
\tabletypesize{\scriptsize} 
\tablecaption{XMM-Newton Counterparts from \cite{haberl2008}  \label{tab:XMM}} 
\tablewidth{0pt}
\tablehead{\colhead{CXOU} & \colhead{$r_{CXO}$/"} & \colhead{$r_{XMM}$/"}    & \colhead{$r_{Total}$/"}   & \colhead{Sep/"}  & \colhead{Remarks} }
\startdata
J004913.5-731138    & 0.53   &        0.55   &  2.01   &   1.29   &    RXJ0049.2-7311, SXP9.13          \\
J004929.7-731058    & 0.49   &        0.81   & 2.09    &  1.04    &    RXJ0049.5-7310, SXP892     \\ 
J004942.0-732314    & 0.45   &        0.23   &   1.93   &   0.07   &   RXJ0049.7-7323, SXP755   \\    
J005044.6-731605    & 0.29   &       0.18    &  1.89    &  0.13    &   RXJ0050.8-7316,  SXP323     \\ 
J005151.9-731033    & 0.44   &       0.59    &  2.00    &  1.19    &    AXJ0051.6-7311,  SXP172    \\
J005215.4-731915    & 0.61   &        0.82   &   2.12    &  1.21   &    Close to SXP 15.3 (Section~\ref{sect:J005215.4})    \\
J005205.6-722604    & 0.77   &       0.14    &  2.02    &  1.17    &    SMCX-3,  SXP7.78                    \\
J005252.2-721715    & 0.48   &       0.21    &  1.93    &  0.66    &   XMMUJ005252.1-721715,   SXP326   \\
J005323.8-722715    & 0.30   &        0.16   &  1.89    &   0.42   &    CXOUJ005323.8-722715,   SXP138   \\   
J005403.9-722633    & 0.43   &       0.27    &  1.93    &  0.27    &   XMMUJ005403.8-722632,   SXP342    \\
J005456.1-722648    & 0.50   &       0.36    &  1.96    &  1.05    &    XTE J0055-724,  SXP59       \\
\enddata
\end{deluxetable}

\begin{deluxetable}{llllll}
\tabletypesize{\scriptsize} 
\tablecaption{Emission-Line Star Counterparts in MA93 Catalog  \label{tab:MA93}} 
\tablewidth{0pt}
\tablehead{\colhead{CXOU} & \colhead{MA93} & \colhead{Sep/"}  & \colhead{$r_{CXO}$/"} & \colhead{$r_{Total}$/"}   & \colhead{Remarks} }
\startdata
J004849.0-731625 &  258 & 2.29 & 4.30 & 5.0  &  Be?  \\
J004929.7-731058 &  300 & 0.62 & 0.49 & 2.7  &  SXP892  \\             
J004942.0-732314 &  315 & 1.05 & 0.45 & 2.6  &  SXP756 \\
J005012.2-731156 &  341 & 0.77 & 1.51 & 3.0 &  Be? \\
J005036.0-731739 &  374 & 0.28 & 0.48 & 2.7  &  Be? \\
J005044.6-731605 &  387 & 0.59 & 0.29 & 2.6  & SXP 323 \\
J005057.1-731008 &  414 & 0.87 & 0.46 & 2.7  &  Be?, RX J0050.9-7310  \\
J005117.0-731606 & 448  & 0.59 & 0.63 & 2.7 & Be?            \\ 
J005151.9-731033 &  504 & 1.10 & 0.44 & 2.6  & SXP172 \\ 
J005153.2-723148 &  506 & 0.20 & 0.95 & 2.8  & SXP8.89 \\
J005205.6-722604 &  531 & 1.37 & 0.77 & 2.7  & SMC X-3 \\
J005214.0-731918 &  552 & 0.65 & 0.71 & 2.7  & SXP15.3 \\
J005237.3-722732 &  590 & 1.09 & 0.93 & 2.8  &  Be?             \\
J005323.8-722715 &  667 & 0.97 & 0.29 & 2.6  & SXP138 \\
J005329.2-723348 &  677 & 1.40 & 2.60 & 3.7  & Be?       \\
J005352.5-722639 &  717 & 0.84 & 0.51 & 2.7  & Be(1) \\
J005446.3-722423 & 798  & 2.56 & 0.56 &2.7   & Be? \\ 
J005456.1-722648 &  810 & 1.45 & 0.50 & 2.7  & SXP59 \\
\enddata
\tablecomments{The matching radius $r_{Total}$ is the quadrature sum of $r_{CXO}$, 0.75" aspect uncertainty and 2.5"  tolerance on the MA93 coordinates. New X-ray plus emission-line counterparts are indicated by ``Be?" in the remarks column. (1) Confirmed Be star, see Section~\ref{sect:J005352}}
\end{deluxetable}

\begin{deluxetable}{llllllllll}
\tabletypesize{\scriptsize} 
\tablecaption{Optical Counterparts V$<$17 in MCPS Catalog \label{tab:MCPS}} 
\tablewidth{0pt}
\tablehead{\colhead{CXOU} &  \colhead{$r_{CXO}$/"}   &  \colhead{Sep/"}  & \colhead{$r_{Total}$/"}  & \colhead{net counts} & \colhead{RA} & \colhead{Dec} & \colhead{Vmag} & \colhead{B-V} &\colhead{ID} }
\startdata
J004849.0-731625 & 4.31 & 1.84 & 4.37 &  9.0 & 12.2057 & -73.27337 & 14.56 & -0.06 &  [MA93]258 \\
J004913.5-731138 & 0.53 & 0.83 & 0.92 & 1178.2 & 12.3070 & -73.19376 & 16.44 & 0.19 & SXP9.13 \\ 
J004922.2-732006 & 3.31 & 1.86 & 3.40 &  7.2 & 12.3409 & -73.33498 & 16.62 & 0.09   &  C \\ 
J004929.7-731058 & 0.48 & 0.43 & 0.89 & 1433.9 & 12.3745 & -73.18284 & 16.15 & 0.20 & SXP892 \\ 
J004941.3-731353 & 1.40 & 0.94 & 1.59 &  9.8 & 12.4215 & -73.23132 & 15.84 & 1.89   &  R \\
J004942.0-732314 & 0.45 & 0.08 & 0.88 & 4116.9 & 12.4251 & -73.38734 & 14.86 & 0.15  & SXP755 \\ 
J004958.8-731634 & 0.97 & 0.43 & 1.23 &  5.8 & 12.4955 & -73.27627 & 14.53 & 0.57     & R  \\ 
J005002.4-731210 & 1.66 & 0.61 & 1.82 &  7.7 & 12.5104 & -73.20317 & 15.74 & 0.94      & R \\ 
J005004.4-731426 & 1.41 & 0.44 & 1.59 &  3.7 & 12.5187 & -73.24079 & 15.50 & 0.06      & C \\ 
J005012.2-731156 & 1.51 & 1.07 & 1.69 &  8.2 & 12.5520 & -73.19892 & 15.31 & 0.16 &  [MA93]341  \\ 
J005035.5-731401 & 0.86 & 0.64 & 1.14 &  4.7 & 12.6485 & -73.23401 & 15.99 & 0.11    &  C   \\ 
J005036.0-731739 & 0.48 & 0.35 & 0.89 & 10.2 & 12.6506 & -73.29428 & 15.61 & 0.08  & [MA93]374\\ 
J005044.6-731605 & 0.29 & 0.31 & 0.80 & 2311.8 & 12.6865 & -73.26814 & 15.48 & -0.11 & SXP323\\ 
J005047.8-731736 & 0.62 & 0.25 & 0.97 &  4.8 & 12.6994 & -73.29346 & 16.58 & 0.01 &  C\\ 
J005047.9-731817 & 0.39 & 0.26 & 0.85 & 41.9 & 12.7001 & -73.30496 & 15.07 & 0.12 & C \\ 
J005057.1-731008 & 0.46 & 0.37 & 0.88 & 700.1 & 12.7382 & -73.16883 & 14.35 & 0.08   & [MA93]414 PSPC\\ 
J005105.7-731312 & 0.96 & 0.77 & 1.22 &  7.7 & 12.7741 & -73.21983 & 15.70 & 0.07 &  C \\ 
J005117.0-731606 & 0.63 & 0.81 & 0.98 &  9.3 & 12.8217 & -73.26843 & 15.00 & 0.18 & [MA93]448\\ 
J005151.9-731033 & 0.44 & 0.65 & 0.87 & 3322.7 & 12.9670 & -73.17608 & 14.45 & -0.07 &  SXP172\\ 
J005153.2-723148 & 0.95 & 0.52 & 1.21 & 183.9 & 12.9716 & -72.53010 & 14.38 & 0.41 &  SXP8.88\\ 
J005205.6-722604 & 0.77 & 0.19 & 1.07 & 76.0 & 13.0237 & -72.43443 & 14.91 & -0.00 & SMC X-3\\ 
J005207.5-722126 & 1.88 & 1.40 & 2.03 & 26.4 & 13.0312 & -72.35704 & 15.20 & -0.07 & C \\ 
J005214.0-731918 & 0.71 & 0.44 & 1.03 & 146.9 & 13.0587 & -73.32179 & 14.49 & 0.14 & SXP15.3 \\ 
J005215.4-731915 & 0.61 & 0.37 & 0.97 & 286.7 & 13.0645 & -73.32086 & 15.90 & -0.14 &  C \\ 
J005230.6-731532 & 5.26 & 4.29 & 5.31 &  6.2 & 13.1313 & -73.25964 & 16.22 & 1.00 &  R \\ 
J005237.3-722732 & 0.93 & 0.48 & 1.19 & 12.0 & 13.1556 & -72.45897 & 14.98 & 0.01 &  [MA93]590\\ 
J005245.1-722844 & 0.91 & 0.98 & 1.18 & 12.0 & 13.1879 & -72.47871 & 14.92 & 0.00 &  C\\ 
J005251.4-731451 & 6.52 & 3.95 & 6.57 &  8.7 & 13.2180 & -73.24753 & 15.85 & 0.87  & R \\ 
J005252.2-721715 & 0.48 & 0.72 & 0.89 & 9292.4 & 13.2177 & -72.28742 & 16.62 & -0.20 & SXP326 \\ 
J005314.6-721848 & 1.83 & 1.77 & 1.98 & 24.0 & 13.3117 & -72.31398 & 16.39 & 0.61 &  R \\ 
J005323.8-722715 & 0.30 & 0.17 & 0.81 & 1232.0 & 13.3496 & -72.45428 & 16.19 & -0.09 & SXP138 \\ 
J005329.2-723348 & 2.60 & 1.00 & 2.71 & 11.5 & 13.3724 & -72.56333 & 14.62 & 0.01 &   [MA93]677 \\ 
J005331.8-721845 & 5.09 & 4.78 & 5.15 &  6.4 & 13.3806 & -72.31389 & 16.03 & -0.17 &  C \\ 
J005349.0-722506 & 0.50 & 0.32 & 0.90 & 13.2 & 13.4544 & -72.41837 & 16.62 & 1.07  & R \\ 
J005355.3-722645 & 0.37 & 0.38 & 0.83 & 62.9 & 13.4808 & -72.44591 & 14.72 & -0.07  & SXP46.6 \\ 
J005403.9-722633 & 0.43 & 0.23 & 0.86 & 36.3 & 13.5163 & -72.44244 & 14.94 & 0.01 & SXP342 \\ 
J005419.2-722049 & 5.78 & 1.92 & 5.83 &  3.6 & 13.5787 & -72.34728 & 16.97 & -0.05 & C \\ 
J005446.3-722523 & 0.56 & 0.55 & 0.94 & 140.4 & 13.6933 & -72.42298 & 15.36 & 0.14 & [MA93]798, P=4693 \\ 
J005456.1-722648 & 0.50 & 0.88 & 0.90 & 398.9 & 13.7344 & -72.44649 & 15.27 & -0.05 & SXP59\\ 
J005507.7-722241 & 0.94 & 0.57 & 1.20 & 114.9 & 13.7827 & -72.37801 & 14.38 & -0.13 & C \\ 
\enddata
\tablecomments{The matching radius $r_{Total}$ is the quadrature sum of $r_{CXO}$, and 0.75" aspect uncertainty. The $ID$ column shows confirmed HMXBs, objects identified with MA93 stars in Table~\ref{tab:MA93}, and the codes $C$=`blue early-type candidate', and $R$=`red-outlier' based on the $B-V$ color index compared to the confirmed HMXBs.}
\end{deluxetable}

\begin{deluxetable}{lllll}
\tabletypesize{\scriptsize} 
\tablecaption{Pulsed Fluxes and Pulse Fractions\label{tab:pfrac}} 
\tablewidth{0pt}
\tablehead{\colhead{Name} & \colhead{P} & \colhead{$\bar{f}$}  & \colhead{$f_p$} & \colhead{PF} \\
\colhead{} & \colhead{(s)} & \colhead{$count ~ ks^{-1}$}  & \colhead{$count ks^{-1}$} & \colhead{$f_{p}/\bar{f}$}}
\startdata
J005437.1              & 10.4258  &  4.26  &  1.95$\pm$0.819  &  0.458 \\
J005331.7                & 131.115  &  3.87  &  2.50$\pm$0.732  &  0.645 \\ 
SXP138 & 138.947  &  12.3  &  5.67$\pm$1.39  &  0.462 \\ 
SXP326 & 326.798  &  91.2  &  41.1$\pm$3.85  &  0.450 \\ 
SXP59   & 58.8336  &  3.86  &  2.72$\pm$0.714  &  0.704 \\ 
 J005446.3                & 4693.43  &  1.44  &  1.16$\pm$0.427  &  0.809 \\ 
SXP8.88& 8.89909  &  1.80  &  1.16$\pm$0.549  &  0.644 \\ 
SXP172 & 171.85  &  30.0  &  6.20$\pm$2.32  &  0.207 \\ 
SXP323 & 317.267  &  22.0  &  8.90$\pm$1.82  &  0.405 \\ 
SXP756 & 746.247  &  39.4  &  10.3$\pm$2.63 &  0.261 \\ 
SXP894 & 894.36  &  13.5  &  6.89$\pm$1.49  &  0.511 \\ 
SXP15.3& 15.2393  &  2.16  &  0.849$\pm$0.610  &  0.393 \\ 
\enddata
\end{deluxetable}


\begin{figure*}
\begin{center}
\includegraphics[angle=0,width=16cm]{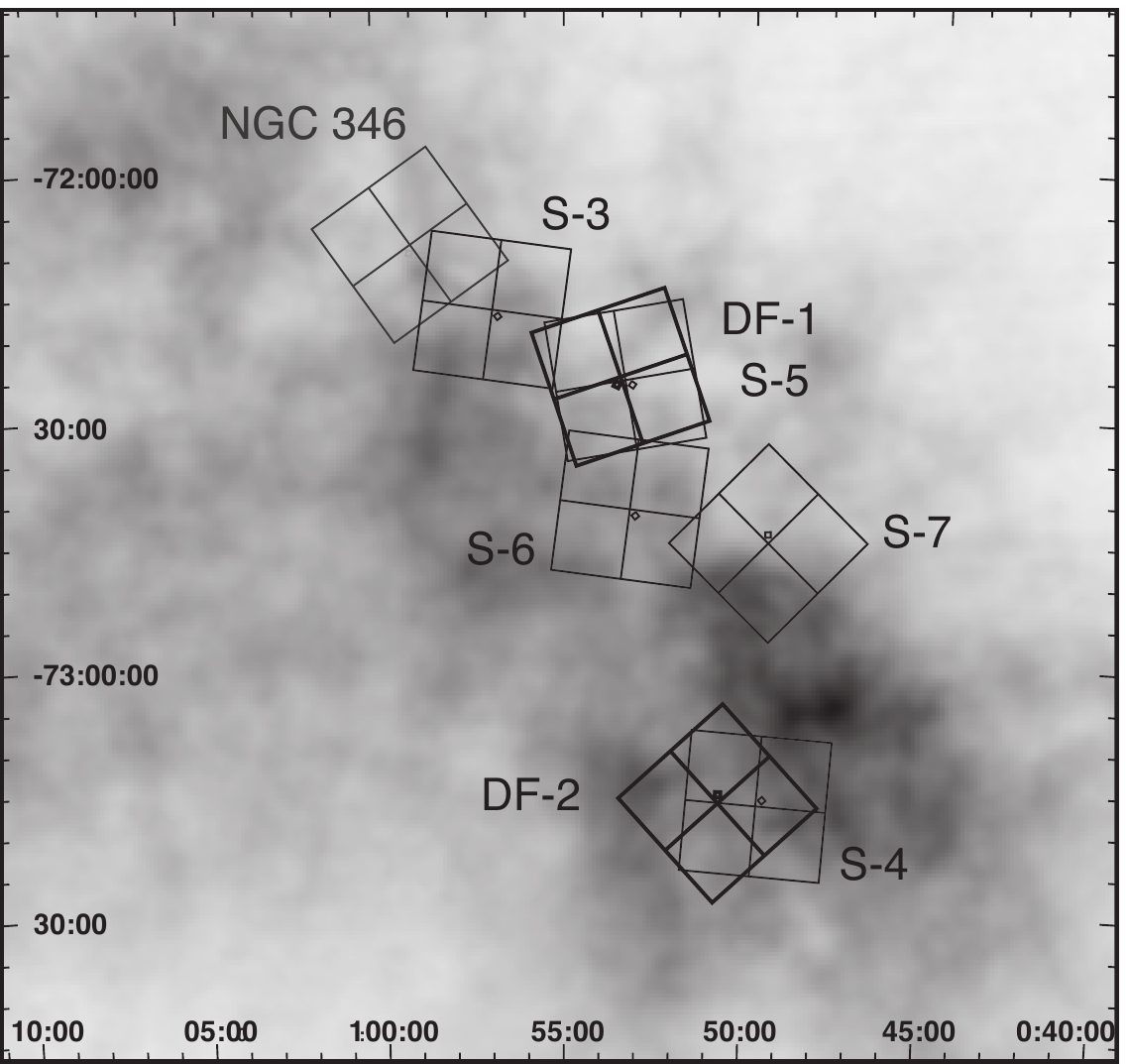}
\caption{{\bf {\it Chandra} observations in the SMC Bar. The 100 ksec fields presented in this paper are labelled DF-1 and DF-2. The shallow (10 ksec) survey of Antoniou et al., 2009 are S-3--7, and the SMC open Cluster NGC 346 (100 ksec)  observed by Naze et al., 2004 is shown in grey. Outlines of the 4 ACIS-I CCDs are plotted, with the appropriate roll-angle, and the {\it Chandra} aim-point is marked. Background is Stanimirovic et al. (1999) HI image.}}
\label{fig:surveyfields}
\end{center}
\end{figure*}

\begin{figure*}
\begin{center}
\includegraphics[angle=0,width=16cm]{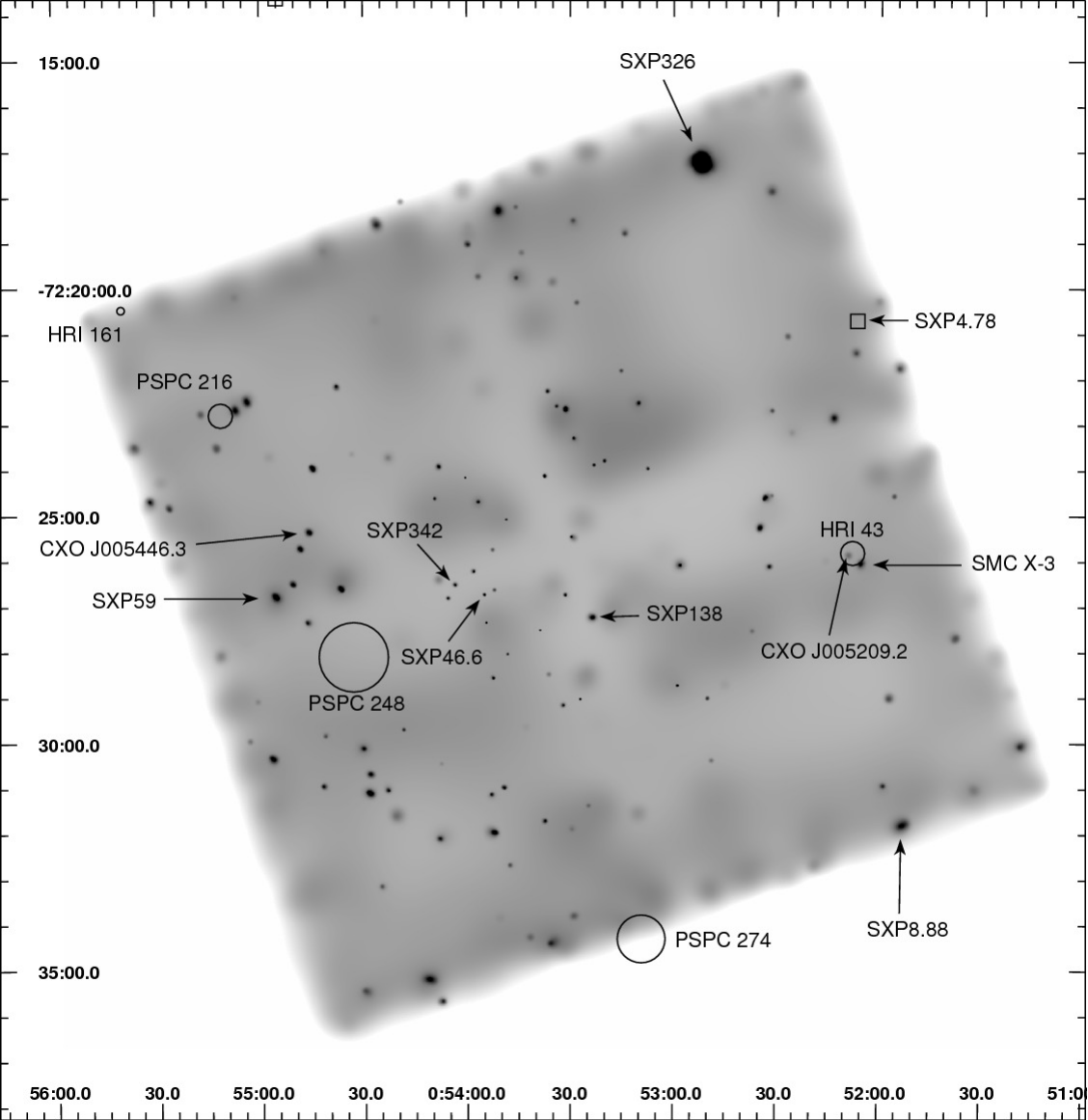}
\caption{{\bf Adaptively smoothed broad-band (0.3-8 keV) {\it Chandra} image of Deep Field 1 (DF1), showing locations of Pulsars, HMXB candidates and unidentified {\it ROSAT} sources.  The ACIS-I field measures 16'$\times$16' on each side. The many unidentified point sources are a mixture of quiescent X-ray binaries (17 {\it Chandra} sources in this field have bright (V$<$17) optical counterparts), stars,  and background AGN.  }}
\label{fig:imageDF1}
\end{center}
\end{figure*}

 \begin{figure*}
\begin{center}
\includegraphics[angle=0,width=16cm]{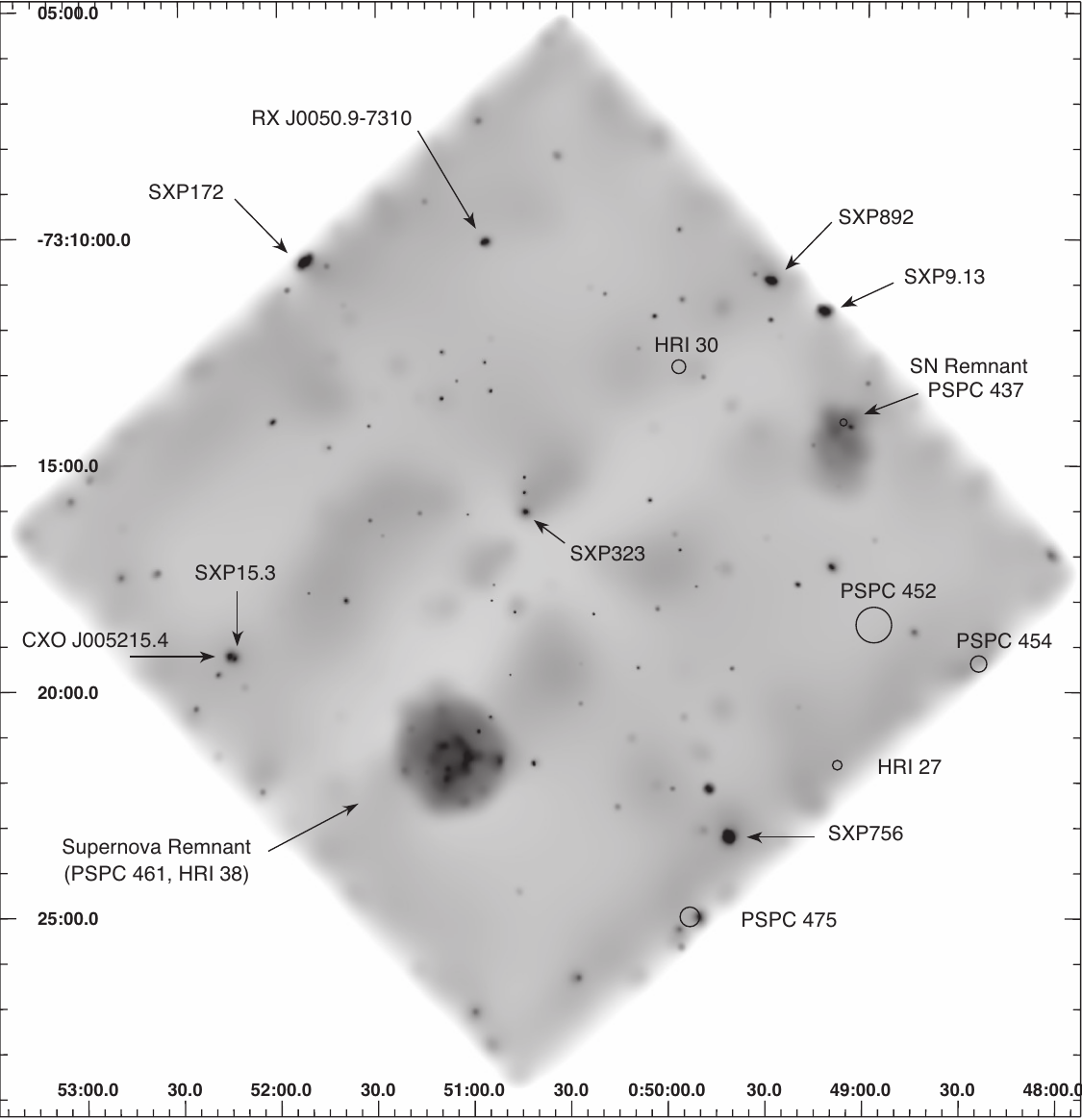}
\caption{{\bf Adaptively smoothed broad-band (0.3-8 keV) {\it Chandra} image of Deep Field 2 (DF2), showing locations of Pulsars, HMXB candidates and unidentified {\it ROSAT} sources.  The ACIS-I field measures 16'$\times$16' on each side.  Note the two supernova remnants, one is the brightest object in the image, the other is much fainter and located to the upper right.  Pulsar SXP15.3 has a close (but resolved) neighbor which is itself a candidate HMXB. The many unidentified point sources are a mixture of quiescent X-ray binaries (23 {\it Chandra} sources in this field have bright (V$<$17) optical counterparts), stars, and background AGN.  }}
\label{fig:imageDF2}
\end{center}
\end{figure*}
\clearpage

\begin{figure}
\begin{center}
\includegraphics[angle=0,width=12cm]{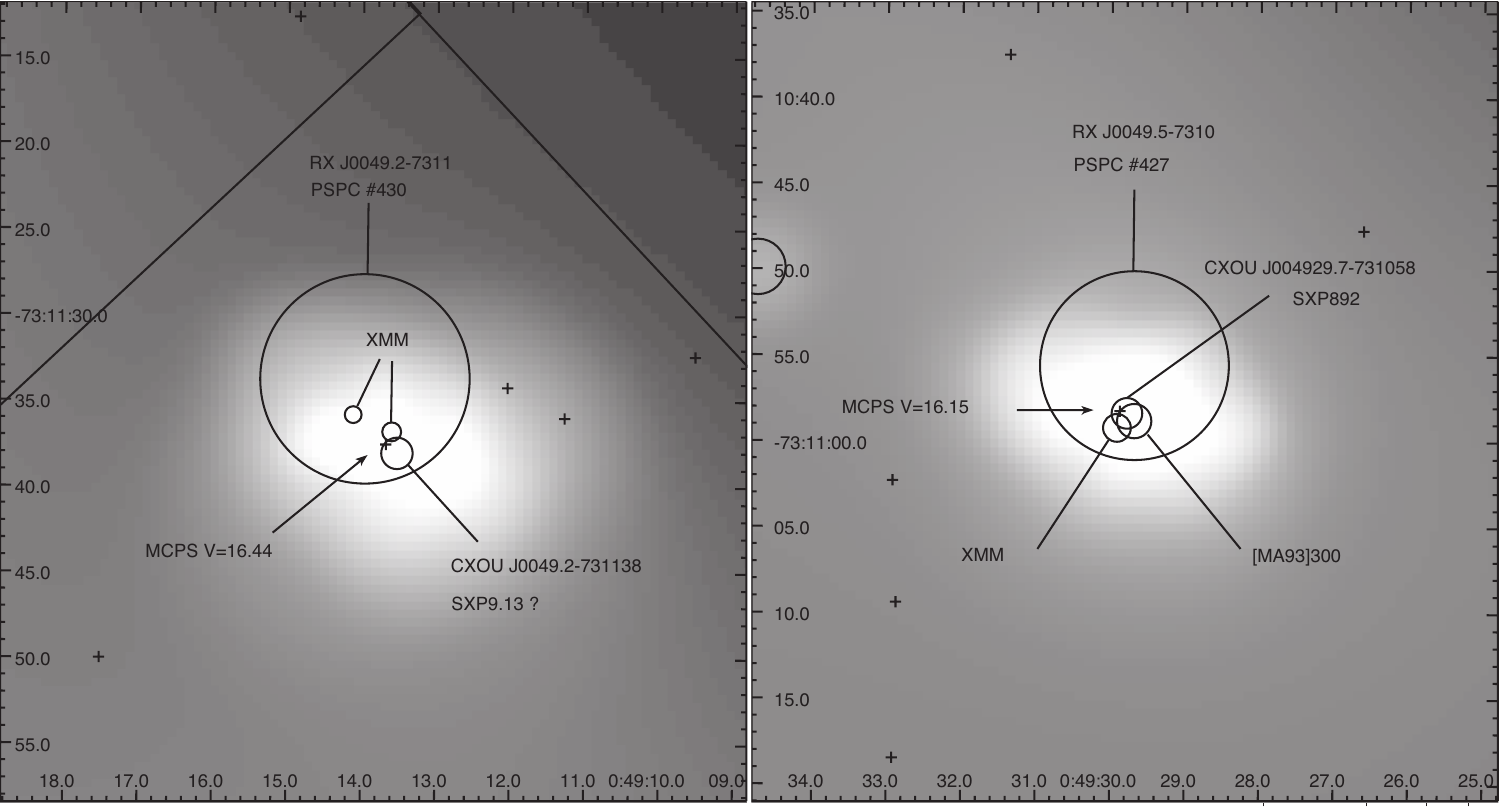}
\caption{{\bf  {\it Chandra} ACIS images of the neighboring bright X-ray sources CXOU J0049.2-731138 = RXJ 0049.2-7311 (Left) and CXOU J004929.7-731058 = RX J0049.5-7310 (Right). Both sources lie inside the ASCA error circle for the pulsar SXP9.13 \citep{filipovic2000}. We find that CXOU J004929.7-731058 is a new 892s pulsar, thus RX J0049.2-7311 is probably SXP 9.13, although some ambiguity remains as the 9.13s pulsations have not been localized to sufficient precision.   Labeled circles indicate positional error radii for Chandra, {\it ROSAT} and {\it XMM-Newton} \citep{haberl2008} detections.  Bright stars (V$<$17) in the MCPS catalog are indicated by crosses, and MA93 emission-line stars are shown as circles with 1" radius. The edge of the ACIS field is visible in left panel, and images are elongated due to distance from Chandra's aimpoint.  }}
\label{fig:imagesxp892}
\end{center}
\end{figure}

\begin{figure}
\begin{center}
\includegraphics[angle=0,width=12cm]{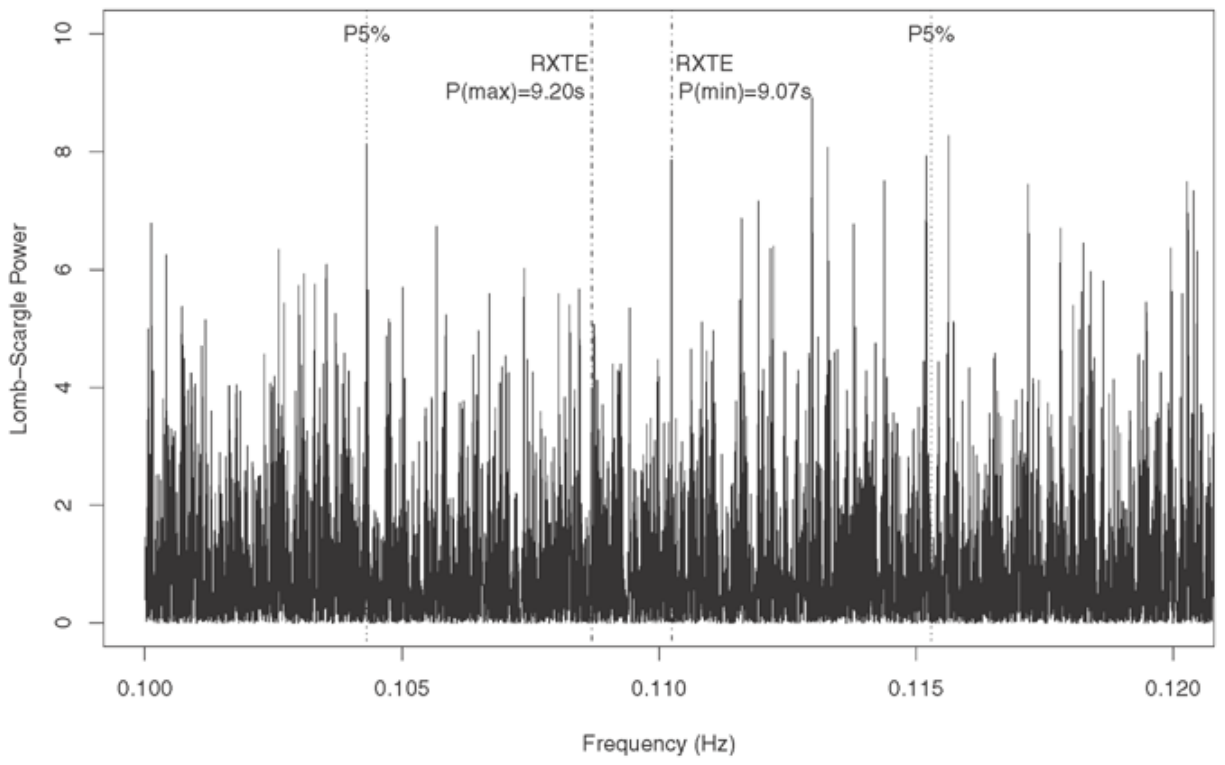}
\caption{{\bf  CXOU J4913.5-731138  = SXP9.13 This known HMXB pulsar was detected with $>$1000 counts, but pulsations were not seen. The power spectrum shows many noise peaks of similar size over the period range P$\pm$5\%. Further restricting the domain of interest to the period range seen by {\it RXTE} does not single-out any identifiable modulation.}}
\label{fig:sxp9.13}
\end{center}
\end{figure}

\begin{figure*}
\includegraphics[angle=0,width=7.5cm]{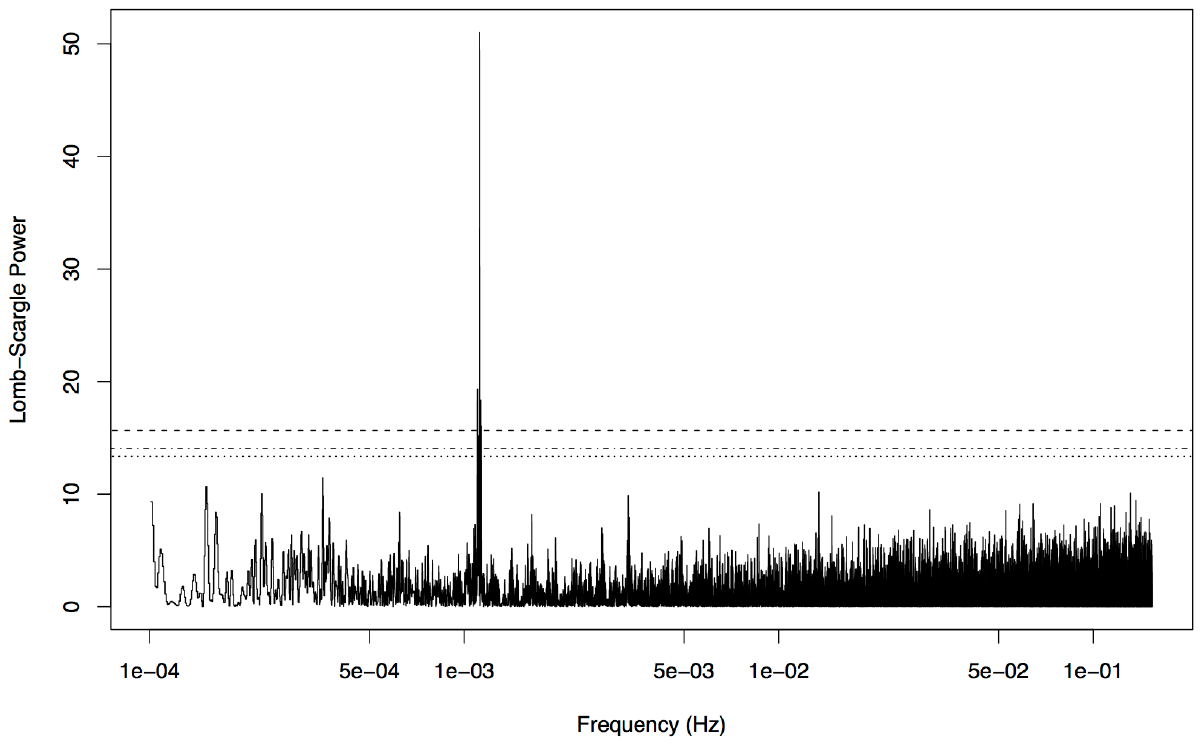}
\includegraphics[angle=0,width=7.5cm]{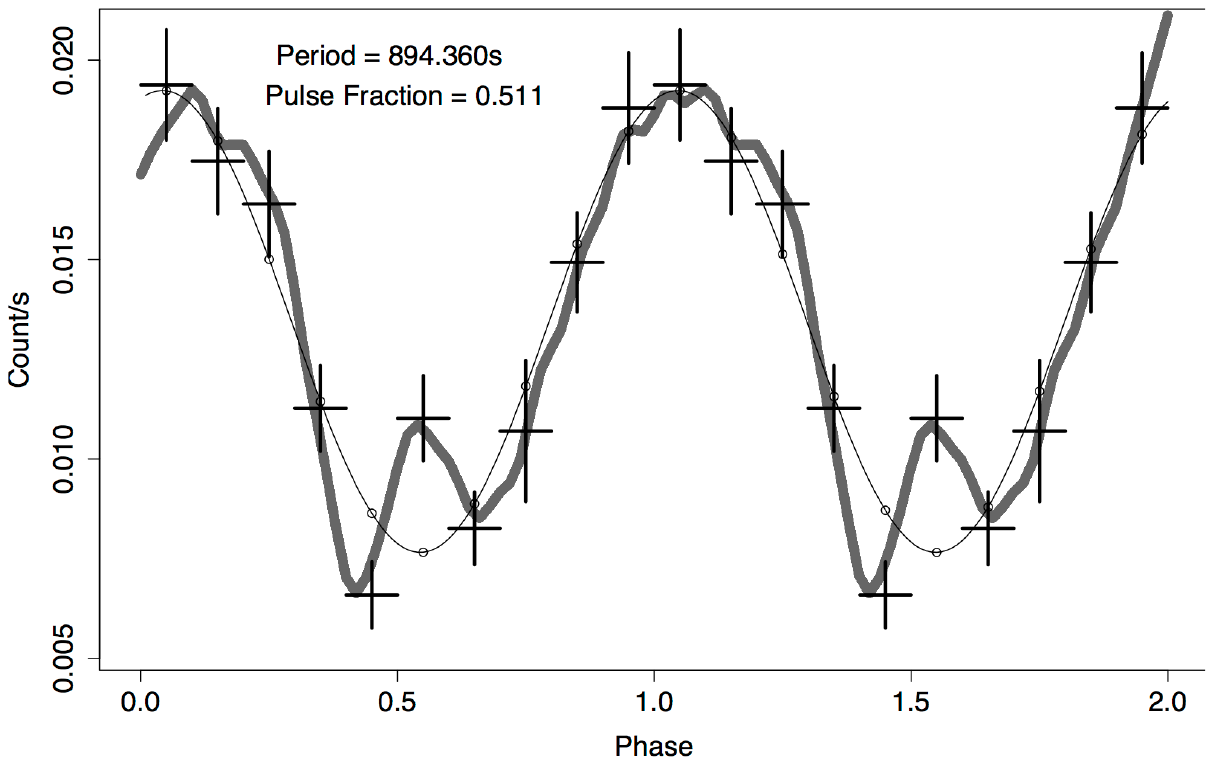}
\caption{{\bf SXP892. New long-period pulsar with P=894.36s and a sinusoidal profile.}}
\label{fig:sxp892}
\end{figure*}

\begin{figure*}
\includegraphics[angle=0,width=7.5cm]{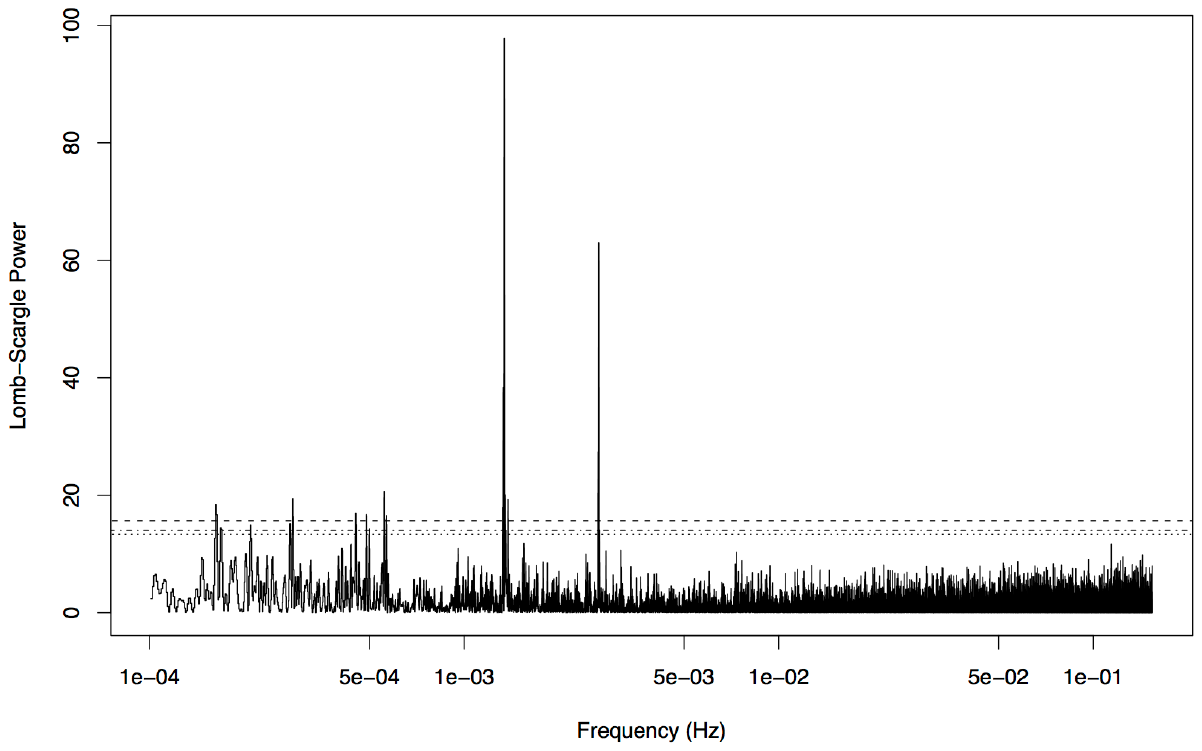}
\includegraphics[angle=0,width=7.5cm]{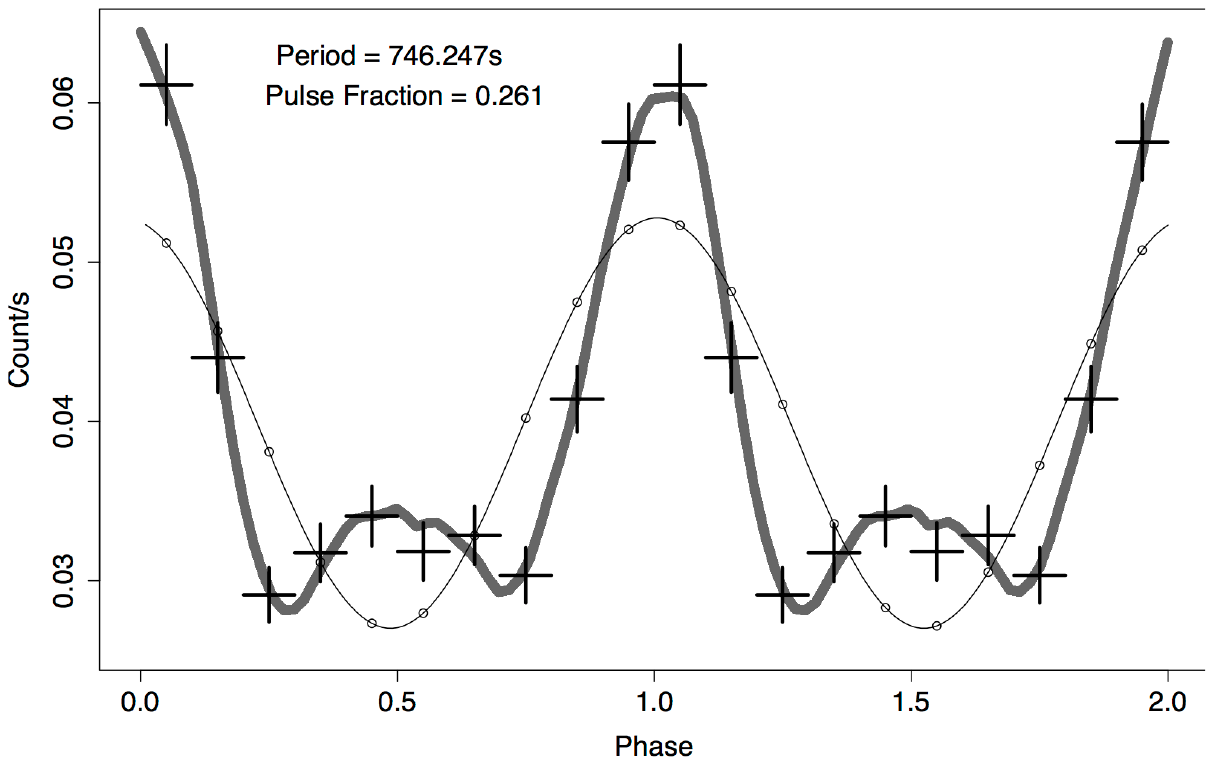}
\caption{{\bf  SXP756. Detected with Period=746.25s. The power spectrum shows a strong first harmonic, indicative of a non-sinusoidal modulation. The folded pulse-profile exhibits a distinct secondary peak 180$\degr$ out of phase with the narrow main pulse.   }}
\label{fig:sxp756}
\end{figure*}

\begin{figure*}
\includegraphics[angle=0,width=7.5cm]{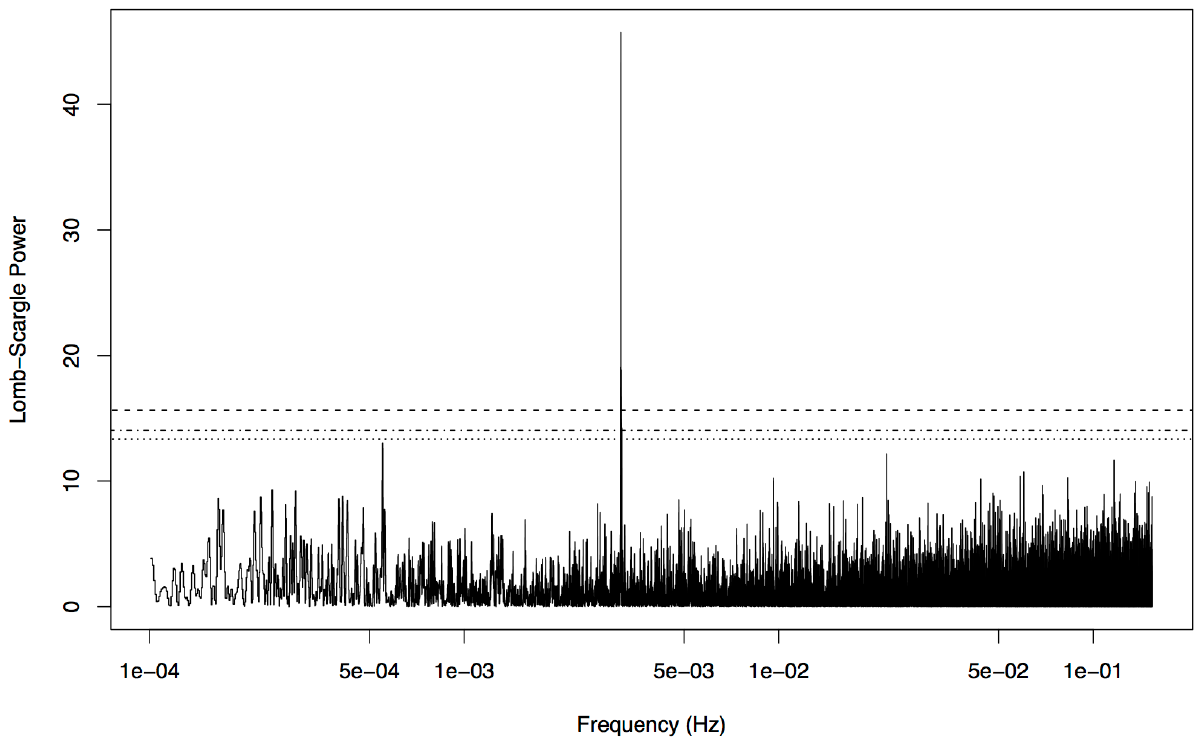}
\includegraphics[angle=0,width=7.5cm]{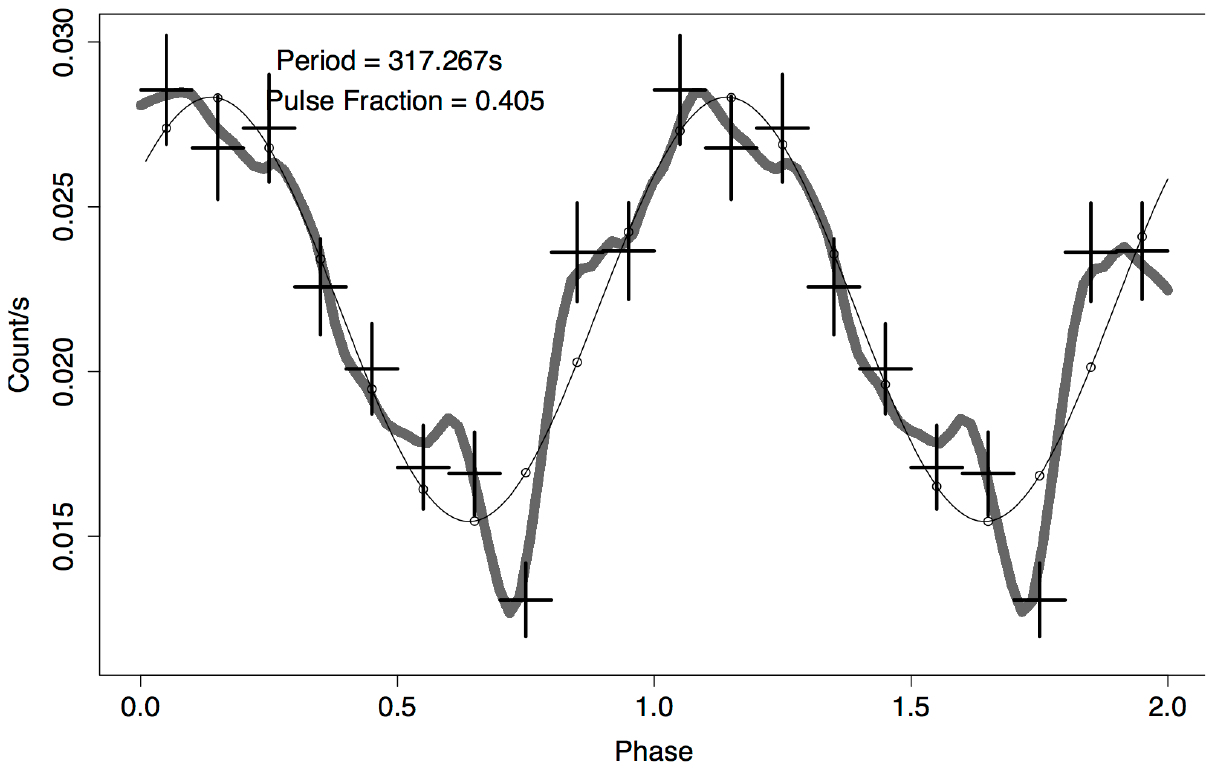}
\caption{{\bf  SXP323. Detected with P=317.3s at the lower boundary of the period range seen by RXTE. The pulse-profile is remarkably similar to that of SXP326.}}
\label{fig:sxp323}
\end{figure*}

\begin{figure*}
\includegraphics[angle=0,width=7.5cm]{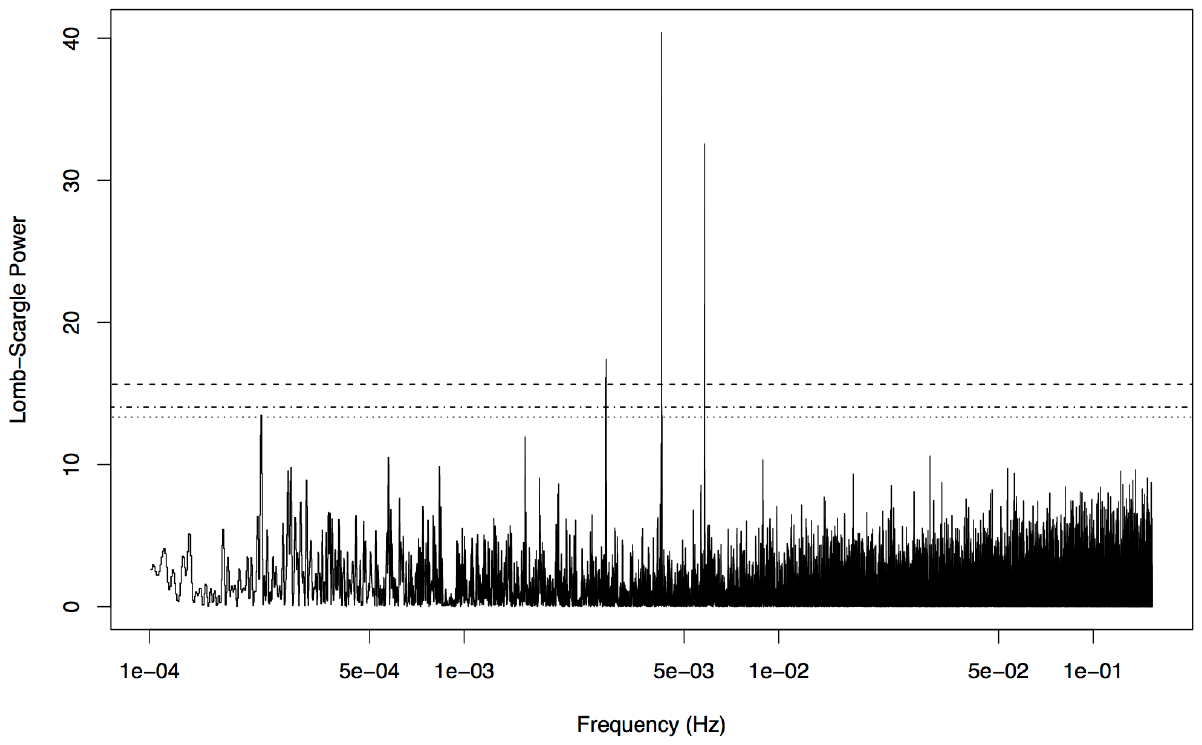}
\includegraphics[angle=0,width=7.5cm]{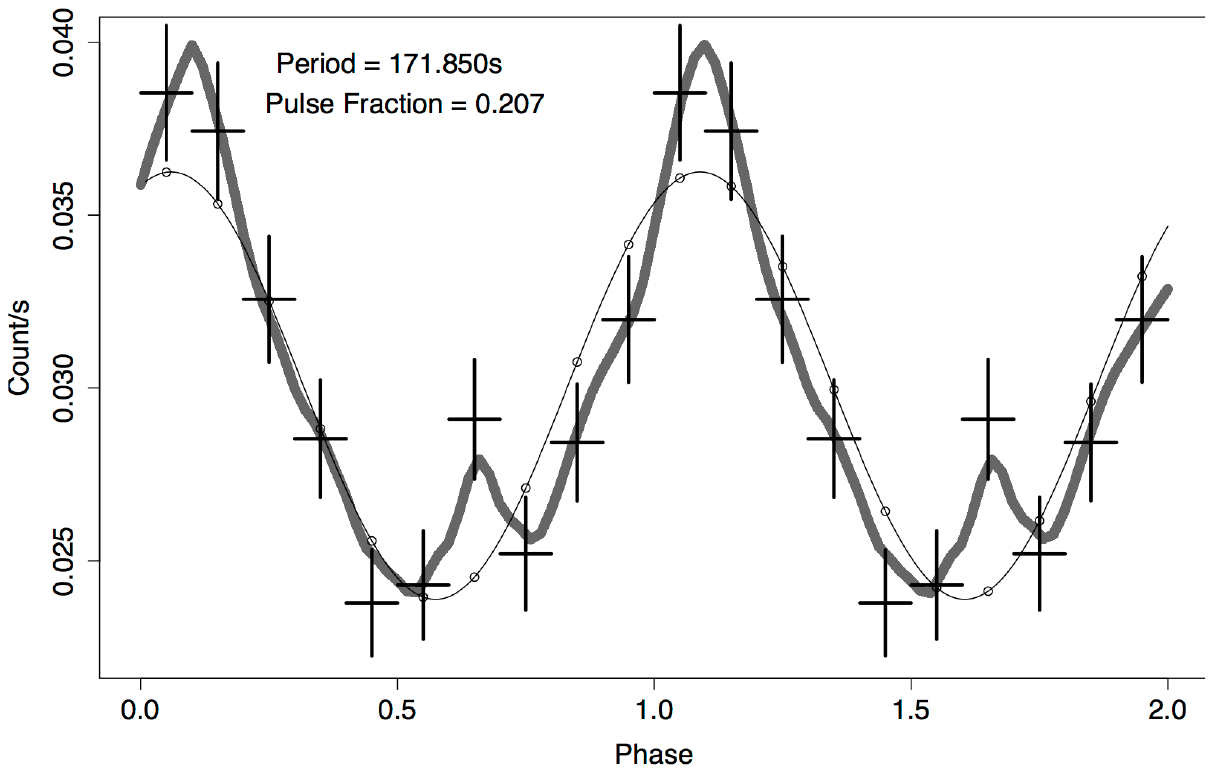}
\caption{{\bf  SXP172.  Detected with pulse period P=171.8s. The source lay close to the outer edge of the ACIS CCD, and the additional peak in the PDS at 235.8s is close to 1/3 the {\em Chandra} ACIS Z-axis dither period of 707s. It could alternatively be a second pulsar within the {\it Chandra} PSF. }}
\label{fig:sxp172}
\end{figure*}

\begin{figure*}
\begin{center}
\includegraphics[angle=0,width=15cm]{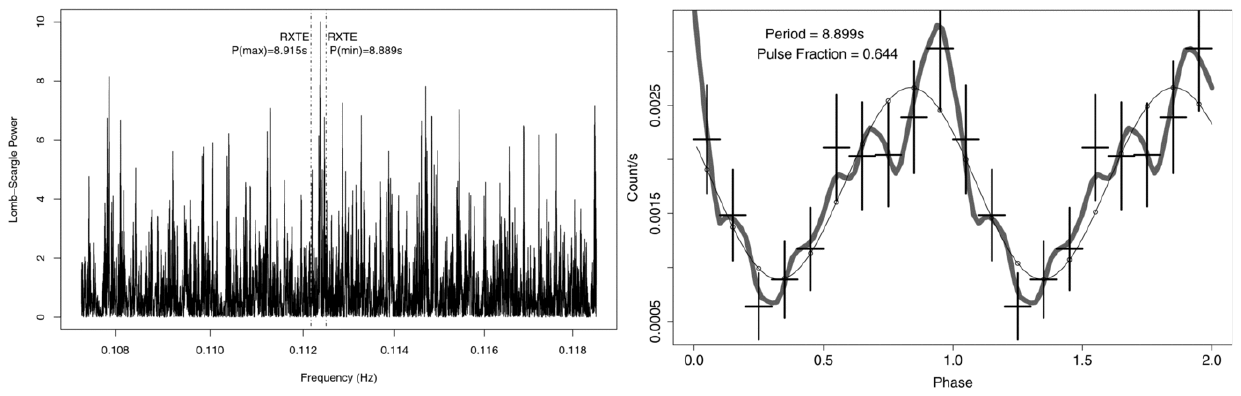}
\caption{{\bf  SXP8.88. The pulsar was detected at P=8.89909s, which is within the narrow range of pulse periods seen by \cite{galache2008} with RXTE, indicated by vertical lines. }}
\label{fig:sxp8.88}
\end{center}
\end{figure*}

\begin{figure*}
\includegraphics[angle=0,width=7.5cm]{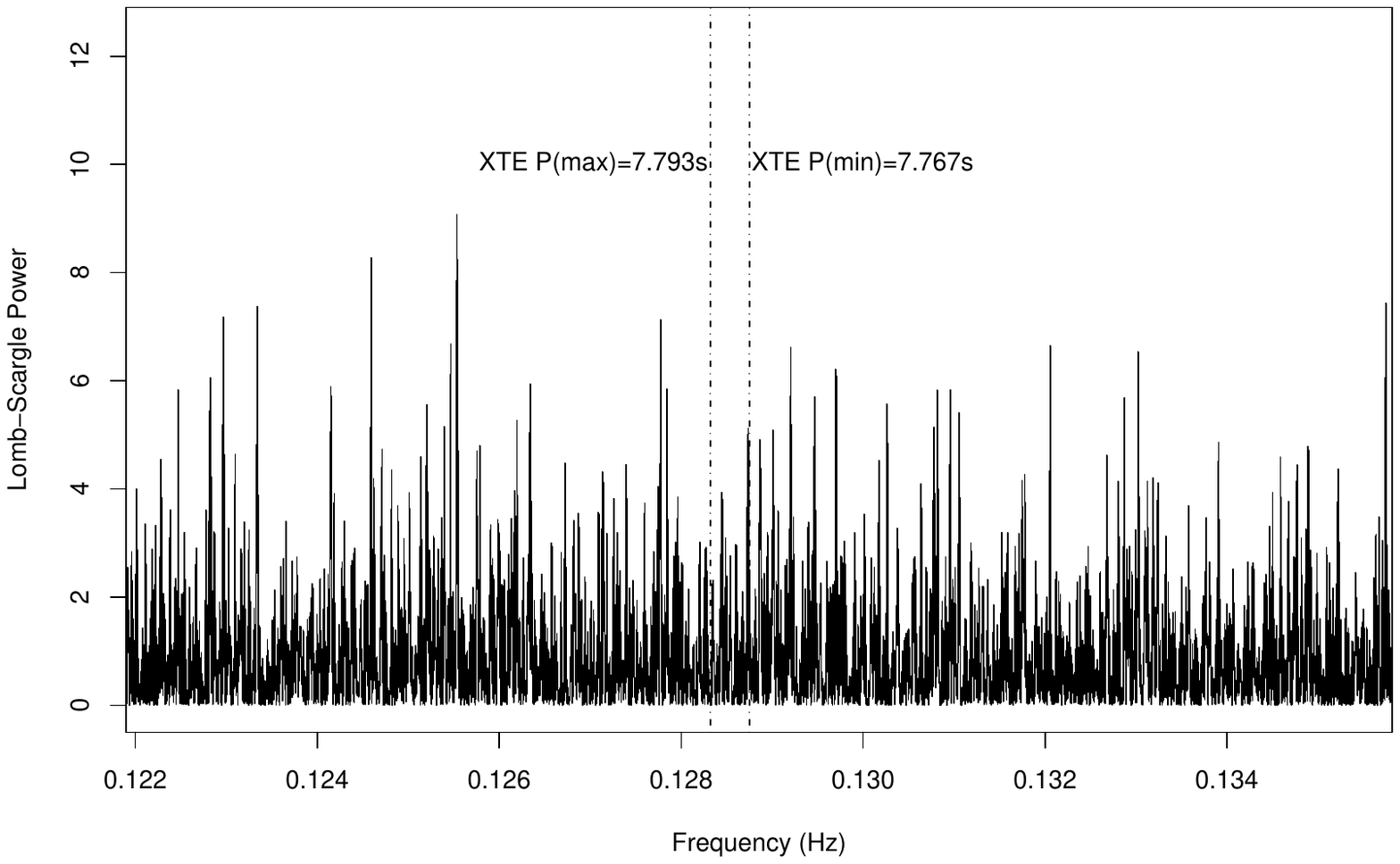}
\includegraphics[angle=0,width=7.5cm]{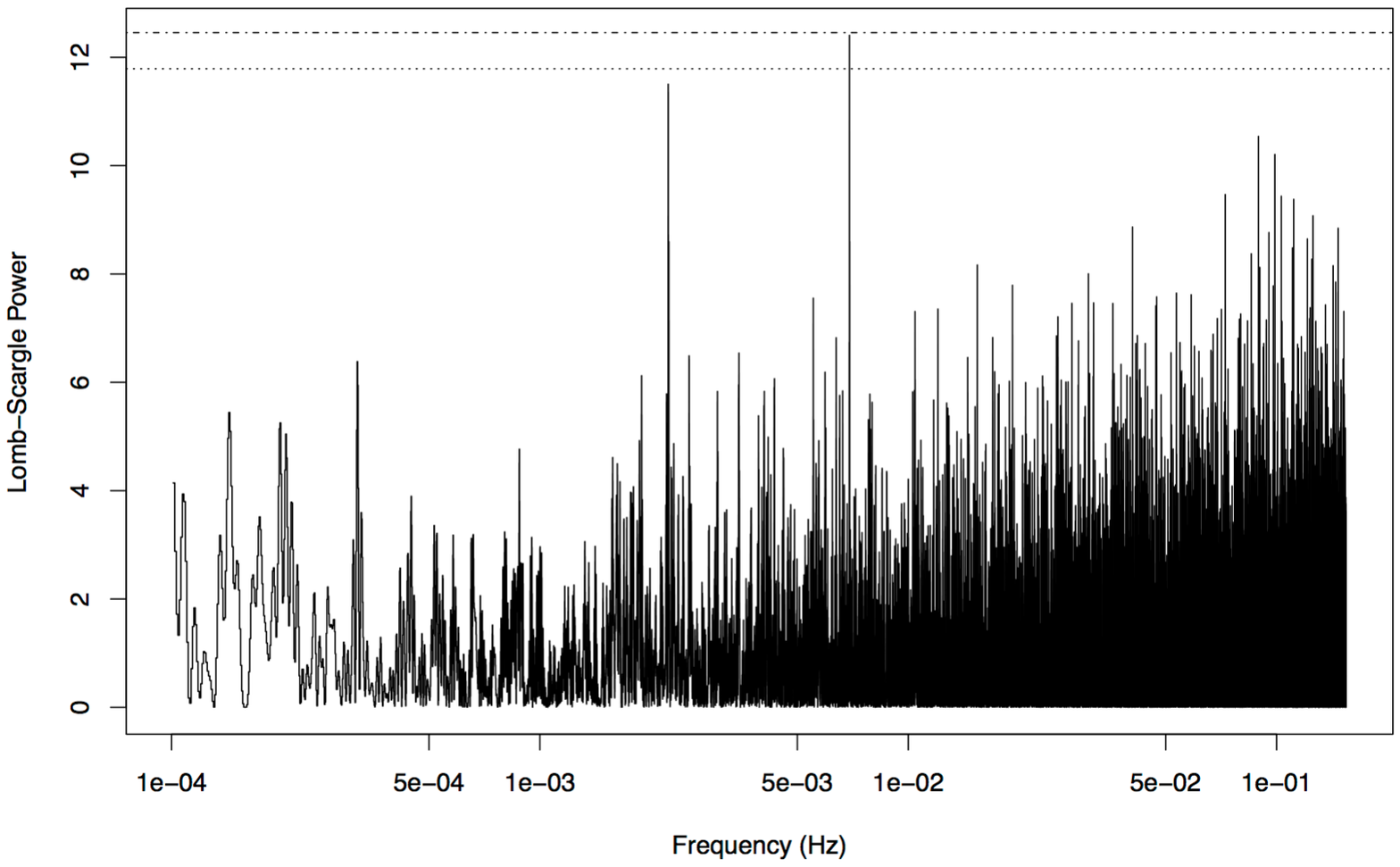}
\caption{{\bf  SMC X-3  =   CXOU J005205.6-722604 = SXP7.78.  Left Panel: This source shows no detectable pulsation within the range of the {\it RXTE} pulse-period history (bounded by vertical lines), the frequency range shown spans 7.78s$\pm$5\%. Right Panel: A blind-search of the light-curve produced a peak at 448s which is probably false, and does not reach the 95\% significance level derived by Monte Carlo.  }}
\label{fig:smcx3}
\end{figure*}

\begin{figure*}
\includegraphics[angle=0,width=7.5cm]{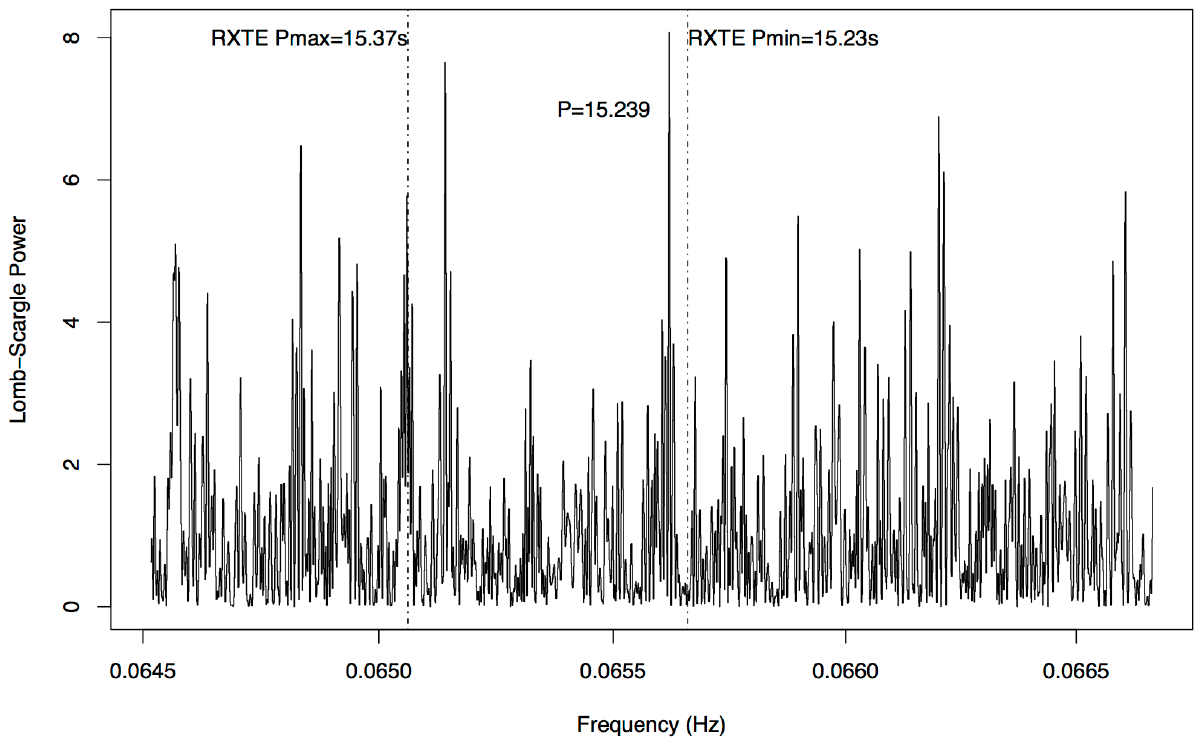}
\includegraphics[angle=0,width=7.5cm]{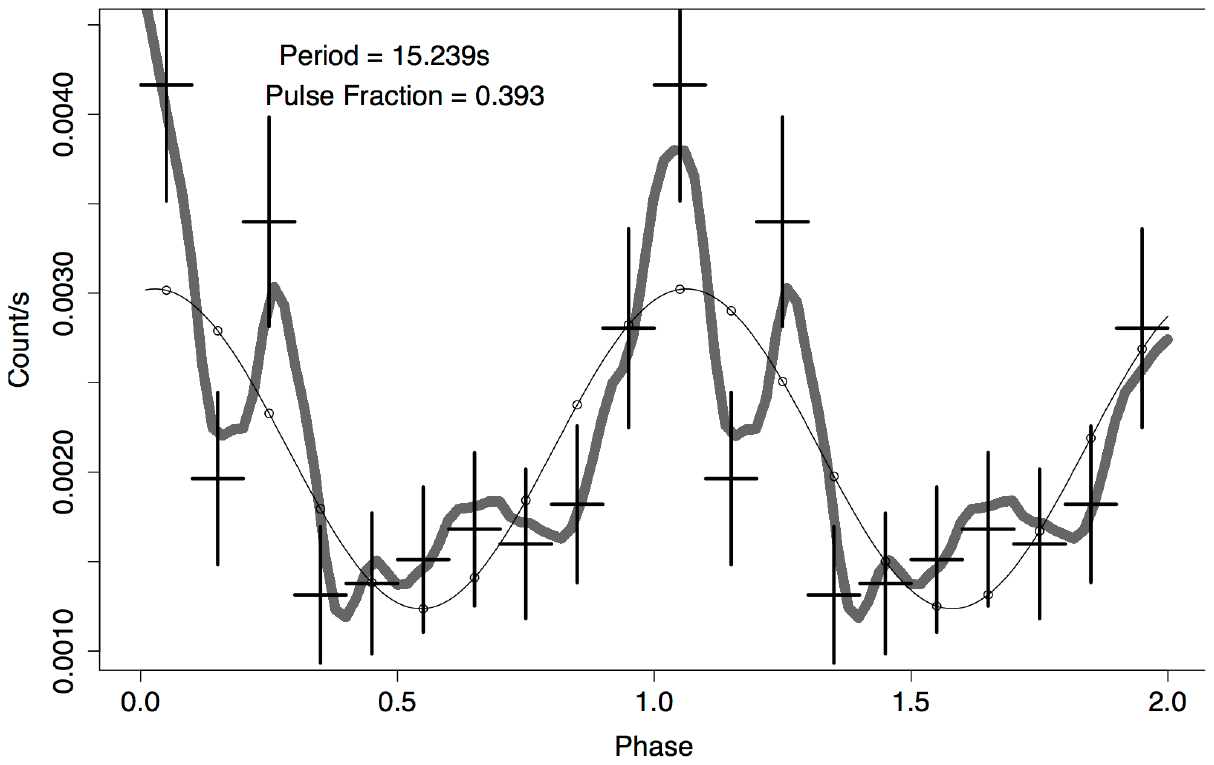}
\caption{{\bf  SXP15.3. We are able to identify weak pulsations thanks to accurate {\it RXTE} measurements of the pulse period. The range of periods seen by \cite{galache2008} with {\it RXTE} is indicated by 2 vertical lines. The {\it RXTE} P(min) value was recorded $\sim$110 days before the {\it Chandra} observation. Prior to this {\it RXTE} showed a 3-year long spin-up from P(max). The peak at P=15.239s is thus consistent with the most contemporaneous {\it RXTE} detection. More conservatively  the data may be interpreted as an upper limit on the pulsed flux.}}
\label{fig:sxp15.3}
\end{figure*}

\begin{figure}
\begin{center}
\includegraphics[angle=0,width=12cm]{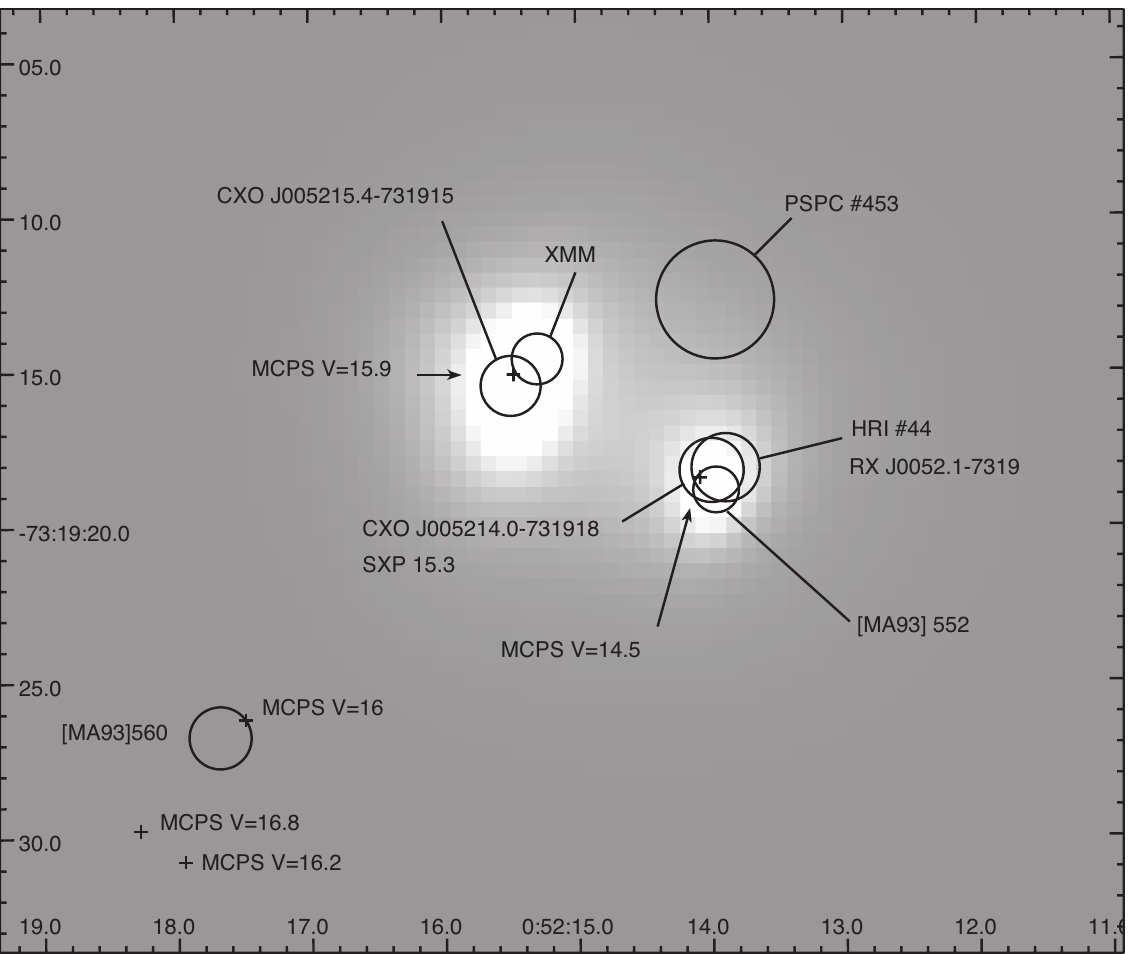}
\caption{{\bf  {\it Chandra} ACIS image of the X-ray Binary SXP15.3 (RX J0052.1-7319), and the nearby source (CXOU J005215.4-731915) which is a new HMXB candidate. Labeled circles indicate positional error radii for Chandra, {\it ROSAT} and {\it XMM-Newton} detections.  The {\it XMM-Newton} source was reported by \cite{haberl2008} who associated it with SXP15.3,  we find it is likely an independent discovery of the new HMXB candidate. Bright stars (V$<$17) in the MCPS catalog are indicated by crosses, and MA93 emission-line stars are shown as circles with 1" radius.  Both SXP15.3 and CXOU J005215.4-731915 have optical counterparts, while only 3 other bright stars appear in this 0.5 arcmin field, indicating the chance of random coincidence is small. }}
\label{fig:imagesxp15}
\end{center}
\end{figure}

\begin{figure*}
\includegraphics[angle=0,width=8cm]{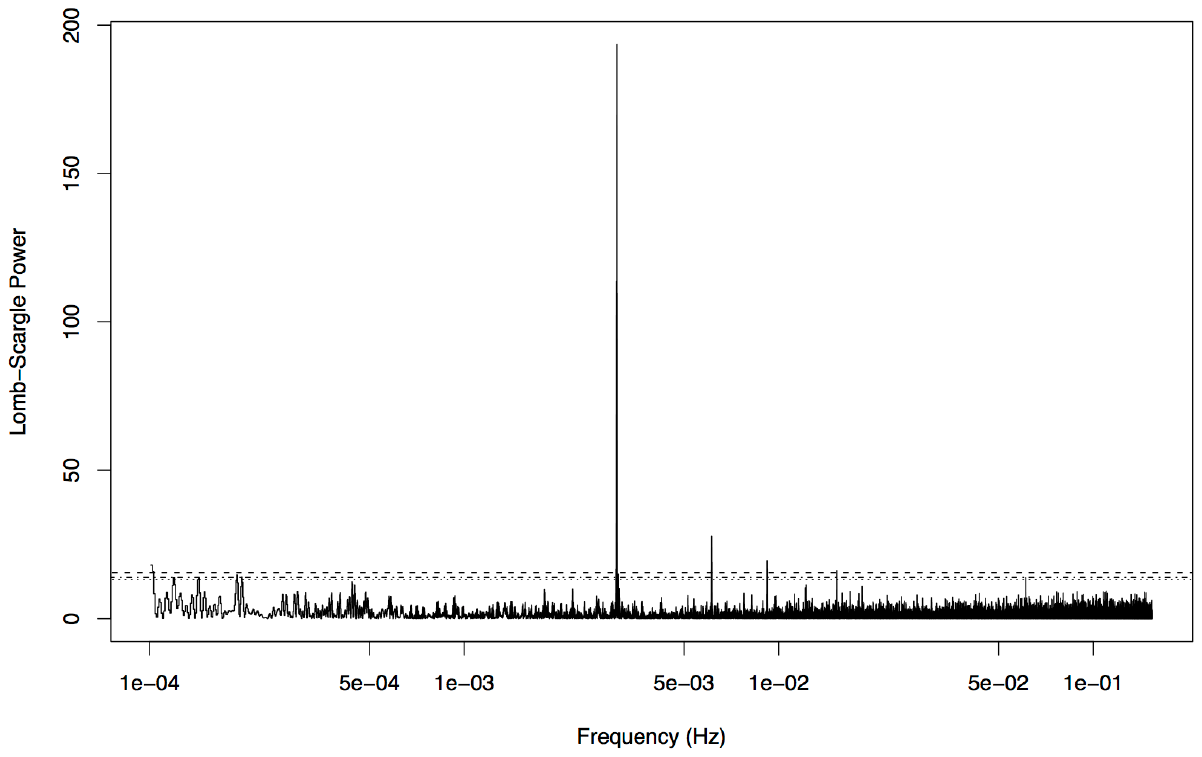}
\includegraphics[angle=0,width=7.5cm]{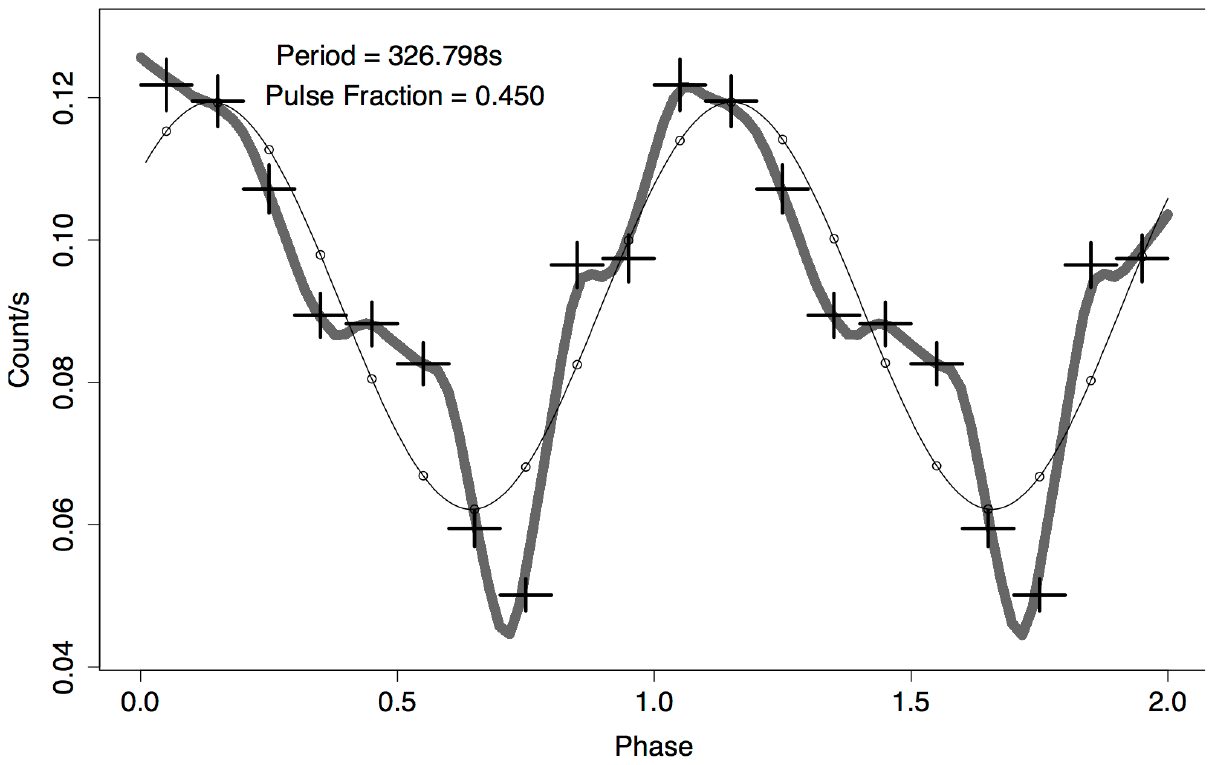}
\caption{{\bf  SXP326. New pulsar, P=326.8s. Several harmonics appear in the power-spectrum and the pulse-profile shows a high degree of structure. }}
\label{fig:sxp326}
\end{figure*}

\begin{figure*}
\includegraphics[angle=0,width=7.5cm]{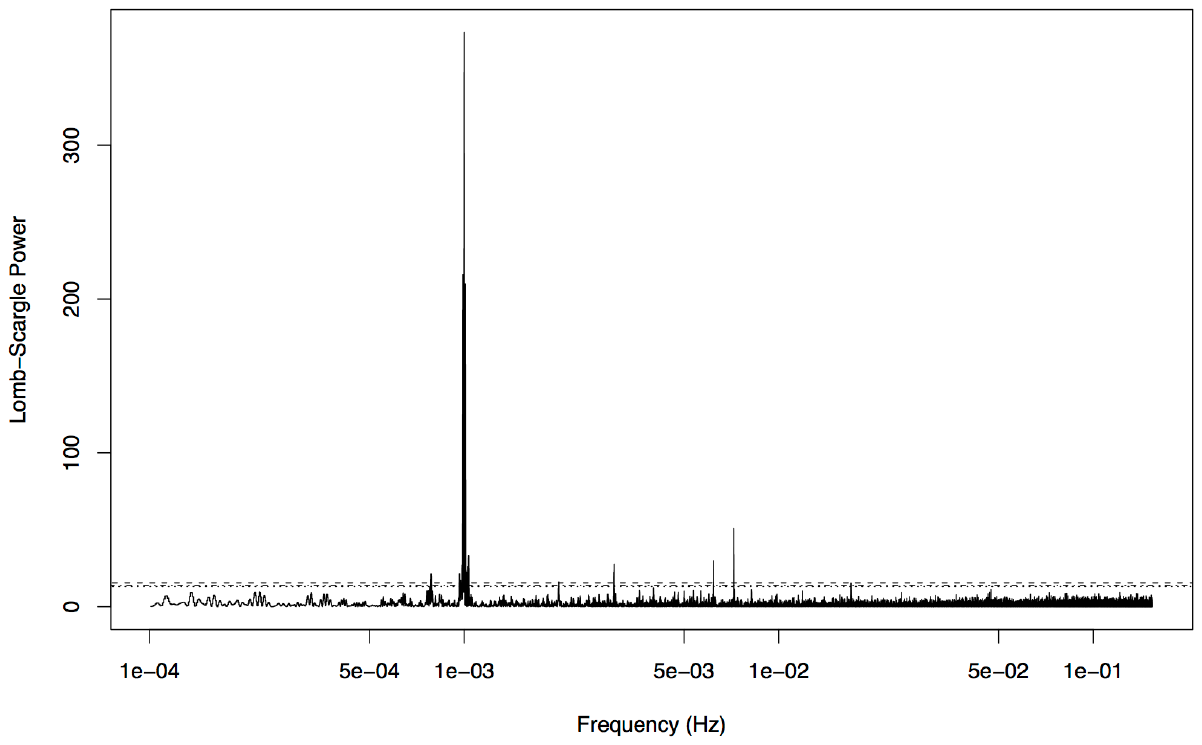}
\includegraphics[angle=0,width=7.5cm]{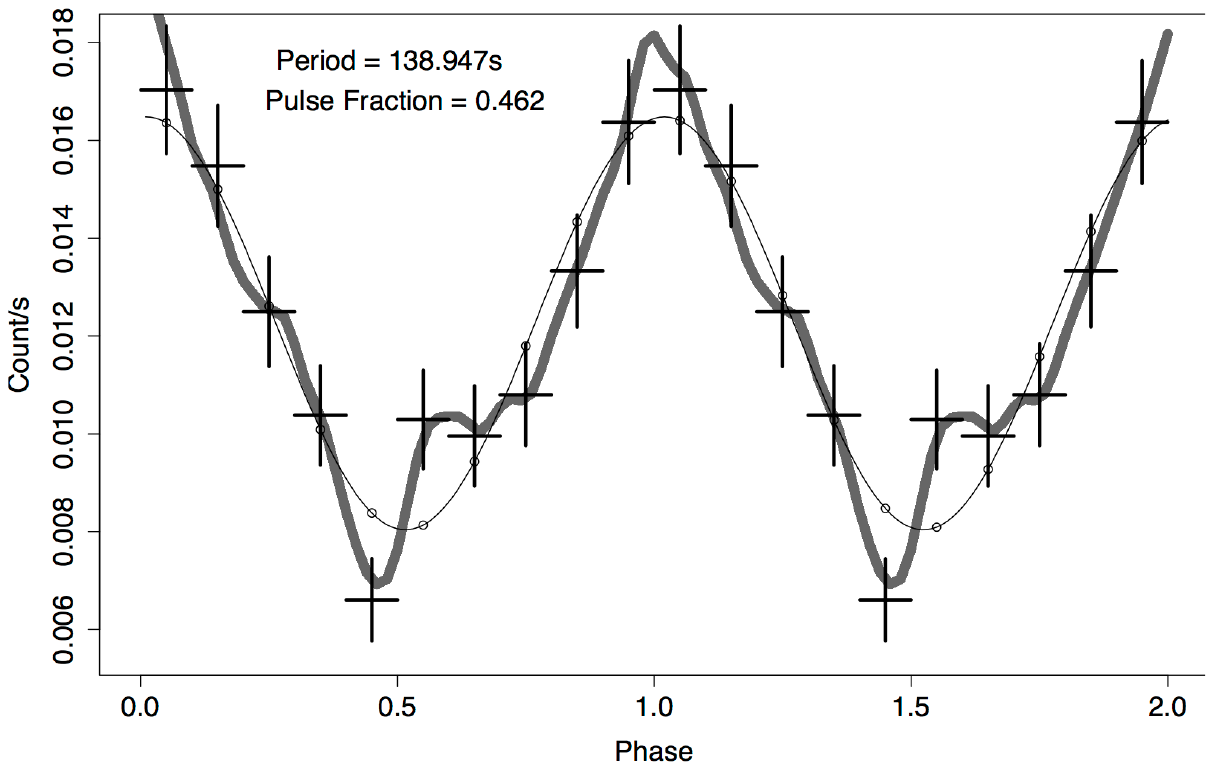}
\caption{{\bf  SXP138.  Detected pulse period = 138.9s. The source was close to a CCD boundary in the ACIS detector, causing the additional peaks in the PDS (e.g. at 1000.0 s) which are artifacts of  {\em Chandra}'s dither pattern.}}
\label{fig:sxp138}
\end{figure*}

\begin{figure*}
\includegraphics[angle=0,width=7.5cm]{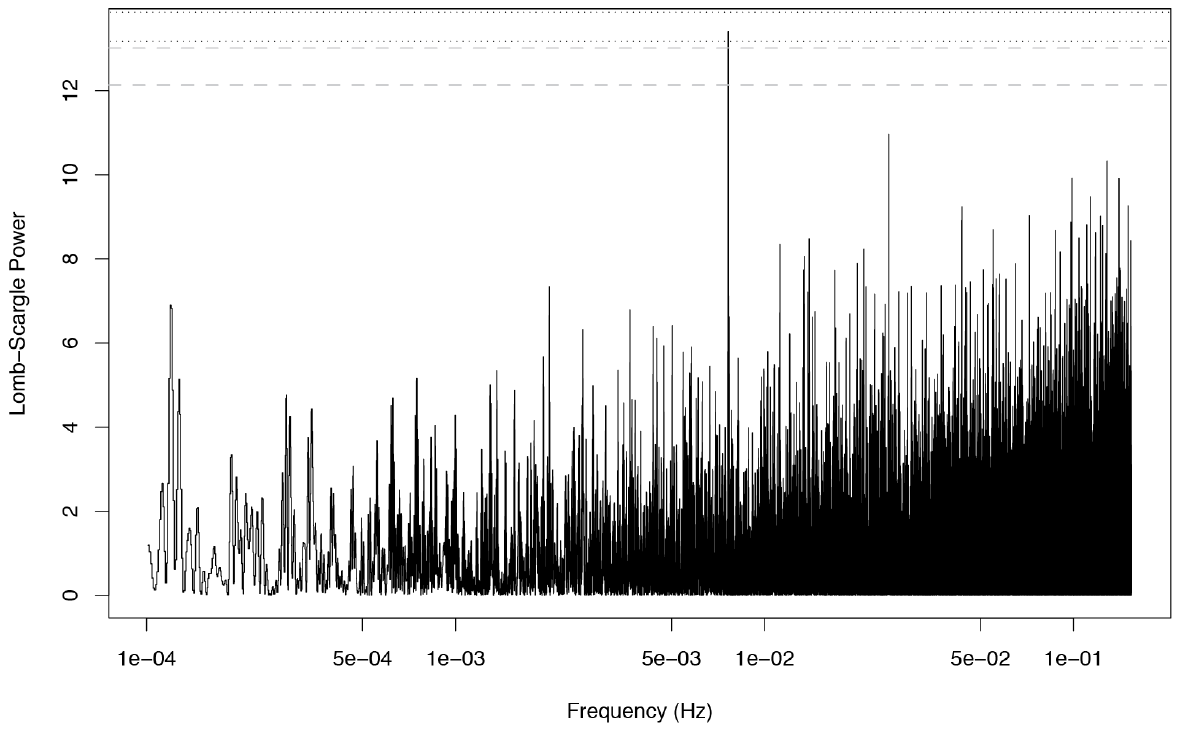}
\includegraphics[angle=0,width=7.5cm]{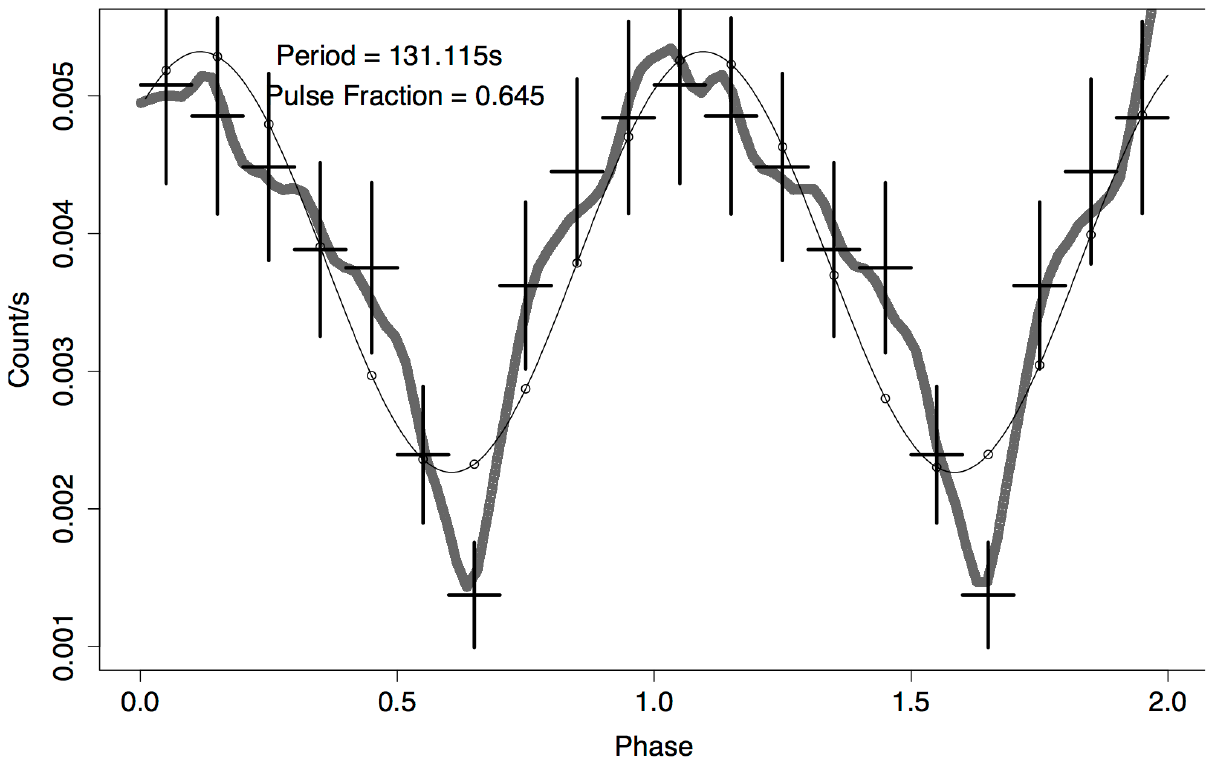}
\caption{{\bf  CXOU J005331.7-722240 exhibits a candidate period of P=131.11s. Detection significance levels 90\% and 95\% derived by Monte Carlo simulation for a blind search over the period range 6s-10,000s are indicated by green lines, while those from the formula of \cite{press1992} are in black.}}
\label{fig:p131}
\end{figure*}

\begin{figure*}
\includegraphics[angle=0,width=7.5cm]{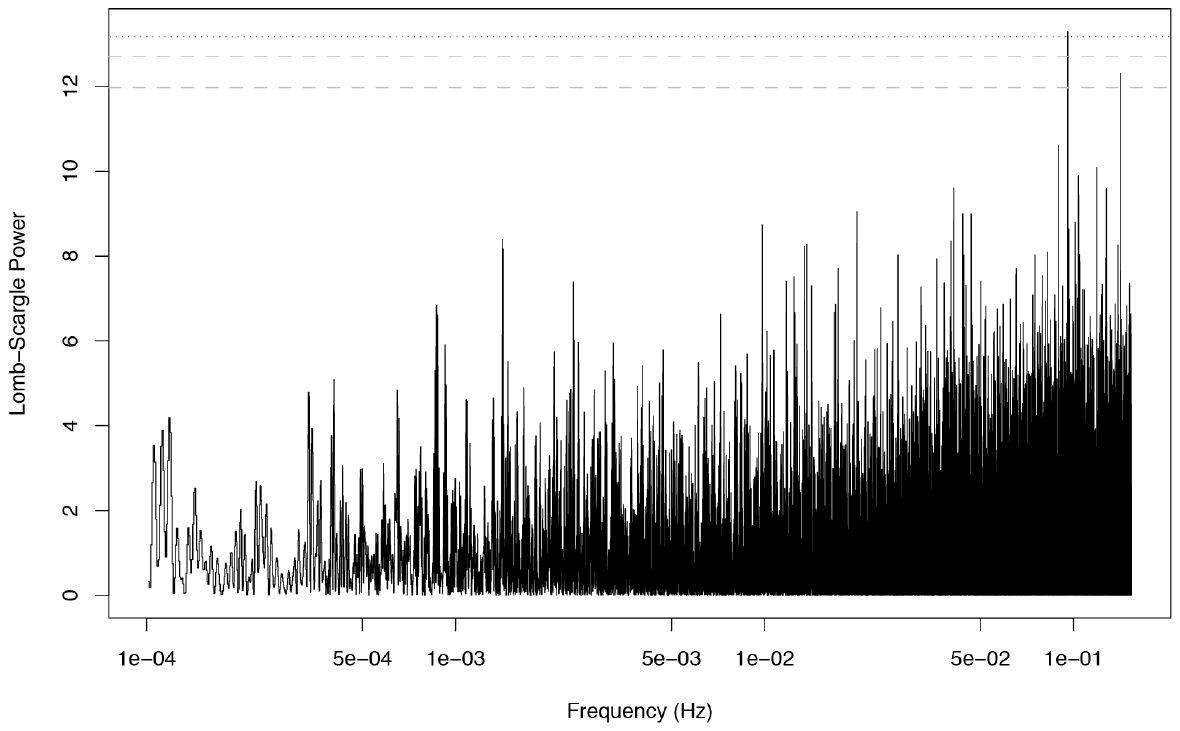}
\includegraphics[angle=0,width=7.5cm]{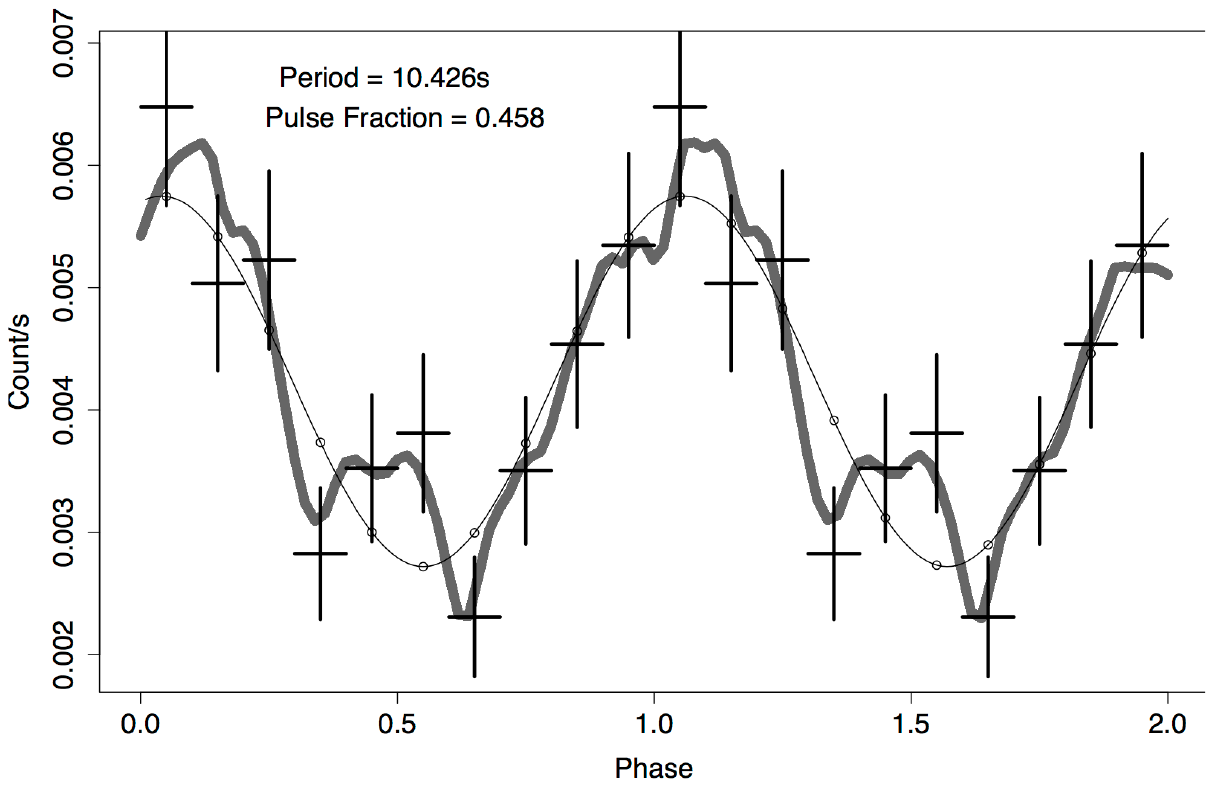}
\caption{{\bf  CXOU J005437.1-722637 exhibits a candidate period of  P=10.426s. Detection significance levels 90\% and 95\% derived by Monte Carlo simulation for a blind search over the period range 6s-10,000s are indicated by green lines, while the 90\% level from the formula of \cite{press1992} is the black dotted line.}}
\label{fig:p10.4}
\end{figure*}

\begin{figure*}
\includegraphics[angle=0,width=7.5cm]{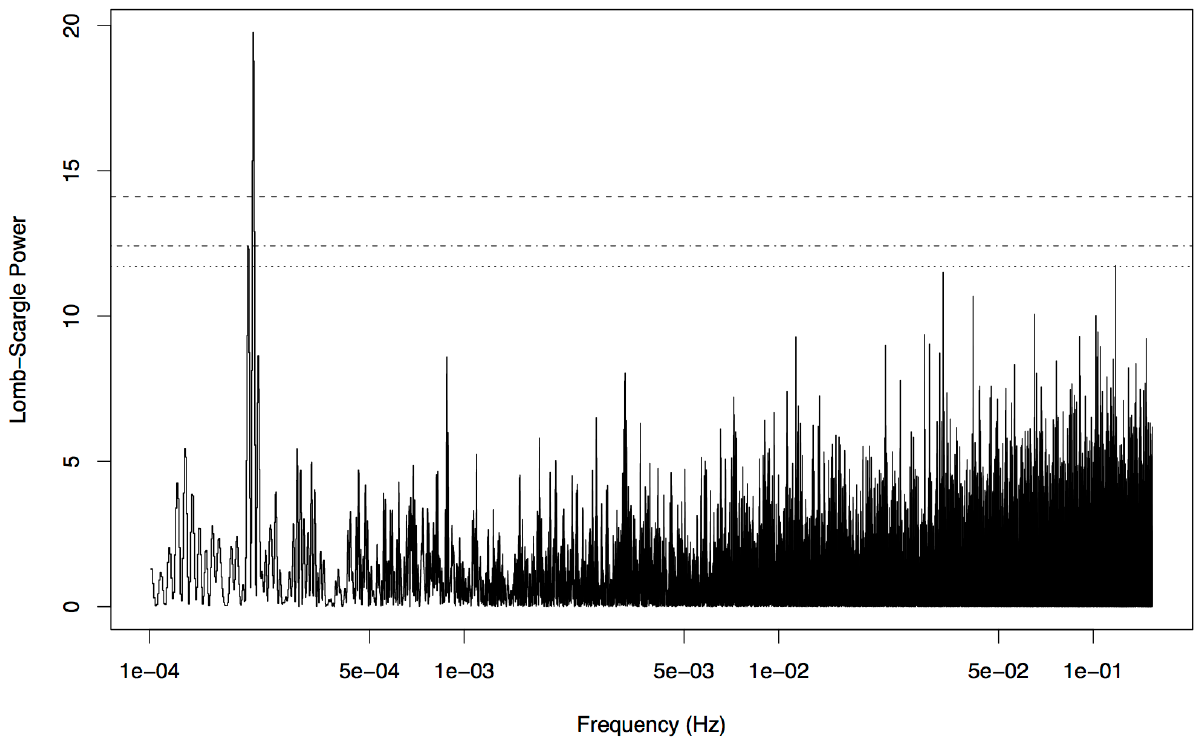}
\includegraphics[angle=0,width=7.5cm]{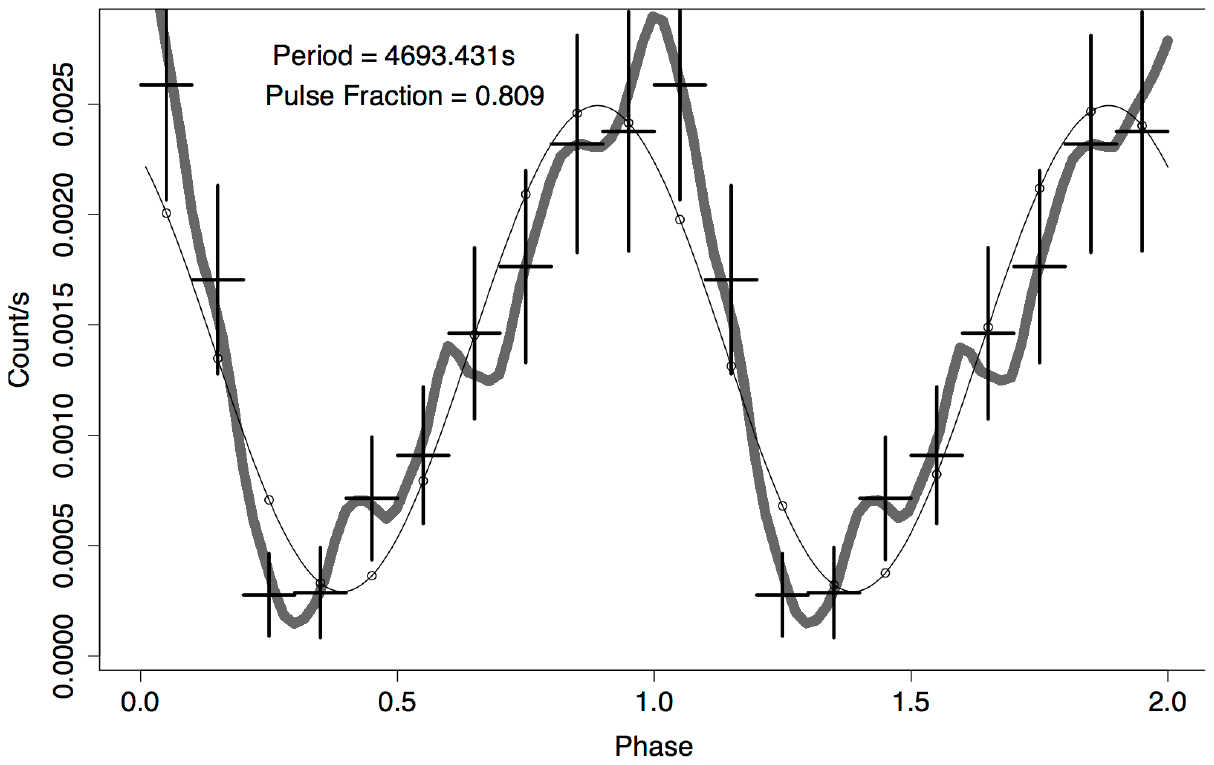}
\caption{{\bf  CXOU J005446.3-722523. The period detected at P=4693s exceeds the 99\% significance level derived by Monte Carlo, and is confirmed by other algorithms. The source has a V=15.36 optical counterpart making it a likely HMXB. If confirmed this would be the longest period pulsar known.    }}
\label{fig:sxp4693}
\end{figure*}

\begin{figure*}
\includegraphics[angle=0,width=7.5cm]{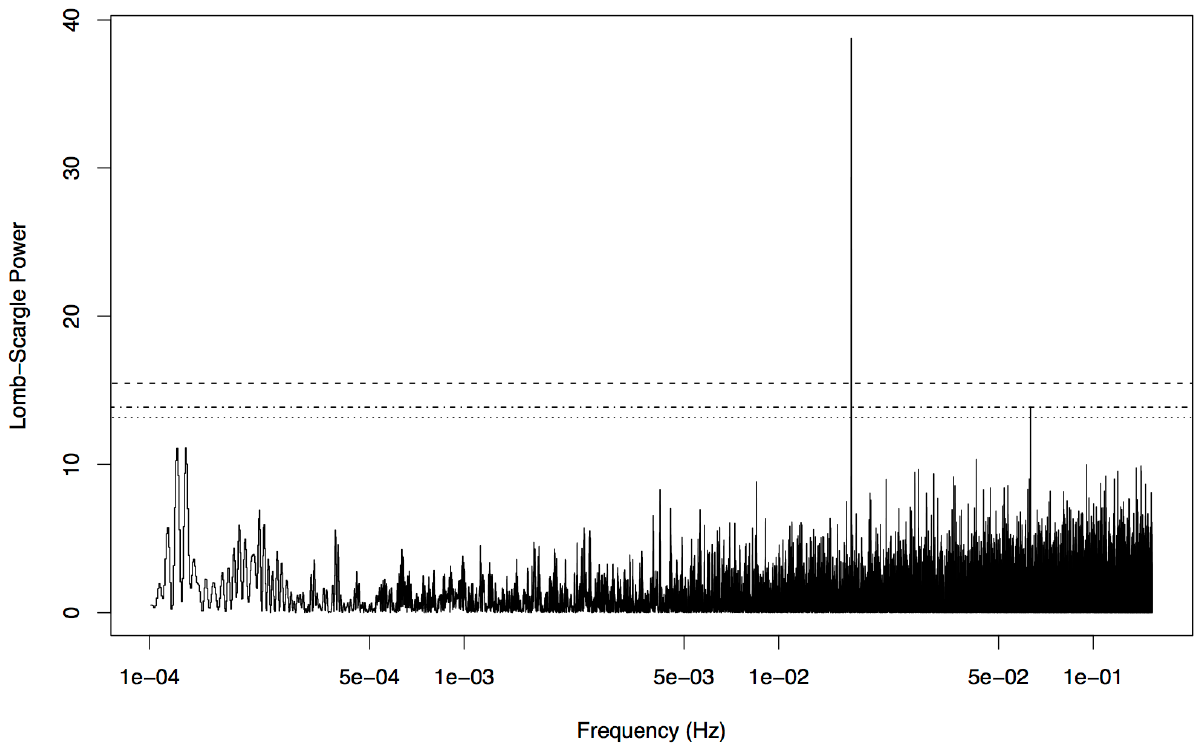}
\includegraphics[angle=0,width=7.5cm]{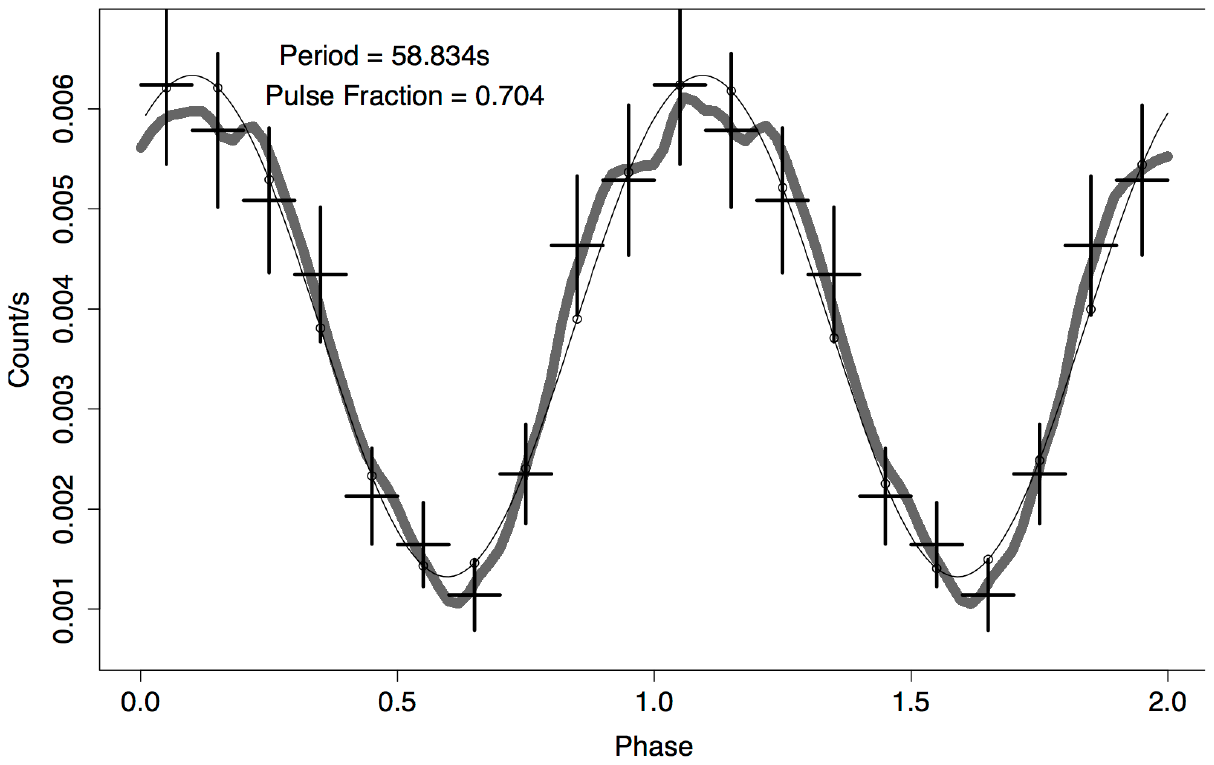}
\caption{{\bf SXP59. Pulse period was  P=58.8s . The profile is very close to perfectly sinusoidal although a harmonic is detected.  }}
\label{fig:sxp59}
\end{figure*}

\begin{figure*}
\begin{center}
\includegraphics[angle=0,width=16cm]{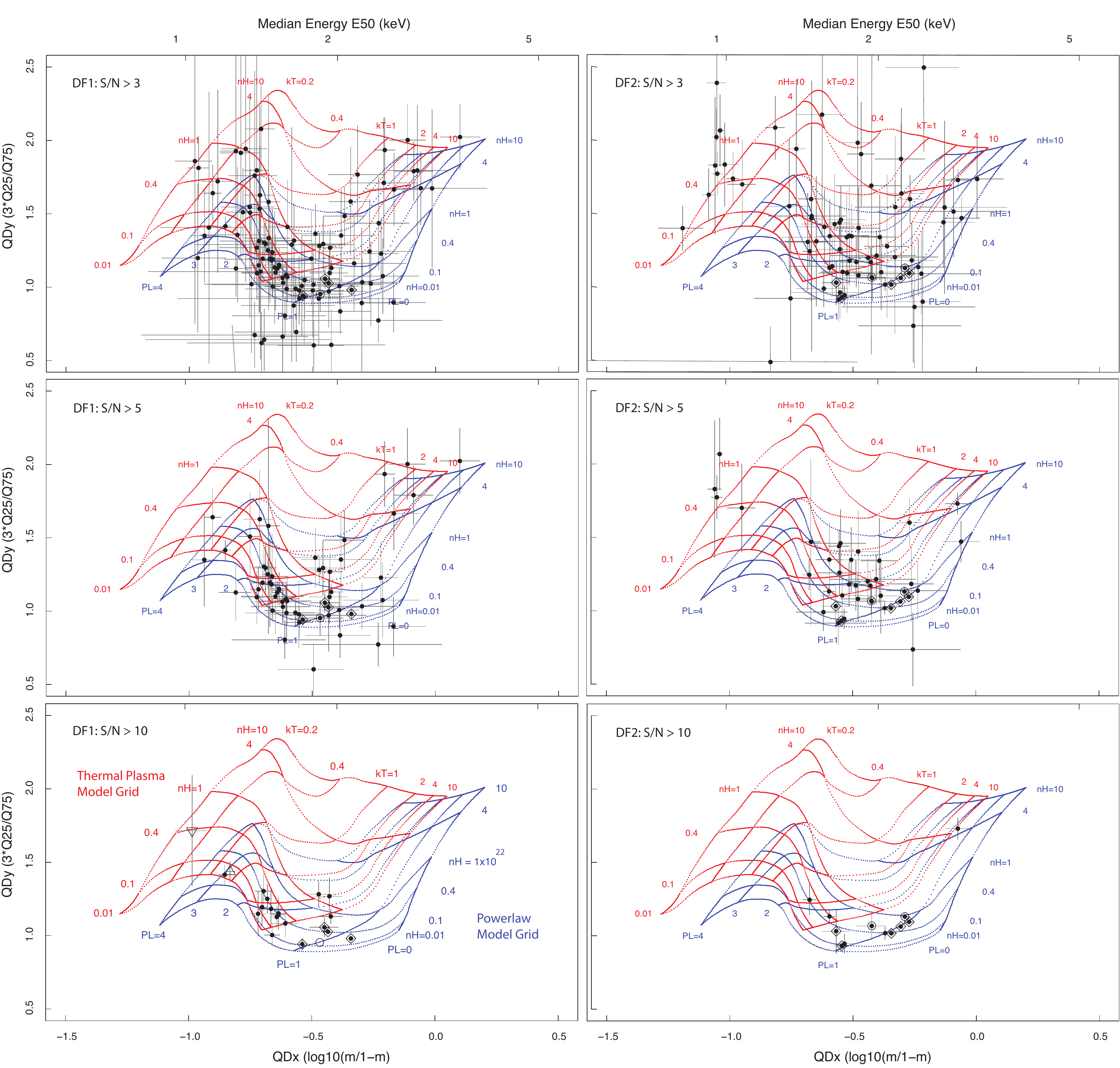}
\caption{{\bf Quantile Diagrams for sources with Signal-to-Noise Ratio of $>$3, $>$5, $>$10 (from top to bottom). Pulsars with detected pulsations in the data are indicated by filled black diamonds, those without detected pulsations are marked by open diamonds.   Spectral model grids are shown for absorbed power-law (Blue) and absorbed thermal bremsstrahlung (TB) (Red). The model grid-lines are labeled with power-law index $\Gamma$, column density $n_H$ in units of 10$^{22}$~cm$^{-2}$, and plasma temperature $kT$ in units of keV. The confirmed pulsars cluster along the $\Gamma$=1.0 contour and are mostly consistent with $N_{H}\leq$10$^{21}$\nh.  In the upper left are a distinct group of sources occupying the TB grid, these are stars with coronal activity. A known foreground flare star is shown by an up-triangle point during a flare and down-triangle in quiescence. }}
\label{fig:quantiles1}
\end{center}
\end{figure*}

\clearpage

\end{document}